\documentclass[aps,prb,superscriptaddress,superscriptaddress,floatfix,twocolumn,tightenlines]{revtex4-2}

\usepackage{bibunits}
\defaultbibliographystyle{apsrev4-2}

\usepackage{subcaption}
\usepackage{adjustbox}
\usepackage{appendix}
\usepackage{mathrsfs}

\makeatletter
\newif\ifsupplementtocentry
\newcommand{\supplementtableofcontents}{%
  \begingroup
  \supplementtocentryfalse
  \def\appendix{\supplementtocentrytrue}%
  \def\numberline##1{\makebox[2.5em][l]{##1}}%
  \def\contentsline##1##2##3##4{%
    \ifsupplementtocentry
      \def\supplement@entrytype{##1}%
      \def\supplement@sectiontype{section}%
      \ifx\supplement@entrytype\supplement@sectiontype
        \def\supplement@entrytitle{##2}%
        \def\supplement@contentstitle{\numberline {}Contents}%
        \ifx\supplement@entrytitle\supplement@contentstitle
        \else
          \par\noindent
          \hyperlink{##4}{##2}\hfill\hyperlink{##4}{##3}\par
        \fi
      \else
        \def\supplement@subsectiontype{subsection}%
        \ifx\supplement@entrytype\supplement@subsectiontype
          \par\noindent\hspace*{2.5em}%
          \hyperlink{##4}{##2}\hfill\hyperlink{##4}{##3}\par
        \fi
      \fi
    \fi
  }%
  \begin{center}
    \large CONTENTS
  \end{center}
  \vspace{1.5em}
  \@starttoc{toc}%
  \endgroup
}
\makeatother

\usepackage{graphicx}  
\usepackage{dcolumn}   
\usepackage{amssymb,amsmath}   
\usepackage{bm}        
\usepackage{url}
\usepackage{layout}
\usepackage{color}
\usepackage{epsfig}
\usepackage{booktabs}
\usepackage{float}
\usepackage{hyperref}
\hypersetup{colorlinks,allcolors=blue}
\usepackage{bm}
\usepackage{makecell}

\def\be{\begin{equation}} \def\ee{\end{equation}}
\def\bp{\begin{pmatrix}} \def\ep{\end{pmatrix}}
\def\bea{\begin{eqnarray}}
\def\eea{\end{eqnarray}}
\def\beaa{\begin{equation}\begin{aligned}}
\def\eeaa{\end{aligned}\end{equation}}
\def\ba{\begin{aligned}}
\def\ea{\end{aligned}}

\def\mb{\mathbf}

\def\g{\tilde{g}} 

\def\be{\begin{equation}}       \def\ee{\end{equation}}
\def\bea{\begin{eqnarray}}      \def\eea{\end{eqnarray}}
\def\bp{\begin{pmatrix}} \def\ep{\end{pmatrix}}
\def\beaa{\begin{equation}\begin{aligned}}
		\def\eeaa{\end{aligned}\end{equation}}

\begin{document}

\title{Space-time crystallography and symmetry-enforced dynamical phenomena in 2+1 dimensions}

\author{Chenhang Ke}
\affiliation{Department of Physics, Fudan University, Shanghai, 200433, China }
\affiliation{Department of Physics, Westlake University, Hangzhou 310024, Zhejiang, China}
\affiliation{New Cornerstone Science Laboratory, Department of Physics, School of Science, Westlake University, Hangzhou 310024, Zhejiang, China}

\author{Congjun Wu}
\email{wucongjun@westlake.edu.cn}
\affiliation{New Cornerstone Science Laboratory, Department of Physics, School of Science, Westlake University, Hangzhou 310024, Zhejiang, China}
\affiliation{Institute for Theoretical Sciences, Westlake University, Hangzhou 310024, Zhejiang, China}
\affiliation{Key Laboratory for Quantum Materials of Zhejiang Province, School of Science, Westlake University, Hangzhou 310024, Zhejiang, China}
\affiliation{Institute of Natural Sciences, Westlake Institute for Advanced Study, Hangzhou 310024, Zhejiang, China}


\date{\today}

\begin{abstract}
 The concept of space group has long served as a fundamental framework for describing the physical properties of crystalline materials, from electronic bands to photonic dispersions. 
Recent advances in spatiotemporal control, including laser-driven lattices, dynamic photonic and phononic crystals, and dynamic optical lattices, 
necessitate the study of symmetries that intrinsically intertwine space and time.   
Standard Floquet theory primarily organizes temporal periodicity and does not by itself provide a general framework for classifying these intertwined space-time symmetries. 
Space-time groups include novel intertwined nonsymmorphic spatial-temporal symmetries such as time-glide reflection and time-screw rotation. 
Here, we establish the complete crystallography of $(2+1)$-dimensional space-time crystals, systematically classifying all 275 space-time groups into 7 crystal systems, 14 Bravais lattices, and 72 arithmetic crystal classes.
We further develop the representation theory of space-time groups and show that these symmetries enforce dynamical phenomena without static counterparts, including handedness-flipping heterodyne responses and a novel “horizontal cone” arising from symmetry-enforced momentum degeneracy in space-time metamaterials. Our results establish space-time crystallography as a general framework for predicting and engineering symmetry-protected phenomena in driven quantum systems and time-modulated metamaterials.
\end{abstract}

\maketitle

\section{Introduction}

crystallography provides one of the most powerful organizing principles in modern physics. By classifying matter according to symmetry, space-group theory provides deep insights into electronic band structures, optical selection rules, and many functional properties of solids. The relation between crystalline symmetry and band topology has further enabled the discovery of topological crystalline materials and symmetry-protected semimetals \cite{fu2011topological,
parameswaran2013topological,
ando2015topological,young2015dirac,
benalcazar2017quantized}.
When the time-reversal (TR) operation is included in the classification, 
a space group is extended to a magnetic space group. 
The complete classifications of space and magnetic space groups make an exhaustive search for materials with targeted properties possible based on symmetry principles   \cite{vergniory2019complete, zhang2019catalogue, tang2019comprehensive}.

The concept of crystallographic symmetries can be generalized to space-time symmetries to describe driven dynamical systems \cite{xu2018space,morimoto2017floquet}.  
These emergent symmetries have recently attracted increasing attention in driven quantum and photonic systems \cite{peng2019floquet,chaudhary2020phonon,chen2021intertwined,mochizuki2020topological,gao2021floquet,jin2021floquet,chen2020topological,engelhardt2021,neufeld2019,lerner2023}, and are closely related to emerging symmetry-protected dynamical topological phases.  
Unlike the conventional Floquet theory, where spatial and temporal degrees of freedom are treated separately, space-time groups incorporate intertwined space-time symmetry operations, including nonsymmorphic symmetries such as time-glide reflection and time-screw rotation.
We refer to systems periodic in $d+1$ dimensions as dynamical crystals.
In general, however, the primitive space-time unit 
cell of a dynamical crystal is not a simple direct product of spatial and temporal periods.
Spatial translation symmetry may be absent at any 
fixed time, and temporal translation symmetry 
may be absent at any fixed spatial position.
The symmetries of each dynamical crystal form a space-time group, which is fundamentally distinct from an ordinary space group with just one more dimension, because space and time are nonequivalent in non-relativistic quantum mechanics.  

These developments raise a fundamental question: What are the crystallographic symmetries of systems ordered simultaneously in space and time? 
Conventional space and magnetic space groups are insufficient because they do not capture symmetries intertwining spatial transformations with fractional time translations. 
Such operations include nonsymmorphic space-time symmetries such as time-glide reflections and time-screw rotations, which can generate physical phenomena absent in static crystals \cite{peng2019floquet,chen2021intertwined,gao2021floquet,jin2021floquet,zhangProjectiveSpacetimeSymmetry2023}.

Here we formulate a systematic crystallographic framework for dynamical crystals, defined as systems with discrete periodicity in $(d+1)$-dimensional spacetime. 
We define space-time groups and perform a complete classification in $2+1$ dimensions: 
275 space-time groups are organized into
7 {\it space-time crystal systems}, 14 {\it space-time Bravais lattices},
and 72 {\it arithmetic crystal classes} (symmorphic crystal classes).
The representation theory of space groups can be extended to space-time groups.
It imposes constraints on dynamical response tensors, yielding handedness-flipping heterodyne responses. 
In space-time metamaterials, we propose the group of frequency-momentum as a generalization of the group of wavevector in static crystals, revealing a novel momentum degeneracy at the boundary of
the frequency-momentum Brillouin zone.


\section{space-time crystallography}
\subsection{Framework}

\noindent
Space-time groups, which are generalizations of space groups, describe the discrete space-time symmetries of time-dependent crystalline systems.
The $d$-dimensional space groups are discrete subgroups of the Euclidean group $E^d$ consisting of all length-preserving transformations, such as point group operations and translations.
Similarly, space-time groups are discrete subgroups of the extended Euclidean group $E^d\times E^1$, where $E^1$ consists of time translation and time-reversal operations.
Throughout this article, we adopt the convention that bold characters denote spatial vectors and spatial rotations, while unbold characters represent space-time actions and vectors.

We start by defining an element $\tilde{g}$ in the continuous  extended Euclidean group $E^d \times E^1$.
The action of $\tilde g$ on a space-time vector $(\mb{r},  t)$ is 
\beaa
\tilde{g} \cdot (\mb r , t) = (\mb R\cdot \mb r + \mb u, s t + \tau),
\eeaa
where $\mathbf{R}$ is a spatial point-group operation, $\mb u$ is a spatial translation, $\tau$ is a time translation, and $s=\pm 1$, with $s=-1$ indicating time-reversal operation. 
We separate $\tilde{g}$ into two parts and denote this space-time operation as 
\begin{equation}
    \tilde{g} = (g|u),
\end{equation}
 where $g$ stands for the point group operation, or, a magnetic point group operation for $s=\pm 1$, respectively, and $u$ represents the space-time translation with spatial and temporal components represented by $\mathbf{u}$ and $\tau$, respectively. 

We denote the $(d+1)$-dimensional space-time group by $G_{st}$.
Similar to the fact that a space group takes its lattice translation group as the invariant subgroup, $G_{st}$ always contains the space-time translation group as its invariant subgroup denoted by $T_L$.
$T_L$ is generated by $d+1$ space-time unit vectors spanning a Bravais lattice in the $(d+1)$-dimensional space–time.

To systematically organize and classify space-time groups, we draw upon concepts from crystallography.
Let us recall the organizing principle of space groups for comparison. 
The 230 space groups are classified into 7 crystal systems — the cubic, tetragonal, orthorhombic, monoclinic, triclinic, trigonal, and hexagonal ones; 32 crystal point group symmetries (geometric classes); 
14 types of Bravais lattices; 73 symmorphic space groups (arithmetic crystal classes).
When considering the $(2+1)$-dimensional space–time groups, the temporal dimension plays a special role which needs to be treated with caution because of the non-equivalence between space and time in non-relativistic quantum mechanics. 

To outline the organizing principle, we first describe the space-time Bravais lattice and space-time crystal system.
Please note that a Bravais lattice is only a special case of a crystal in which each lattice point is featureless. 
In addition to the space-time translation symmetry $T_L$, a space-time Bravais lattice also possesses magnetic point group symmetries.
Each space-time crystal system could contain different types of space-time Bravais lattices if they share the same set of magnetic point group symmetries.
Conversely, a magnetic point group may be compatible with a set of space-time Bravais lattices, including both the primitive type and various centered ones.

Next we introduce the arithmetic crystal class of space-time crystals.
A space-time crystal structure is obtained by decorating the lattice points with identical objects.
These objects typically possess only a subgroup of the magnetic point symmetry of the underlying space-time {\it Bravais lattice}. 
The combination of the crystal’s actual magnetic point group and its Bravais lattice defines a {\it space-time  arithmetic crystal class}. 

Each space-time arithmetic crystal class contains a set of non-equivalent space-time crystal structures.
If the magnetic point group $M$ is also the actual symmetry group of the space-time crystal, then it is a  {\it space-time symmorphic crystal}, and the corresponding space-time group is defined as a symmorphic one. 
A space-time arithmetic crystal class typically also contains {\it space-time nonsymmorphic crystals}.
Such a crystal is not invariant under the magnetic point-group operation alone but is invariant under a combined operation $(g|u)$, where $u$ is a fractional space-time translation in terms of the space-time unit vectors. 

Different space-time groups belonging to the same space-time arithmetic  crystal class are organized as a group using group cohomology: The symmorphic space-time group corresponds to the identity of the cohomology group, and each nonsymmorphic one maps to its element labeled by the fractional translation $u$. 
The product of two elements in the cohomology group follows the addition of fractional translations.

\subsection{  Classification of the $(2+1)$D space-time groups}


\begin{figure*}
\includegraphics[width=1\linewidth]{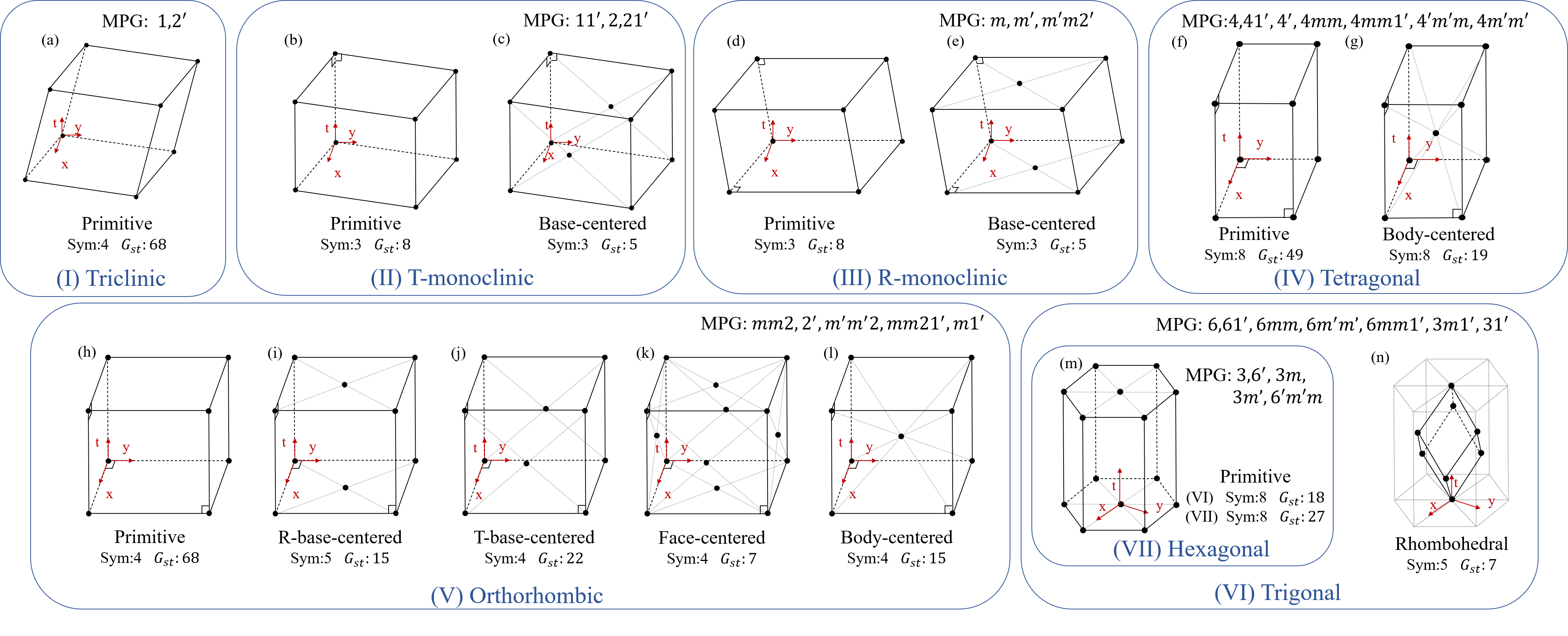}
\caption{List of fourteen $(2+1)$D space-time Bravais lattices grouped into seven crystal systems. 
The red arrows represent the space-time coordinates, $x$, $y$ and $t$.
The magnetic point groups (MPG) belonging to different crystal systems are listed at the top of each box. 
The numbers of space-time groups ($G_{st}$) and the symmorphic ones ($\mathrm{Sym}$) associated with each Bravais lattice
are listed.
The 3D monoclinic crystal system splits into the T- and R-monoclinic crystal systems in $2+1$ dimensions. 
The orthorhombic crystal system features two different base-centered Bravais lattices depending on whether the centering point lie in a spatial plane or a space-time plane.
The hexagonal primitive lattice is able to exhibit MPGs from both crystal systems (VI) and (VII).}
\label{fig.Bravais_lattice}
\end{figure*}

\noindent
Following the procedure outlined above, we classify the $(2+1)$D space-time groups starting from the magnetic point groups.
The $(2+1)$D space-time Bravais lattices are analyzed and the group cohomology technique is applied to construct $(2+1)$D space-time groups.
The main results are presented in Fig. \ref{fig.Bravais_lattice} and the detailed analysis is given in the Supplemental Material (S. M. I,II). 

There exist 7 space-time crystal systems as summarized in Fig. \ref{fig.Bravais_lattice}.
Although the total number of crystal systems equals that of the 3D case, there exist significant differences: 
The cubic lattice system does not exist in $2+1$ dimensions since the lengths of spatial and temporal directions cannot be compared. 
Furthermore, there exist two non-equivalent monoclinic space-time crystal systems in $2+1$ dimensions.
In monoclinic crystal systems, the $c$-axis is perpendicular to the $ab$-plane, while the $a$ and $b$ axes are non-orthogonal.
Depending on whether the $c$-axis is chosen along the temporal direction or a spatial direction, we obtain the T-monoclinic or R-monoclinic space-time crystal systems, respectively. 
In each of the T- and R-monoclinic space-time crystal systems, there exist both primitive and centered space-time Bravais lattices as shown in Fig. \ref{fig.Bravais_lattice} (II) and (III), respectively. 

Although both the $(2+1)$D  space-time crystal systems and their 3D 
counterparts include the orthorhombic system, they possess different types of Bravais lattices. 
In 3D orthorhombic crystal systems, there exist primitive, base-, face- and body-centered Bravais lattices.
As for the $(2+1)$D space-time case, the base-centered Bravais lattice is further divided into R-base-centered and T-base-centered ones.
In the R-base-centered case, center points are  added in the spatial plane, and in the T-base-centered case, center points are added in the spatial-temporal plane as shown in Fig. \ref{fig.Bravais_lattice} (V).

\paragraph{Time-glide reflection}
In a space group, a glide reflection refers to a symmetry operation that combines a mirror reflection with a fractional translation in space parallel to the mirror plane. 
However, in space-time crystals, this concept can take on new forms.
In 1+1D, the fractional translation can be directed along the time axis, leading to a symmetry operation known as time-glide reflection symmetry \cite{xu2018space}.
Furthermore, in $2+1$ dimensions, glide reflections can involve mixed space-time fractional translations, where the translation component has both spatial and temporal contributions.
For example, a time-glide reflection operation $(m_x|T_y^{\frac{1}{2}}T_t^{\frac{1}{2}})$ consists of a mirror  reflection $m_x$ reversing the direction of 
the $x$-axis with a half-period translation in both space and time. 

\paragraph{Glide time-reversal}
The time-reversal operation $m_t$ could be combined with a fractional spatial translation to form a symmetry operation.
In magnetic space groups for static crystals, say, a N\'eel ordered antiferromagnet, such a symmetry also exists. 
For example, $(m_t|T_x^{\frac{1}{2}})$ represents $m_t$ followed by a half unit vector translation along the $x$-axis, $T_x^{\frac{1}{2}}$.
The difference is that in space-time crystals, the reference time point for time-reversal operation typically can only be chosen discretely. 

\paragraph{Time-screw rotation}
In conventional space groups, a symmetry operation of screw rotation exists only in 3 dimensions, which consists of a rotation combined with a fractional translation along the rotation axis.
In $(2+1)$D space-time crystals, the counterpart to a screw rotation is termed a time-screw rotation, which consists of a rotation in the 2D plane combined with a fractional translation along time. 
For instance, a time-screw rotation, 
$(R_{\pi/2}|T^{1/4}_t)$, describes an operation of rotation through the angle of $\pi/2$ accompanied by a time translation of one quarter of the period. 

We have established the fundamental group-theoretic framework for classifying and understanding 
space-time crystals in $2+1$ dimensions.
The complete classification yields 275 space-time groups, 203 of which are nonsymmorphic.
They are organized into seven crystal systems.
In comparison with the 3D space groups, the cubic crystal system is absent while the monoclinic crystal system splits into two due to the non-equivalence between spatial and temporal directions. 
Novel space-time nonsymmorphic symmetries include time-glide reflection and time-screw rotation. 
The power of this new formalism will be demonstrated through 
two distinct physical applications below.

\section{ Symmetry constraints in the theory of dynamical response}

The consequences of space-time symmetry can be explored using pump–probe spectroscopy: a pump field coherently drives the system out of equilibrium, and the response to a subsequent weak probe is measured.
Unlike static systems, space-time crystals exhibit heterodyne responses, whose signal frequencies differ from the probe frequency.
The space-time symmetries provide guidance on the nature of these response functions. 
Previous works have investigated the selection rules for laser fields that possess space-time symmetries \cite{neufeld2019,lerner2023,engelhardt2021}.

However, the tensor structures of the response functions of space-time symmetric dynamical systems have not yet been systematically investigated.
Distinct from the response tensors for static systems, which are constrained solely by the (magnetic) point group, the nonsymmorphic space-time symmetries determine
the structure of the heterodyne responses.
To illustrate, we study the current response of a pumped space-time crystal possessing the time-screw symmetry $(R_{2\pi/3}|T_t^{1/3})$ to a probe electric field.
The response exhibits a handedness-flipping behavior, as dictated by the space-time representation theory. 
Such a symmetry imposes the following constraint on the Hamiltonian:
\begin{equation}
\begin{aligned}
H(\mathbf{r},t) = H(R_{\frac{2\pi}{3}}\mathbf{r}, t + T/3),
\end{aligned}
\end{equation}
which indicates that the system at time $t$ is identical to the system at $t+\frac{T}{3}$ after the rotation $R^{-1}_{\frac{2\pi}{3}}$.
The response current is related to the probe electric field via the retarded conductivity tensor $\sigma^{\mathrm{ST}}_{R,ij} (t,t')$, defined by
\begin{equation}
\begin{aligned}
J_i(t) = \int_{-\infty}^{\infty} dt'
\sigma^{\mathrm{ST}}_{R,ij}(t,t') E_j(t'),
\end{aligned}
\end{equation}
where causality requires $\sigma^{\mathrm{ST}}_{R,ij}(t,t') = 0$ for $t < t'$.
This conductivity kernel satisfies the periodicity $\sigma^{\mathrm{ST}}_{R,ij}(t,t') = \sigma^{\mathrm{ST}}_{R,ij}(t+T,t'+T)$.
For a monochromatic probe electric field $\mb{E}(t) = \tilde{\mb{E}}(\omega)e^{-i\omega t}+\tilde{\mb{E}}(-\omega)e^{i\omega t}$, it is convenient to study the conductivity tensor in the frequency domain,
 \begin{equation}
    \begin{aligned}
        \tilde{J_i} (\omega+n\Omega) = \tilde{\sigma}^{\mathrm{ST}}_{ij}(n\Omega,\omega)\tilde{E_j}(\omega),
    \end{aligned}
    \label{eq:conductivity0}
\end{equation}
where $\Omega = 2\pi/T$ is the driving frequency and $n\Omega$ is the shifted frequency representing the heterodyne
response at $n\neq 0$. 
The time and frequency domain conductivity tensors are related by
\begin{equation}
\sigma^{\mathrm{ST}}_{R,ij}(t,t') = \sum_n \int d\omega \ \tilde{\sigma}^{\mathrm{ST}}_{ij}(n\Omega,\omega) e^{-i(n\Omega+\omega)t + i\omega t'},
\end{equation}
in which $\omega$ is the frequency conjugate to
$t^\prime-t$.
Response currents can appear at frequencies shifted by integer multiples of $\Omega$ relative to the probe frequency $\omega$.



Now consider how the conductivity tensor transforms 
under the time-screw rotation, $(R_{2\pi/3}|T_t^{1/3})$.
As shown in the Methods, it transforms as a direct product representation of $\mathbf{\tilde{J}}(\omega+n\Omega)$ and $\mathbf{\tilde{E}}^*(\omega)$, yielding
\begin{equation}
\begin{aligned}
\tilde{\sigma}'^{\mathrm{ST}}_{\alpha\beta} (n\Omega,\omega) =
D_{\alpha i}(e^{-i n \frac{2}{3} \pi}  D_{\beta j})^* \ \tilde{\sigma}^{\mathrm{ST}}_{ij}
(n\Omega,\omega),
\label{eq:conductivity}
\end{aligned}
\end{equation}
where $\tilde{\sigma}'^{\mathrm{ST}}$ is the transformed tensor and 
$D$ is the rotation matrix of $R_{2\pi/3}$.
The phase factor in Eq. [\ref{eq:conductivity}] arises from the fractional time-shift, which is the key difference from the transformation of susceptibility tensors in static crystals.
Furthermore, in a Floquet system without a nonsymmorphic space-time symmetry, the phase factor is also absent. 


By Neumann's principle, the conductivity tensor $\tilde{\sigma}^{\mathrm{ST}}_{ij}(n\Omega,\omega)$ must remain invariant under symmetry transformation.
We expand $\tilde\sigma^{\mathrm{ST}}=
\sigma_0 \tau_0 +
\sigma_1 \tau_1 +
\sigma_3 \tau_3+
\sigma_2 i\tau_2
$, where $\tau_0$ is the $2\times 2$ identity matrix and $\tau_{1,2,3}$ are Pauli matrices.
As for $n=3p$ with $p$ an integer, the phase factor is trivial, and then the
rotation part of $(R_{2\pi/3}|T_t^{1/3})$
requires that
\begin{equation}
\tilde\sigma^{\mathrm{ST}}(3 p \Omega,\omega)=
      a_{3p}(\omega)\tau_0+
    b_{3p}(\omega)  i\tau_2,
\end{equation}
where $a_{3p}$ and $b_{3p}$ are the isotropic longitudinal and transverse Hall conductivities, 
respectively. 
Remarkably, for the cases of $n=3p\pm 1$,
the nontrivial phase factor in Eq.[\ref{eq:conductivity}] yields non-trivial responses,
\begin{equation}
\tilde\sigma^{\mathrm{ST}}\bigl((3 p\pm1) \Omega,\omega\bigr)= 
a_{3p\pm 1}(\omega) (\tau_3 \mp i\tau_1),
%
\label{eq:3p+1}
\end{equation}
where $\tau_3\mp i \tau_1$
transforms as two 1D representations of the $C_3$
group.
Remarkably, Eq. [\ref{eq:3p+1}] indicates that the rotating probe electric fields generate heterodyne response currents with the opposite handedness at $n = 3p\pm 1$.

To illustrate, the counterclockwise and clockwise rotating fields and currents are defined as
\begin{eqnarray}
\mb{E_{\pm,\omega}}(t) = E \mb{e_{\pm,\omega}} (t), \ \
\mb{J_{\pm,\omega}}(t) = J \mb{e_{\pm,\omega}} (t), 
\label{eq:circular}
\end{eqnarray}
respectively,
where $\mb{e_{\pm,\omega}} (t)= \mb{e_x}\cos \omega t  \pm \mb{e_y} \sin \omega t$ defines the rotating frame. 
In this basis, the response currents for two probe 
fields can be organized as
\begin{equation}
\begin{aligned}
\mb{J_{\mp ,\omega_\pm^\prime}}\bigl( t)=\sigma_{3p\pm 1} R_z(\theta_{3p\pm 1} ) \mb{E}_{\pm,\omega}(t),
\end{aligned}
\label{eq:chiralcurrent}
\end{equation}
where $\omega_\pm^\prime=\omega+ (3p\pm 1)\Omega$;
$a_{3p\pm 1}(\omega)=\sigma_{3p\pm 1} e^{ \mp i\theta_{3p\pm1}}$ with $\sigma_{3p\pm 1}$ the 
magnitudes of the conductivity and $\theta_{3p\pm1}$ 
the phase differences between the currents and fields; $R_z(\theta )$ means the rotation around the $z$-axis at the angle of $\theta$.

The handedness-flipping behavior in Eq. [\ref{eq:chiralcurrent}] can be understood from the symmetry group generated by $(R_{2\pi/3}|T_t^{1/3})$.
Its irreducible 1D representations are 
labeled by pseudo angular momenta $0,\pm 1$ (defined modulo 3), respectively. 
The fractional temporal translation enables the heterodyne conductivities 
$\tilde\sigma^{\mathrm{ST}}\bigl(n\Omega,\omega\bigr)$
with $n=3p\pm1$ to carry a pseudo angular momentum $\pm 1$, respectively.
Similarly, $\mb{E}_\pm$ and currents $\mb{J}_\pm$ each carry pseudo angular momentum $\pm1$, respectively. 
Hence, the pseudo angular momenta on both sides of Eq. [\ref{eq:3p+1}] are conserved modulo 3, with chiralities exchanged. 

This result demonstrates that the heterodyne response of space-time crystals is intrinsically structured by their nonsymmorphic space-time symmetries, leading to a separation of response channels associated with different chiralities. 
Our analysis shows that the interplay between temporal periodicity and spatial symmetry enables new routes for controlling and detecting symmetry-protected dynamical phenomena in driven systems.

\section{ Momentum degeneracy and group of frequency-momentum in space-time metamaterials}
\label{sec:Momentum_degeneracy}

The concept of energy (frequency) gaps plays a central role in the band theory of condensed matter physics and classical waves.
An energy gap in a Bloch spectrum indicates the absence of states within an energy interval.
Closing such a gap often leads to the Dirac cone structure, which refers to the 
linear dispersion around the closing point. 
Dirac materials, exemplified by graphene, have become a major research focus in condensed matter physics  \cite{Dirac_cone1,Dirac_cone2,Dirac_cone3,Dirac_cone4,Xu2022}.

\begin{figure*}[t]
   \centering
    \includegraphics[width=\linewidth]{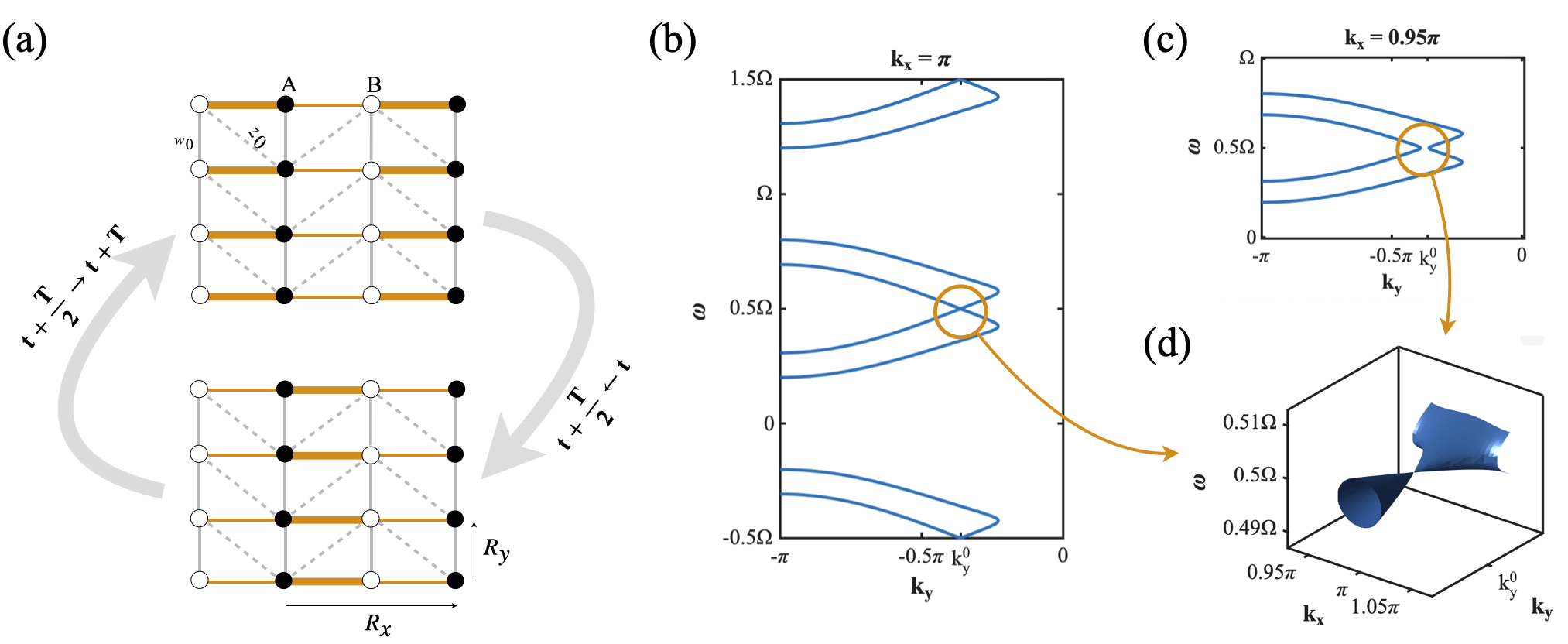}
    \caption{(a) Space-time diagram of a tight-binding model on a rectangular lattice with the time-glide reflection symmetry $(m_x|T_t^{\frac{1}{2}})$.
    Each unit cell contains two sites, $A$ (black) and $B$ (white).
    The thick and thin yellow bonds represent the time-dependent hopping amplitudes $u(t)$ and $v(t)$, respectively.
    The gray and dashed bonds denote the static hopping amplitudes $z_0$ and $w_0$, respectively.
    The onsite potentials on two sublattices are $\mu\pm m$, respectively.
    (b) and (c) The quasi-energy spectra at $k_x=\pi$  and $k_x=0.95\pi$, respectively.
    The values of parameters are $r_0=1$, $a=0.6$, $m=2$, $w_0=5$, $z_0=2$, $\Omega=5$, and $\mu=10$.
    (d) The horizontal cone plotted as the $(2+1)$D spectrum, corresponding to the regions indicated by the orange circles in (b,c).
    }
    \label{fig:momentum_cone}
\end{figure*}

In contrast, a momentum gap, or, 
``$k$-gap'', {\it i.e.}, a gap in momentum (wavevector) space, has recently become a focus of research in space-time metamaterials \cite{M_gap1,M_gap2,M_gap3,M_gap4,M_gap5,M_gap6,M_gap7,
kgap_phonon1,kgap_phonon2,kgap_phonon3,kgap_phonon4,
kgap_phonon5,kgap_phonon6}.
It corresponds to the absence of classical wave modes (e.g., electromagnetic and acoustic waves) within a certain region of wavevector space.
It is natural to ask whether analogous degenerate Dirac-like points can arise by closing a momentum gap, where Bloch states located on opposite sides of the $k$-gap become connected.
Below, we construct a minimal lattice model [Fig. \ref{fig:momentum_cone}(a)] and show that its spectrum exhibits a cone structure [Fig. \ref{fig:momentum_cone}(d)], dubbed the horizontal cone, protected by the time-glide reflection symmetry.

To analyze such degeneracies, we generalize the group of wavevector from static crystals to space-time crystals. 
Unlike static crystals, whose Brillouin-zone boundaries are only along momentum directions, the frequency-momentum Brillouin zone (FMBZ) of a space-time crystal has additional boundaries along the direction of quasi-frequency.
Consequently, symmetry-protected degeneracies may arise at quasi-frequency boundaries and lie beyond the scope of the conventional formalism of group of wavevector.

\paragraph{Group of frequency-momentum}
 
The group of frequency-momentum is defined 
as the set of all space-time symmetry operations that leave a high-symmetry point in the FMBZ invariant.
Its representation theory predicts a new class of symmetry-enforced momentum degeneracies located at high-symmetry points in the FMBZ. 
Such a degeneracy structure is intrinsically spatiotemporal and is protected by the nonsymmorphic space-time symmetry, which cannot be realized in systems with only spatial modulation or only temporal modulation.

We take a classical wave system to demonstrate the 
momentum degeneracy.
Similar to electron systems, Bloch's theorem is generalized to the Bloch-Floquet form \cite{xu2018space},
     $\psi_\kappa(\mathbf r,t)
     =e^{i(\mathbf{k}\cdot\mathbf r-\omega t)}
     u_\kappa(\mathbf r,t)$,
where $\kappa=(\mathbf{k},\omega)$ is the crystal momentum-frequency vector.
However, the classical wave equations involve a second-order derivative with respect to time expressed as
 \begin{equation}
     \begin{aligned}
         \left[ \partial_t^2+H(\mb{r}   ,t) \right]\psi_\kappa(\mb{r},t)=0.
     \end{aligned}
     \label{eq:EOM}
 \end{equation}
We consider the case in which $H(\mb{r},t)$ is a real Hermitian operator whose derivatives are only with respect to $\mb{r}$.
Unlike electron states, classical wave modes 
are essentially real: The complex-conjugate state $\psi_{-\kappa}=\psi_\kappa^*$ is also a solution.
If $H(\mb{r},t)$ is time-dependent,  
$\mbox{Re}\psi_{\kappa}$
and $\mbox{Im}\psi_{\kappa}$ generally cannot be related by a time shift, hence, they represent two independent real solutions.



The lattice model of classical waves 
is constructed to illustrate the horizontal cone structure as shown in Fig. \ref{fig:momentum_cone} (a):
The model exhibits two sublattices, labeled $A$ and $B$, with periodically modulated hopping amplitudes.
These sites represent small resonators in a photonic crystal or spherical particles in a phononic crystal, and each site supports a single mode. 
The unit cells are labeled by two integers 
$n_1$ and $n_2$,
\begin{equation}
    \begin{aligned}
        \mb{R_n}=n_1\mb{R_x}+n_2\mb{R_y}, 
    \end{aligned}
\end{equation} 
where $\mb{R_x}$ and $\mb{R_y}$ are the unit vectors.
The local modes at sites $A$ and $B$ in a unit cell are represented by $\phi^A(\mb{r -R_n})$ and $\phi^B(\mb{r-R_n})$, respectively, with $\mb{n}=(n_1,n_2)$.

We employ the tight-binding approximation, which is widely adopted in the study of metamaterials \cite{TB1,TB2,TB3,TB4,TB5,Monomode}.
This system can be viewed as an array of coupled one-dimensional diatomic chains.
Each chain extends along the 
$x$-direction 
with hopping amplitudes  $u(t)=r_0+a\cos \Omega t$ and $v(t)=r_0-a\cos \Omega t$.
Here $\Omega$ denotes the driving frequency while $r_0$ and $a$ denote the averaged and the oscillating parts of the hopping amplitude, respectively. 
They satisfy $u(t)=v(t+T/2)$, giving rise to the time-glide reflection symmetry, $(m_x|T_t^{\frac{1}{2}})$, which transforms a space-time coordinate $(x,t)_{A/B}$ to $(-x,t+T/2)_{A/B}$.
The chemical potentials on the $A/B$ sublattices are given by $\mu\pm m$, respectively.
The nearest-neighbor amplitude along the $y$-direction is denoted as $z_0$; the next-nearest-neighbor one $w_0$ connects sites on opposite sublattices.
The Bloch-wave bases for $A$ and $B$-sublattices are defined as
$        \psi_{\mb{k}}^{A/B}(\mb{r}) = \frac{1}{\sqrt{N}}\sum_{\mathbf{R}_n}  e^{i \mb{ k\cdot R_n}} \phi^{A/B} (\mb{ r-R_n}),
$
where $N$ is the total number of unit cells.
In this basis, $H(\mb{k},t)$ is a $2\times 2$ matrix expanded in terms of the Pauli matrices $\sigma_{i}$ with $i=1\sim 3$, 
\beaa
H(\mb{k},t)= \mu-2 z_0 \cos{k_y} + 
h_1(t) \sigma_1 + h_2 (t)\sigma_2 + m\sigma_3,
\label{eq:Fourier}
\eeaa
where $h_1(t)= u(t)+v(t)\cos{k_x} 
+ w_0 \left[\cos{k_y}+\cos{\left(k_y-k_x\right)}\right]$
and $h_2(t)= -v(t)\sin{k_x} 
+ w_0 \left[\sin{k_y}+\sin{\left(k_y-k_x\right)}\right]$.



We numerically solve the spectrum $\omega(\mathbf{k})$ of Eq. [\ref{eq:EOM}] using standard methods \cite{M_gap1,M_gap6} applied to the momentum-space representation of $H$ given in Eq. [\ref{eq:Fourier}].
Due to the periodic modulation, the quasi-frequency is folded such that $\pm \Omega/2$ are identified denoted as $\omega_0$.
In Fig. \ref{fig:momentum_cone} (b,c), the spectra $\omega(\mathbf{k})$ are shown as functions of $k_y$ at fixed values $k_x=\pi$ and $0.95\pi$, respectively, which exhibit the horizontal cone dispersion with its axis along the direction of $k_y$.
The touching point is located at $\kappa_0=(\pi,k_y^0, \Omega/2)$, while $k_y$ is determined by energetics. 
When $k_x$ deviates from $\pi$, the touching point splits, as shown in Fig. \ref{fig:momentum_cone}(c), and
the spectrum near the touching point is shown in Fig. \ref{fig:momentum_cone}(d).

The origin of the horizontal cone is rooted in the group of frequency-momentum symmetries.
For the present model, we focus on the horizontal cone located at $M$, defined by $(k_x,\omega)=(\pi,\Omega/2)$.
The symmetries leaving $M$ invariant are generated by the time-glide reflection $(m_x|T_t^{1/2})$ and time-reversal operation $m_t$.
The time-glide reflection reverses $k_x$.
Since $k_x=\pi$ and $k_x=-\pi$ are identified modulo a reciprocal lattice vector, $(m_x|T_t^{1/2})$ leaves the $M$ point invariant.
The inclusion of $m_t$ requires further explanation.
Based on the reality of the wave equation Eq. [\ref{eq:EOM}], complex conjugation $K$ can be viewed as a symmetry mapping $\kappa$ to $-\kappa$.
Combining $K$ with the conventional time-reversal operation $\mathcal T$ defines
$m_t=K\mathcal T$ for  the present classical system.
Applying $m_t$ to $\kappa$ yields
$\kappa
\rightarrow 
\kappa_{\mathrm{TR}}=(k_x,k_y,-\omega)
$, {\it i.e.}, reversing the sign of $\omega$ while keeping $\mathbf{k}$ invariant. 
Correspondingly, its action on the wavefunction is just $t\to -t$, {\it i.e.}, 
\begin{equation}
\begin{aligned}
m_t\psi_\kappa(\mathbf r,t)
&=
\psi_{\kappa_{\mathrm{TR}}}(\mathbf r,t)
=
\psi_\kappa(\mathbf r,-t).
\end{aligned}
\end{equation}
At the boundary of the FMBZ, $\Omega/2$ and $-\Omega/2$ are identified modulo the driving frequency $\Omega$.
Therefore, $m_t$ leaves the $M$ point invariant and belongs to its group of frequency-momentum.


The above two generators of the group of frequency-momentum for the $M$-point satisfy the relation
\beaa
m_t (m_x|T^{\frac{1}{2}}_t)
= (I|T^{-1}_t) (m_x|T^{\frac{1}{2}}_t) m_t,
\label{eq:comm}
\eeaa
where $(I|T^{-1}_t)$ denotes a full-period time translation.
At $\omega_0 = \Omega/2$, $(I|T^{-1}_t)$ reduces to a phase factor $-1$.
Consequently, the group representation becomes projective. 
Its irreducible representation is therefore 2-dimensional, leading to the momentum degeneracy.
The proof is briefly outlined as follows:
We assume a solution of Eq. [\ref{eq:EOM}] denoted as $\psi_{1,\kappa}(\mathbf{r},t)$ which is also an eigenstate of $m_t$ with the eigenvalue $\eta$.
Since $(m_t)^2=I$, $\eta=\pm 1$. 
Then we define $\psi_{2,\kappa}(\mathbf{r},t)= (m_x|T^{\frac{1}{2}}_t)\psi_{1,\kappa}(\mathbf{r},t)$, which is also a solution of Eq. [\ref{eq:EOM}].
According to Eq. [\ref{eq:comm}], 
$m_t \psi_{2,\kappa}(\mathbf{r},t)=-\eta \psi_{2,\kappa}(\mathbf{r},t)$.
As shown in Methods, $\langle \psi_{1,\kappa} |\psi_{2,\kappa} \rangle=0$ by the inner product defined in Eq. [\ref{eq:inner_prod}].
Furthermore, $\psi_{1,\kappa} (\mathbf{r},t)$ and $\psi_{2,\kappa} (\mathbf{r},t)$ are not complex conjugation pairs, hence, they represent independent states.


The proof above can be verified explicitly in terms of Eq. [\ref{eq:EOM}] with the Hermitian operator given in Eq. [\ref{eq:Fourier}].
At the $M$-point with $k_x=\pi$, it is cast into the form of
\begin{eqnarray}
[-\partial_t^2+V_0(t)]\psi_{\kappa}=f(k_y) \psi_{\kappa},
\label{eq:time-sch}
\end{eqnarray}
with $f(k_y)=\mu-2z_0\cos k_y$ and 
\begin{eqnarray}
V_0(t)=-\lambda \left(
\cos\theta\,\sigma_3 +2\sin\theta\cos\Omega t\,\sigma_1
\right),
\end{eqnarray}
where
$\lambda=\sqrt{a^2+m^2}$ and $\sin\theta=a/\lambda$. 
If $t$ is reinterpreted as a spatial coordinate, Eq. [\ref{eq:time-sch}] becomes the stationary Schr\"odinger equation with $f(k_y)$ as the eigenvalue to be determined. 
The problem is recast as that of an electron propagating in an effective magnetic field  $\mathbf{B}$, consisting of a uniform component $B_z = -\lambda \cos\theta$ and  a periodic component  $B_x = -\lambda \sin \theta \cos \Omega t$.
This model possesses a spin-space-group symmetry combining a spin rotation about the $z$ axis at the angle of $\pi$ with a half-period translation.
The corresponding symmetry operation is denoted by $g=\{R^{\mathrm{spin}}_z(\pi)|T/2\}$ and is inherited from the time-glide reflection symmetry of the original model.
The anti-commutation of $g$ and $m_t$  at the Brillouin zone boundary $\omega_0=\pm\Omega/2$ yields the degeneracy.

Although the effective ``Hamiltonian'' ${\cal H}_0^{eff}=-\partial^2_t+V_0(t)$ has infinitely many eigenvalues, the physically relevant one $f(k_y)$ is restricted to the interval $[\mu-2z_0,\mu+2z_0]$. 
For the parameter values (specified in Fig. \ref{fig:momentum_cone}), 
only one degenerate pair occurs at the temporal Brillouin-zone boundary, {\it i.e.}, $\omega_0=\pm \frac{\Omega}{2}$.
At leading order in perturbation theory, we retain the frequency components at $\pm \frac{\Omega}{2}$.
This yields the doubly degenerate eigenvalue $f(k_y^0)\approx\lambda+\Omega^2/4$, with solutions at $\pm k_y^0$, where $k_y^0\approx-0.35\pi$.
Correspondingly, the eigen-solutions are given as 
\begin{eqnarray}
\phi_{1}=\cos {\frac{\Omega t}{2}}
\left( \begin{array}{c}
  \sin\frac{\theta}{2}\\ 
- \cos\frac{\theta}{2}
\end{array}
\right), 
\ \ \
\phi_{2}=\sin {\frac{\Omega t}{2}}
\left( \begin{array}{c}
  \sin\frac{\theta}{2}\\ 
  \cos\frac{\theta}{2}
\end{array}
\right). 
\label{eq:eigen_solution}
\end{eqnarray}
Restoring the wavevector, we obtain Floquet-Bloch wavefunctions as
$\psi_{1,2,\kappa_0}=(-)^{n_1}
e^{i k_y^0 n_2} \phi_{1,2}$.





By analogy with the $k\cdot p$ perturbation theory, the wave equation near the $M$-point is expanded
${\cal H}^{eff}={\cal H}_0^{eff}+\delta V(t)$ with
{$\delta V(t)= \delta h_1 \sigma_1 +\delta h_2 \sigma_2$, where $\delta h_1=w_0 \sin{k_y^0} \delta k_x$ and $\delta h_2=[v(t)+w_0 \cos k^0_y ]\,\delta k_x$}.
Consider the eigen-solutions of ${\cal H}^{eff} \psi_\kappa(t)= f(k_y)\psi_\kappa(t)$ with the quasi-frequency $\omega=\frac{\Omega}{2}+\delta \omega$ and momentum
$k_x=\pi +\delta k_x$.
In the limit of $\Omega\gg \lambda\sin\theta$, the leading order perturbation theory yields the eigenvalue 
{$f(k_y)=f(k_y^0)\pm \sqrt{
(w_0 \sin k_y^0 \sin\theta\, \delta k_x)^2
+
(\Omega \cos\theta\,\delta\omega)^2
}$}, leading to $k_y=k_y^0\pm \delta k_y$ with 
\begin{equation}
    \begin{aligned}
        \delta k_y= \frac{1}{2z_0 } \sqrt{
(w_0  \sin\theta)^2\,( \delta k_x)^2
+
(\frac{\Omega}{\sin k_y^0} \cos\theta)^2\,(\delta\omega)^2
}.
    \end{aligned}
\end{equation}
As for the case of $\delta \omega=0$ and 
$\delta k_x\neq 0$, the eigensolutions remain
the expressions in Eq. [\ref{eq:eigen_solution}]
but the degeneracy in $k_y$ is broken by the splitting 
of $\frac{w_0}{z_0} \sin\theta |\delta k_x|$
as shown in Fig. \ref{fig:momentum_cone} (d).



 
\section{ Discussion}
In summary, we have established the fundamental group-theoretic framework for classifying and understanding 
space-time crystals in 2+1 dimensions.
The complete classification yields all 275 space-time groups, including 203 nonsymmorphic ones.
They are organized into 7 crystal systems.
In comparison to the 3D space groups, the cubic crystal system is absent while the monoclinic crystal system splits into two due to the non-equivalence between spatial and temporal directions. 
Novel space-time nonsymmorphic symmetries include time-glide reflection and time-screw rotation. 
The power of this new formalism has been demonstrated through 
its application to two distinct phenomena.
The space-time symmetries enable non-trivial handedness-flipping heterodyne response rules that can be applied to optical response theory in dynamical systems. 
In space-time metamaterials, a novel phenomenon—the ``horizontal cone''—is found: An unusual momentum degeneracy at fixed quasi-frequency is protected by nonsymmorphic space-time symmetry without a static-crystal counterpart.

Our findings confirm that the space-time group 
is not merely a mathematical extension of the space group, but also provides a fundamental principle governing unique physical phenomena in driven systems.
The theoretical framework presented here provides a powerful tool for the future design and discovery of exotic states in dynamic matter, active metamaterials, and other non-equilibrium systems.



\vspace{0.5\baselineskip}


\section*{Acknowledgments} 
 We thank Shenglong Xu for his contributions to the initial stages of this work.
We are grateful to Zhixing Lin, Zheng-Xin Liu, Wei Yan, Zhiming Pan, Lun-hui Hu, Zhuang Qian, Huan Wang, Yue Wang and Xiangyu Zhang for valuable discussions.
C. W. is supported by the National Natural Science Foundation of China under the Grants No. 12234016 and No. 12174317.
This work has been supported by the New Cornerstone Science Foundation.

 \bibliography{Submission_PNAS}

\clearpage
\onecolumngrid
\section*{Supplimentary information for  ``Space-time crystallography and symmetry-enforced dynamical phenomena in 2+1 dimensions''}

\begin{appendix}

\renewcommand{\theequation}{S\arabic{equation}}
\setcounter{equation}{0}
\setcounter{figure}{0}
\setcounter{table}{0}


\supplementtableofcontents

\section{Crystallography in space-time crystals}

In the main text, we have outlined
the classification of space-time groups, which parallels the crystallographic hierarchy: crystal system, geometric crystal class (GCC), arithmetic crystal class (ACC), and space-time group.
Here we provide the definitions underpinning this classification.

\subsection{Definitions}
To highlight the special role of time-reversal (TR) operation, which is anti-unitary in quantum mechanics, we introduce the concept of the invariant unitary subgroup.
Each space-time group $G_{st}$ contains an invariant subgroup, $K$, consisting of all the elements that do not involve time-reversal, which is referred as the maximal invariant unitary subgroup of $G_{st}$.
To systematically organize space-time groups,
we define equivalence relations based on  symmetries and subgroup properties.

\medskip\noindent{Space-time group}
A $(d+1)$-dimensional space-time group $G_{st}$ is a group of affine transformations
$\widetilde g=(g|u[g])$ that contains a full-rank translation lattice $T_L$ as a normal subgroup.
The quotient $G_{st}/T_L$ is the magnetic point group (MPG)
\begin{equation}
    P_M =G_{st}/T_L,
\end{equation}
and the corresponding maximal invariant unitary group  $\bar K$ of $P_M$ is can be obtained by 
\begin{equation}
    \bar K =K/T_L\triangleleft P_M,
\end{equation}
where ``$\triangleleft$" means that
$\bar K$ is a normal subgroup of $P_M$.
Nevertheless, $P_M$ and $\bar K$ do not necessarily belong to the symmetry group of $G_{st}$.

Two space-time groups $G_1$ and $G_2$ are considered as equivalent if they are related by
\beaa
\g\, G_1\, \g^{-1} = G_2,  
\g  K_1\, \g^{-1} = K_2,
\eeaa
where $K_{1,2}$ are the maximal invariant unitary subgroups of $G_{1,2}$, respectively; $\g=(g|u[g])$ is an affine transformation with $g$ 
and $u[g]$ a space--time translation.
$g$ is a square integer matrix with determinant $1$, {\it i.e.}, the special unimodular matrix.  
The 2nd condition is based on the fact that time is a direction distinct from spacial ones in non-relativistic physics, which should be treated separately.

\medskip\noindent{\it Arithmetic Crystal Class (ACC).}
There is a one-to-one correspondence between each ACC and a symmorphic space-time group by disregarding the fractional translations carried by nonsymmorphic operations.
Two space-time groups $G_1,G_2$ belong to the same ACC if their MPGs $P_{M_{1,2}}$ and their maximal invariant unitary subgroups $\bar K_{1,2}$ are related by
\beaa
g\,P_{M_1}\, g^{-1} = P_{M_2}, \qquad g\,\bar K_1\, g^{-1} = \bar K_2,
\label{eq:Z_equi}
\eeaa
for a special unimodular matrix $g$.

Taking the primitive translation vectors of the Bravais lattice as the lattice bases, each symmetry operation is represented by an integer matrix, because it maps every lattice vector to an integer linear combination of the basis vectors.
An ACC can therefore be regarded as an equivalence class of such integral matrix representations under special unimodular changes of the lattice basis, with the invariant unitary subgroup preserved.

\medskip\noindent{\it Bravais lattice.}
\label{subsec:Bravais lattice}
The translation symmetry of a $(d{+}1)$-dimensional space-time group is determined by its underlying Bravais lattice.
A Bravais lattice is generated by $d{+}1$ linearly independent primitive translation vectors.
The corresponding Bravais group contains all symmetry operations that leave the lattice invariant.
A Bravais lattice determines a Bravais group, whose integral representation belongs to an ACC; the lattice points themselves are featureless.
Two Bravais lattices are distinct if their Bravais groups belong to different ACCs.

\medskip\noindent{\it Geometric Crystal Class (GCC).}
The magnetic point group of a space-time group $G_{st}$ corresponds to a GCC.
As for ACCs (symmorphic space-time groups), they are grouped into the same GCC
if they share the same magnetic point groups.
More precisely, two ACCs belong to the same GCC, if their magnetic point groups $P_{M_{1,2}}$ and the associated maximal invariant unitary subgroups $\bar K_{1,2}$ are related by
\beaa
g\,P_{M_1}\, g^{-1} = P_{M_2}, \qquad g\,\bar K_1\, g^{-1} = \bar K_2,
\eeaa
where $g$ is an arbitrary similarity transformation in 
$(d+1)$ dimensions. 
 
\medskip\noindent{\it Crystal system.}
A type of magnetic group (GCC) is compatible with a set of different Bravais lattices, e.g., primitive and  different types of centered lattices.
Different GCCs sharing with the same set of Bravais lattices are grouped together into one crystal system.


\subsection{Classification of
    space-time groups}
We classify space-time group $G_{st}$ starting from its magnetic point group $P_M$ together with its maximal invariant unitary subgroup $\bar K $.
The first task is to determine the space-time Bravais lattices compatible  with the symmetry of $P_M$.
Their symmorphic combination generates the corresponding ACCs, i.e., the symmorphic space-time groups.
Second, for every ACC, we enumerate all non-equivalent non-symmorphic extensions obtained by assigning fractional space-time translations to the magnetic point-group 
$P_M$.

\subsubsection{Construction of space-time Bravais lattices and ACCs}

To construct all the space-time Bravais lattices and ACCs, we take the advantage of the crystallography in 3D.
Seemingly, a $(2+1)$D magnetic point group corresponds to a 3D crystallographic point group by treating time as the 3rd dimension.
However, taking into account the difference between time and spatial directions, a single 3D spatial ACC may split into several distinct space-time ACCs.

We present the above procedure with a magnetic group $P_M=m1'$ and its invariant subgroup $\bar K$
\begin{eqnarray}
m1'=\{1,m_x,m_t,m_xm_t\}, \qquad
\bar K=\{1,m_x\}.
\end{eqnarray}
When $t$ is treated as the 3rd dimension, it is isomorphic to the point group $mm2=\{1,\sigma_x,\sigma_y,C_{2z}=\sigma_x\sigma_y\}$.
There exist two maps $\phi_{1,2}$ from $m1'$ to $mm2$:
\begin{equation}
    \begin{array}{c|ccc}
               & m_x      & m_t      & m_xm_t \\ \hline
        \phi_1 & \sigma_x & \sigma_y & C_{2z} \\
        \phi_2 & \sigma_y & \sigma_x & C_{2z}
    \end{array}
    \label{eq:mm2_two_mappings}
\end{equation}
Then we consider the ACCs of $mm2$ in 3D and construct space-time ACC from them. 
The illustration is performed through two examples, {\it i.e.} $Amm2$ and $Cmm2$, both of which are constructed from $mm2$ acting on the base-centered orthorhombic Bravais lattice.
For each of them, the maps of $\phi_{1,2}$ may lead to two different space-time ACCs.

\begin{figure}[h]
    \caption{The lattice orientations for $Amm2$ and $Cmm2$.
        Both of them contain the reflection planes in the $x$-$z$ and $y$-$z$ faces.
    }
      \includegraphics[width=0.3\linewidth]{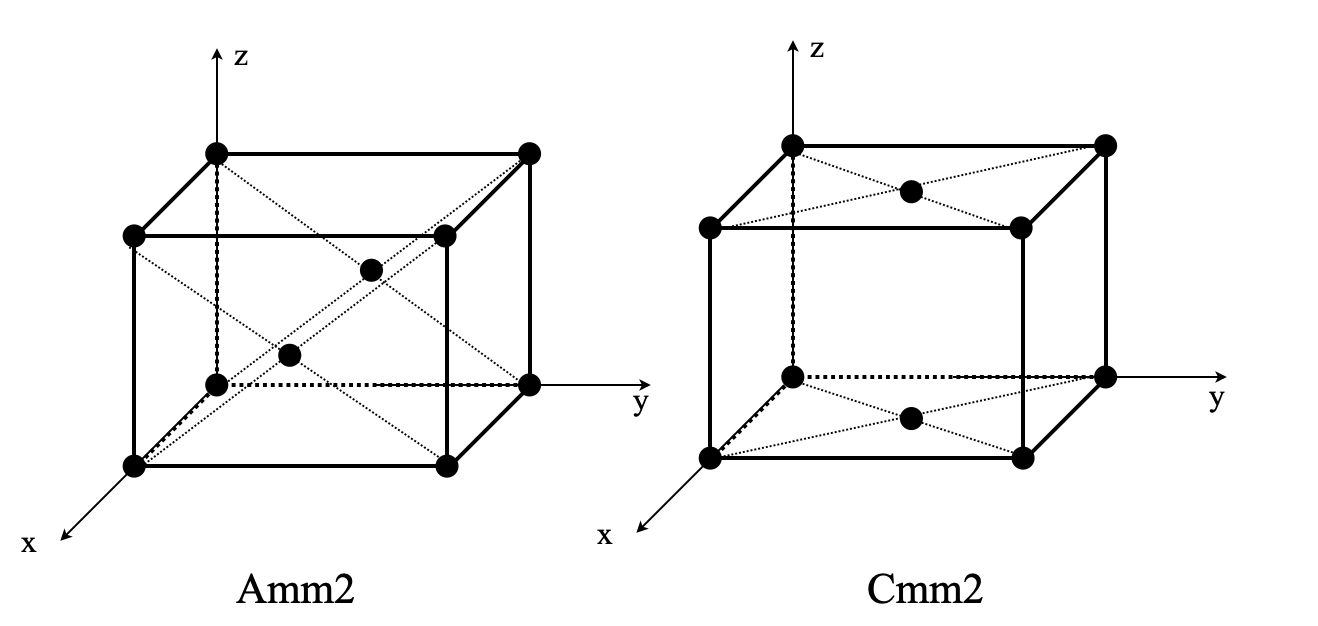}
    \label{mm2}
\end{figure}

Based on $Amm2$, it generates two space-time ACCs from $m1'$.
Due to the centering point at $y-z$ face, $x,y$ directions are non-equivalent as shown in Fig.\ref{mm2}.
Hence, two different mappings of $\phi_{1,2}$ are non-equivalent.
To be more rigorous, we first derive the matrix forms of $\sigma_x$ and $\sigma_y$ under the primitive lattice bases,
\begin{equation}
    a^A_1=(1,0,0),\quad
    a^A_2=\tfrac12(0,1,1),\quad
    a^A_3=\tfrac12(0,-1,1),
\end{equation}
yielding
\begin{equation}
    \begin{aligned}
        \phi_1(m_x)=\sigma_x
          =\begin{pmatrix}-1&0&0\\0&1&0\\0&0&1\end{pmatrix},\quad
        \phi_2(m_x)=\sigma_y
           =\begin{pmatrix}1&0&0\\0&0&1\\0&1&0\end{pmatrix}.
    \end{aligned}
\end{equation}
It can be shown that there does not exist a unimodular matrix $a$
satisfying the following relation for each $g\in P_M$,
\begin{equation}
    a\, \phi_1(g)\, a^{-1}
    =\phi_2(g),
    \label{eq:mapping_equivalence}
\end{equation}
which proves the non-equivalence of two mappings $\phi_{1,2}$.
Hence, $\phi_1$ gives rise to the space-time ACC $m_x1'A$ (ACC No.~25), shown in the lower-left panel of Table~\ref{tb:tbase_orthg} and associated with the T-base-centered orthorhombic lattice.
In contrast,
$\phi_2$ leads to $m1'C$ (ACC No.~21), shown in the upper-left panel of the same table and associated with the R-base-centered orthorhombic lattice.

For $Cmm2$, two spatial directions $x$ and $y$ are equivalent as shown in Fig.\ref{mm2}.
Therefore, the maps of $\phi_{1,2}$ yield the same ACC.
Again, the primitive bases are chosen as
\begin{equation}
    a_1=\tfrac12(1,1,0),\quad
    a_2=\tfrac12(-1,1,0),\quad
    a_3=(0,0,1).
\end{equation}
Then the representations of spatial operations
$\sigma_{x,y}$ are given as,
\begin{equation}
    \begin{aligned}
        \sigma_x
        =\begin{pmatrix}0&1&0\\1&0&0\\0&0&1\end{pmatrix},
        \ \ \,
        \sigma_y
        =\begin{pmatrix}0&-1&0\\-1&0&0\\0&0&1\end{pmatrix},
    \end{aligned}
\end{equation}
together with $\phi_1(m_x)=\sigma_x, \phi_2(m_x)=\sigma_y$.
We find that $\phi_{1,2}$ are related by the unimodular matrix,
\begin{equation}
    a=\begin{pmatrix}0&-1&0\\1&0&0\\0&0&1\end{pmatrix},
\end{equation}
satisfying
\begin{equation}
    a\, \phi_1(g)\, a^{-1}
    =\phi_2(g),
    \label{eq:mapping_equivalence2}
\end{equation}
for each $g\in P_M$.
The two mappings $\phi_{1,2}$
are therefore equivalent, and $Cmm2$ gives rise to only one space-time ACC, denoted $m_y1'A$ (ACC No.~24), as shown in the lower-left panel of Table~\ref{tb:tbase_orthg}.

Following the above process of mapping a (2+1)D magnetic point group to a 3D point group, we derive all ACCs for space-time groups.

\subsubsection{Group extensions and cohomological classifications}

Once all the compatible space-time Bravais lattices are found, the remaining problem is to construct all non-equivalent groups $G_{st}$ based on
each ACC.
This is a problem of group-extension ~\cite{Howard1921,Ascher1965,Ascher1968}.
The translation subgroup $T_L$ is a normal subgroup of $G_{st}$ with the corresponding
quotient magnetic point group $P_M$, then each space-time group fits into the exact sequence
\begin{equation}
    1\longrightarrow T_L\longrightarrow G_{st}
    \stackrel{}{\longrightarrow}P_M\longrightarrow 1.
    \label{eq:group_extension_exact}
\end{equation}
The trivial extension leads to the symmorphic space-time group, which is the semi-direct product
\begin{equation}
    G_{st}^{\mathrm{sym}}=T_L\rtimes P_M.
\end{equation}
In contrast, non-trivial extensions represent nonsymmorphic space-time groups.

Normally, this problem is solved by the 2nd cohomology group $H^2(P_M,T_L)$.
However, the first-cohomology, which is isomorphic to $H^2(P_M,T_L)$ for this problem, is more convenient for crystallographic calculations and then constructed below.
For a legitimate fractional translation $u[g]$ assigned to each magnetic point-group operation $g$, it must satisfy
\begin{equation}
    u[gh]-u[g]-g u[h]=0 \mod T_L.
    \label{eq:one_cocycle}
\end{equation}
for all $g,h\in P_M$.
Such a solution of $u$ is called a
{\it vector system}, and all vector systems form a group, called the 1-cocycle $Z^1(P_M,V/T_L)$, where $V=\mathbb R^{d+1}$ denotes the continuous space-time translation vector space.
For the trivial solution of $u[g]=0$, 
it just gives rise to the symmorphic extension.

The vector systems in the 1-cocycle exhibit redundancy in determining space-time groups.
The shift of the origin by an arbitrary translation $\eta$  changes the vector system and
leads to a new legitimate vector system $u'$ reprsented by
\begin{equation}
    u'[g]=u[g]+\eta[g],\quad\eta[g]= g\eta-\eta.
    \label{eq:origin_shift}
\end{equation}
Such a redundancy should be eliminated.
We find that the shift $\eta[g]$ itself is also a vector system.
Such shifts form a subgroup of 1-cocycle denote as the 1-coboundary $B^1(P_M,V/T_L)$.
The vector systems modulo origin shifts are classified by the first cohomology group
\begin{equation}
H^1(P_M,V/T_L)=Z^1(P_M,V/T_L)/B^1(P_M,V/T_L).
\label{eq:H1}
\end{equation}

For completeness, we sketch the proof that the first/second cohomology group $H^1(P_M,V/T_L)$ and $H^2(P_M,T_L)$ are isomorphic.
Algebraically, the equivalence follows from the long exact sequence in group cohomology \cite{homological}, applied to the short exact sequence of $P_M$-modules
\begin{equation}
    0\longrightarrow T_L\longrightarrow V
    \longrightarrow V/T_L\longrightarrow 0.
    \label{eq:module_exact}
\end{equation}
Then the long exact cohomology sequence induced by Eq.~\eqref{eq:module_exact} gives rise to
\begin{equation}
    H^2(P_M,T_L)\cong H^1(P_M,V/T_L).
    \label{eq:H2_H1}
\end{equation}
Hence, the vector system method contains precisely the extension information encoded in $H^2(P_M,T_L)$.


\subsubsection{Final identification under affine equivalence}

The elements of $H^1(P_M,V/T_L)$ classify the extensions associated with each ACC.
However, distinct elements of the cohomology group do not necessarily correspond one-to-one to inequivalent crystallographic groups.
For example, in two-dimensional wallpaper groups, consider the two non-symmorphic groups
\begin{equation}
Pmg=\langle m_x,(m_y|T_x^{1/2})\rangle
\end{equation}
and
\begin{equation}
Pgm=\langle (m_x|T_y^{1/2}),m_y\rangle,
\end{equation}
where $T_{x,y}^{1/2}$ denote half translations along the $x$ and $y$ directions, respectively.
These two groups are equivalent under an interchange of the lattice directions
($x\rightarrow y$, $y\rightarrow -x$), although they correspond to different elements of the cohomology group.
Such equivalent vector systems must therefore be identified.
This redundancy is characterized by the orientation-preserving arithmetic normalizer (or normalizer for brevity) of $P_M$, which is the set of special unimodular matrices that map $P_M$ onto itself under conjugation, i.e., $A P_M A^{-1}=P_M$, for $A$ belonging to the normalizer.
Also, the elements in the normlizer should map the maximal invariant unitary subgroup to itself.
The normalizer can relate different elements of $H^1(P_M,V/T_L)$ into one another, and such elements should be identified.
After removing this redundancy, we obtain all inequivalent space-time groups.


\subsection{Example: ACC  $m_y1'A$ (No.~24)}

We now illustrate the above cohomological procedure by using the space-time ACC 
$m_y1'A$ (No.~24), which generates a set of space-time groups listed in the lower-left panel of Table~\ref{tb:tbase_orthg}.
It possesses the magnetic point group $P_M=m1'$ and a T-base-centered orthorhombic space-time Bravais lattice.
Its primitive space-time lattice vectors are chosen as,
\begin{equation}
    \begin{aligned}
        a_1^{\text{TB}}  =(x_0,0,0),      \quad      
        a_2^{\text{TB}}  =\tfrac12(0,y_0,t_0),  \quad
        a_3^{\text{TB}}  =\tfrac12(0,-y_0,t_0).
    \end{aligned}
    \label{eq:acc24_basis}
\end{equation}
Here $x_0$, $y_0$, and $t_0$ set the lattice scales along the $x$, $y$, and $t$ directions, respectively.
The magnetic point group is
\begin{equation}
    \begin{aligned}
        P_M    & =\langle m_y,m_t\mid m_y^2=m_t^2=1,\
        m_ym_t=m_tm_y\rangle.
    \end{aligned}
\end{equation}
In terms of the primitive bases of Eq.~\ref{eq:acc24_basis}, the integral representations of generators are,
\begin{equation}
    \begin{aligned}
        D_A(m_y)  =\begin{pmatrix}1&0&0\\0&0&1\\0&1&0\end{pmatrix},   \quad
        D_A(m_t)  =\begin{pmatrix}1&0&0\\0&0&-1\\0&-1&0\end{pmatrix}.
    \end{aligned}
    \label{eq:acc24_point_matrices}
\end{equation}
The unitary reflection $m_y$ exchanges $a_2^{\text{TB}}$ and $a_3^{\text{TB}}$, whereas time-reversal $m_t$ exchanges $a_2^{\text{TB}}$ and $-a_3^{\text{TB}}$.

\paragraph{Cocycles and vector systems.}
We attach fraction translations
$u$ to the magnetic point group operations,
\begin{equation}
    u[m_y]=(a,b,c), ~
    u[m_t]=(d,e,f).
\end{equation}
Here $a,b,c,d,e,f$ are initially undetermined parameters, defined modulo integer lattice translations.
They are constrained by the group composition rule of $P_M$ and reduced by the origin shiftings as well.
The corresponding affine representatives are presented in the bases of Eq. (\ref{eq:acc24_basis}),
\begin{equation}
    \begin{aligned}
        \widetilde m_y & =
        \begin{pmatrix}
            1 & 0 & 0 & a \\0&0&1&b\\0&1&0&c\\0&0&0&1
        \end{pmatrix}, 
        ~~
        \widetilde m_t=
        \begin{pmatrix}
            1 & 0 & 0 & d \\0&0&-1&e\\0&-1&0&f\\0&0&0&1
        \end{pmatrix}.
    \end{aligned}
\end{equation}

Now we check the composition relations of the magnetic point group $P_M$.
As for $m_y^2=m_t^2=1$, they must hold modulo lattice translations, yielding
\begin{equation}
    \begin{aligned}
    [I+D_A(m_y)]u[m_y]  =(2a,b+c,b+c),  \quad
    [I+D_A(m_t)]u[m_t]  =(2d,e-f,f-e), 
    \end{aligned}
    \label{eq:acc24_square_constraints}
\end{equation}
where $2a,2d,b+c$, and $e-f$ correspond to integer translations. 
In contrast, as for $m_y m_t =m_t m_y$,
it does not introduce additional conditions.
Consequently, we solve
\begin{equation}
    \begin{aligned}
           a,d=0 ~\mbox{or}~ \frac{1}{2};\qquad
           b=-c,~ e=f\pmod{\mathbb Z}.
    \end{aligned}
    \label{eq:acc24_cocycle_solutions}
\end{equation}
The remaining parameters of $b$ and $e$ are continuous at the level of $Z^1$.

\paragraph{Coboundaries and $H^1$.}

Let the origin be shifted by $s=(x_1,x_2,x_3)$
where $x_{1\sim 3}$ are defined modulo three basis vectors, respectively. 
The corresponding coboundaries are
\begin{equation}
    \begin{aligned}
        [D_A(m_y)-I]s  =(0,x_3-x_2,x_2-x_3),   \quad
        [D_A(m_t)-I]s =(0,-x_2-x_3,-x_2-x_3).
    \end{aligned}
\end{equation}
No matter what values of $b$ and $e$ are taken, they can always be canceled by a properly chosen set of 
$s$ without affecting $a$ and $d$,
which means $b$ and $e$ can be simply removed.

The generators of $H^1(P_M,V/T_L)$ for the ACC $m_y1'A$ (No.~24) are
\begin{equation}
    \begin{aligned}
        u_1^{(24)}:&\qquad
        u_1^{(24)}[m_y]=(0,0,0),
        &u_1^{(24)}[m_t]&=(\tfrac12,0,0)
        &&\text{(glide time-reversal)},\\
        u_2^{(24)}:&\qquad
        u_2^{(24)}[m_y]=(\tfrac12,0,0),
        &u_2^{(24)}[m_t]&=(0,0,0)
        &&\text{(glide reflection)}.
    \end{aligned}
    \label{eq:acc24_H1_generators}
\end{equation}
Here, glide time-reversal means a time-reversal operation followed by a fractional translation in space.
Both generators in the cohomology group contain the half-translation of $(\tfrac12,0,0)=\tfrac12a_1^{\text{TB}}$.
The resulting nonsymmorphic symmetry is a glide reflection or a glide time-reversal operation, depending on whether it is attached to a reflection or to time-reversal.
Therefore,
\begin{equation}
    H^1(P_M,V/T_L)\cong\mathbb Z_2\oplus\mathbb Z_2.
\end{equation}

\paragraph{Normalizer and space-time groups.}

The normalizer can be derived easily as,
\begin{equation}
\begin{aligned}
\text{Normalizer} 
=
\Bigg\{
\begin{pmatrix}
1&0&0\\
0&1&0\\
0&0&1
\end{pmatrix},
\begin{pmatrix}
1&0&0\\
0&-1&0\\
0&0&-1
\end{pmatrix},
\begin{pmatrix}
-1&0&0\\
0&0&1\\
0&1&0
\end{pmatrix},
\begin{pmatrix}
-1&0&0\\
0&0&-1\\
0&-1&0
\end{pmatrix}
\Bigg\}.
\end{aligned}
\label{eq:acc24_linear_normalizer}
\end{equation}

A direct examination of their actions 
that the four elements remain inequivalent under the normalizer.
All of them represent distinct space-time groups listed under ACC No.~24 in the lower-left panel of Table~\ref{tb:tbase_orthg}.

\clearpage
\section{Complete list of 2+1D space-time groups\label{2+1D}}
Our classification yields 72 ACCs and 275 non-equivalent space-time groups in 2+1 dimensions.
The complete results are presented in Tables~\ref{tb:triclinic}--\ref{tb:p_hex2}.

To explain the notation used in these tables, we take the ACC $21'P$ (No.~5) in the left column of Table~\ref{tb:ptmnclnc} as an example.
The following one illustrates the classification for the ACC $21'P$ (No.~5). Here, $1$ denotes the identity element of the cohomology group, representing the symmorphic space-time group, while $u_{1,2}$ and $u_1\cdot u_2$ correspond to the nonsymmorphic ones.
\begingroup
\small
\setlength{\arraycolsep}{3pt}
\setlength\extrarowheight{2.85pt}
\begin{equation*}
    \begin{array}{@{}llc@{}}
        \toprule
        \multicolumn{3}{l}{\mbox{\textbf{ACC No.~5: }}21'P} \\
        \multicolumn{3}{l}{P_M=21',\quad H^1(21',V/T_L)=\mathbb Z_2^3} \\[2pt]
        P112/m     & R_\pi,\ m_t                         & 1             \\
        P112_1/m   & R_\pi T_t^{1/2},\ m_t               & u_1           \\
        P112/a     & R_\pi,\ g_t                         & u_2           \\
        P112_1/a   & R_\pi T_t^{1/2},\ g_t               & u_1\cdot u_2 \\
        \bottomrule
    \end{array}
\end{equation*}
\endgroup
In this example, the first line gives the ACC number, No.~5, and its symbol, $21'P$. 
In the ACC symbol, the Bravais lattice symbol $P$ is placed at the end to distinguish $21'P$ from a space-time group symbol, e.g., $P112/m$.
The second line gives its magnetic point group $P_M$ and first cohomology group $H^1$.
Each subsequent row gives one non-equivalent space-time group, 
where the 1st, 2nd, and 3rd columns present the group symbol, the symmetry generators, and the 
corresponding element of the cohomology group, respectively.
For compactness in the classification tables, we abbreviate a combined operation $(g|u[g])$ as $gT_\alpha^{1/n}$ (or a product of such factors), where $T_\alpha^{1/n}$ denotes the fractional translation specified for $\alpha=x,y,t$ in the corresponding table caption.

For this ACC, the fractional translations assigned by the two cohomology group generators are
\[
\begin{aligned}
u_1^{(5)}[R_\pi]&=\frac12a_3^{\text{TP}},
&
u_1^{(5)}[m_t]&=0,\\
u_2^{(5)}[R_\pi]&=0,
&
u_2^{(5)}[m_t]&=\frac12a_1^{\text{TP}}.
\end{aligned}
\]
The superscript labels the index of ACC, and the subscript marks the generator of the cohomology group. 
The argument of $u$ specifies the magnetic point-group operation to which the fractional translation is assigned.

For the space time group $P112/m$, $P$ denotes a primitive lattice, while $2/m$ in the third symmetry position represents the twofold spatial rotation $R_\pi$ and the temporal reflection $m_t$.
Replacing $2$ by $2_1$ leads to $P112_1/m$, in which $2_1$ denotes the time-screw rotation $R_\pi T_t^{1/2}$.
Replacing $m$ by $a$ gives rise to  $P112/a$, in which $a$ denotes the glide time-reversal operation $g_t=m_tT_x^{1/2}$ with the glide translation along the $x$-direction.
Accordingly, $P112_1/a$ contains both the time-screw rotation and the glide time-reversal operation.
For brevity, the ACC superscript is suppressed within each table, such that $u_j^{(i)}$ is written as $u_j$ for ACC No.~$i$.

\paragraph{Space-time Hermann--Mauguin notation.}
We construct the space-time Hermann--Mauguin notation through the correspondence
\begin{equation}
    (x,y,z)_{\mathrm{3D}}
    \longleftrightarrow(x,y,t)_{\mathrm{ST}},
    \label{eq:HM_coordinate_correspondence}
\end{equation}
where the 3rd spatial coordinate is replaced by time.

For example, in $P11m$ (ACC No.~4), listed in the left column of Table~\ref{tb:ptmnclnc}, the last $m$ denotes time-reversal $m_t$ rather than a spatial reflection.
Similarly, in $P121$ (ACC No.~9), listed in the left column of Table~\ref{tb:prmnclnc}, the middle $2$ denotes $m_xm_t$, namely reflection about the $y$-axis combined with time-reversal, rather than a twofold rotation entirely in space.

Glide symbols involving the temporal direction also acquire a space-time interpretation.
In $P112/a$ (ACC No.~5), listed in the left column of Table~\ref{tb:ptmnclnc}, the final $a$ denotes the glide-time-reversal operation $m_tT_x^{1/2}$, in which time-reversal is combined with half a lattice translation along the 
$x$ -direction.
In $Pmc2_1$ (ACC No.~16), listed in the left column of Table~\ref{tb:po2}, the middle $c$ denotes the time-glide reflection $m_yT_t^{1/2}$, in which the spatial reflection $m_y$ is combined with half a temporal period.

The detailed classification for each crystal system is presented below.
For each Bravais lattice, the lattice basis vectors are introduced first.
The ACC associated with the maximal magnetic point group of the lattice is then discussed, together with 
associated nonsymmorphic symmetries.
The magnetic point-group symmetry is subsequently lowered, and the corresponding ACCs are presented.

\subsection{Triclinic crystal system}
\begin{figure}[h] 
\caption{The basis vectors (blue arrows) for the primitive lattice in the space-time triclinic crystal system.}    
\includegraphics[width=0.4\linewidth]{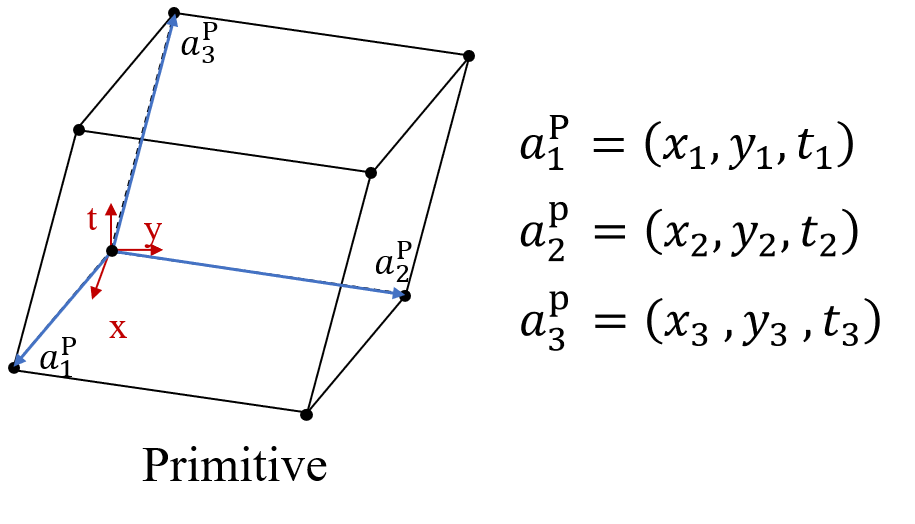}
\label{fig:triclinic_crystal_system}
\end{figure}

The space-time primitive triclinic lattice requires no constraints on its basis vectors as illustrated in Fig. \ref{fig:triclinic_crystal_system}.
We write the three basis vectors shown in the figure as
\begin{equation}
    a_1^{\text{P}}=(x_1,y_1,t_1),\qquad
    a_2^{\text{P}}=(x_2,y_2,t_2),\qquad
    a_3^{\text{P}}=(x_3,y_3,t_3),
\end{equation}
where they are linearly independent and otherwise unconstrained.
Therefore, the allowed point operation is only $R_\pi m_t$ which reverse all three 
space-time coordinates and it cannot carry a nontrivial fractional translation. 
Together with the trivial magnetic point, this crystal system yields two ACCs and two space-time groups, as summarized in Table~\ref{tb:triclinic}.

\begin{table}[h]
    \caption{2+1 dimensional space-time groups for the primitive triclinic
    lattice. This crystal system contains 2 ACCs and 2 space-time groups.}
    \centering
    \setlength{\extrarowheight}{1.9pt}

    \begin{tabular}{ccc}
        \toprule
        \multicolumn{3}{c}{\textbf{Primitive Triclinic}} \\[4pt]

        \multicolumn{3}{l}{\textbf{ACC No. 1:} $1P$} \\
        \multicolumn{3}{l}{$P_M=1,\qquad H^1(1,V/T_L)=\mathbb{I}$} \\[3pt]

        $P1$ & & $1$ \\[5pt]

        \multicolumn{3}{l}{\textbf{ACC No. 2:} $2'P$} \\
        \multicolumn{3}{l}{$P_M=2',\qquad H^1(2',V/T_L)=\mathbb{I}$} \\[3pt]

        $P\bar{1}$ & $R_{\pi}m_t$ & $1$ \\
        \bottomrule
    \end{tabular}
    \label{tab:triclinic}
    \label{tb:triclinic}
\end{table}
\clearpage

\clearpage
\subsection{ T-monoclinic crystal system}
\subsubsection{Primitive lattice}

The primitive T-monoclinic lattice exhibits two spatial lattice basis vectors, which span a two-dimensional oblique lattice, and a third lattice basis vector along the temporal direction.
The basis vectors are chosen as
\begin{equation}
    a_1^{\mathrm{TP}}=(x_1,y_1,0),\qquad
    a_2^{\mathrm{TP}}=(x_2,y_2,0),\qquad
    a_3^{\mathrm{TP}}=(0,0,t_0),
\end{equation}
illustrated in Fig. \ref{fig:T_monoclinic_crystal_system}(a).
There exist 3 ACCs $2P$, $11'P$, and $21'P$ (NO. $3\sim 5$), giving rise to 2, 2, and 4 non-equivalent space-time groups, respectively, as summarized in the left column of Table~\ref{tb:ptmnclnc}.

The maximal magnetic point group compatible with the primitive T-monoclinic Bravais lattice is $21'$, generated by the twofold spatial rotation $R_\pi$ and time-reversal $m_t$.
The corresponding ACC is $21'P$
(NO. 5).
The nonsymmorphic operations are listed below:
$R_\pi$ can be associated the fractional translation $a_3^{\text{TP}}/2$, producing a time-screw rotation, and 
$m_t$ can carry $a_1^{\text{TP}}/2$ or $a_2^{\text{TP}}/2$, producing glide-time-reversal operations along the two spatial lattice-basis directions,
respectively.
The first cohomology group of the ACC $21'P$ (No.~5) is generated by
\be
\begin{aligned}
     & u_{1}^{(5)}[R_{\pi}]=a_{3}^{\text{TP}}/2, &  & u_{1}^{(5)}[m_{t}]=0                      \\
     & u_{2}^{(5)}[R_{\pi}]=0,            &  & u_{2}^{(5)}[m_{t}]=\frac{1}{2}a_{1}^{\text{TP}}  \\
     & u_{3}^{(5)}[R_{\pi}]=0,            &  & u_{3}^{(5)}[m_{t}]=\frac{1}{2}a_{2}^{\text{TP}}.
\end{aligned}
\ee
The generators $u_{2,3}$ represent glide-time-reversal operations along two spatial lattice-basis directions, respectively, which are equivalent under relabeling of the two spatial lattice basis vectors. 

By symmetry reduction based on the magnetic point group $21'$, we arrive at the ACCs $2P$ (No.~3) and $11'P$ (No.~4) by preserving $R_\pi$ and $m_t$, respectively.
Their nonsymmorphic operations are inherited from those associated with the ACC $21'P$:
The corresponding cohomology-group generators are
\be
\begin{aligned}
     & u_{1}^{(3)}[R_\pi]=\frac{1}{2}a_{3}^{\text{TP}}; \\
     & u_{1}^{(4)}[m_t]=\frac{1}{2}a_{1}^{\text{TP}}    \\
     & u_{2}^{(4)}[m_t]=\frac{1}{2}a_{2}^{\text{TP}}.
\end{aligned}
\ee
For $11'P$, the two generators are equivalent.

\subsubsection{Base-centered lattice}
\begin{figure}[h]
        \caption{The basis vectors (blue arrows) for the  (a) primitive and (b) base-centered    Bravais lattices in the space-time T-monoclinc crystal system.}
 \includegraphics[width=\linewidth]
 {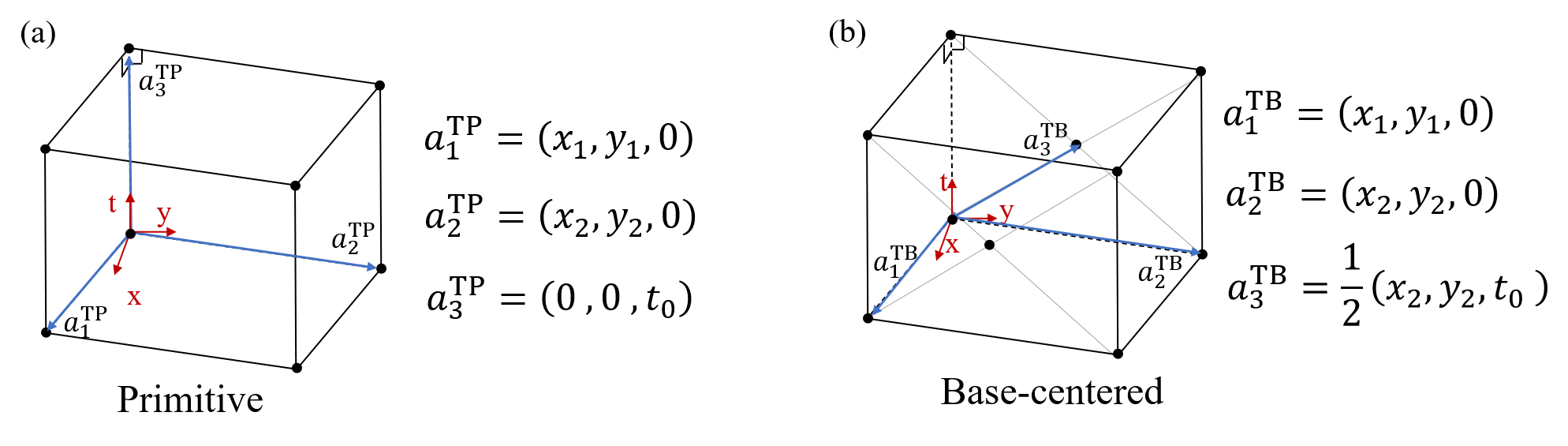}

    \label{fig:T_monoclinic_crystal_system}
\end{figure}

The base-centered T-monoclinic lattice is obtained by adding a center point at the surface spanned by the second and third basis vectors of the primitive lattice
as shown in Fig. \ref{fig:T_monoclinic_crystal_system}(b).
The basis vectors, written in terms of the primitive basis, are
\begin{equation}
    a_1^{\mathrm{TB}}=a_1^{\mathrm{TP}},\qquad
    a_2^{\mathrm{TB}}=a_2^{\mathrm{TP}},\qquad
    a_3^{\mathrm{TB}}=\tfrac12\bigl(a_2^{\mathrm{TP}}+a_3^{\mathrm{TP}}\bigr).
\end{equation}
There exist 3 ACCs $2A$, $11'A$, and $21'A$ (NO. $6\sim 8$), giving rise to 1, 2, and 2 non-equivalent space-time groups, respectively, as summarized in the right column of Table~\ref{tb:ctmncln}.

The maximal magnetic point group for the base-centered T-monoclinic lattice remains $21'$.
The corresponding ACC is $21'A$ (No.~8).
When applying the time-screw rotation for the primitive lattice to the centered one, 
it becomes symmorphic since the translation turns out to be integer due to the added lattice point. 
Similarly, the glide time-reversal operation
along the direction of $a_2^{\text{TP}}$ for the primitive lattice becomes symmorphic when applied to the centered lattice. 
Consequently, the only nonsymmorphic operation is a glide-time-reversal, which carries the translation $a_1^{\text{TP}}/2$.

We consider the magnetic point subgroups of  $21'$ generated by $R_\pi$ and $m_t$, respectively, and the corresponding ACCs 
are $2A$ (No.~6) and $11'A$ (No.~7).  
The ACC $2A$ can only give rise to the symmorphic space-time group, while the ACC $11'A$ possesses only one non-symmorphic space-time group by inheriting the glide-time-reversal operation of $21'A$.

The cohomology-group generators for ACCs $11'A$ (No.~7) and $21'A$ (No.~8) are
\be
\begin{aligned}
     & u_1^{(7)}[m_t]=\frac{1}{2}a_1^{\text{TP}};                                       \\
     & u_1^{(8)}[R_\pi]=0,
     &                                     & u_1^{(8)}[m_t]=\frac{1}{2}a_1^{\text{TP}}.
\end{aligned}
\ee

 
\begin{table}[h]
   \caption{2+1 dimensional space-time groups for the T-monoclinic crystal
    system. This crystal system contains 6 ACCs and 13 space-time groups. In the primitive panel, the glide reflection $g_t$ and fractional temporal translation $T_t^{1/2}$ act as
    $g_t(x,y,t)=(x,y,-t)+a_1^{\mathrm{TP}}/2$ and
    $T_t^{1/2}(x,y,t)=(x,y,t)+a_3^{\mathrm{TP}}/2$ respectively.}
    \centering
    \setlength{\extrarowheight}{2pt}
    \setlength{\tabcolsep}{5pt}
    \begin{tabular}{@{}lll@{\hspace{3.5em}}lll@{}}
        \toprule
        \multicolumn{3}{c}{\textbf{Primitive T-monoclinic}}
        & \multicolumn{3}{c}{\textbf{Base-centered T-monoclinic}} \\[2pt]
        \cmidrule(r){1-3}\cmidrule{4-6}

        \multicolumn{3}{l}{\textbf{ACC No. 3:} $2P$}
        & \multicolumn{3}{l}{\textbf{ACC No. 6:} $2A$} \\
        \multicolumn{3}{l}{$P_M=2,\qquad H^1(2,V/T_L)=\mathbb{Z}_2$}
        & \multicolumn{3}{l}{$P_M=2,\qquad H^1(2,V/T_L)=\mathbb{I}$} \\[3pt]
        $P112$   & $R_{\pi}$          & $1$
        & $A112$ & $R_{\pi}$          & $1$ \\
        $P112_1$ & $R_{\pi}T_t^{1/2}$ & $u_1$
        &        &                     &     \\

        \addlinespace[5pt]
        \multicolumn{3}{l}{\textbf{ACC No. 4:} $11'P$}
        & \multicolumn{3}{l}{\textbf{ACC No. 7:} $11'A$} \\
        \multicolumn{3}{l}{$P_M=11',\qquad H^1(11',V/T_L)=\mathbb{Z}_2^2$}
        & \multicolumn{3}{l}{$P_M=11',\qquad H^1(11',V/T_L)=\mathbb{Z}_2$} \\[3pt]
        $P11m$ & $m_t$ & $1$
        & $A11m$ & $m_t$ & $1$ \\
        $P11a$ & $g_t$ & $u_1$
        & $A11a$ & $g_t$ & $u_1$ \\

        \addlinespace[5pt]
        \multicolumn{3}{l}{\textbf{ACC No. 5:} $21'P$}
        & \multicolumn{3}{l}{\textbf{ACC No. 8:} $21'A$} \\
        \multicolumn{3}{l}{$P_M=21',\qquad H^1(21',V/T_L)=\mathbb{Z}_2^3$}
        & \multicolumn{3}{l}{$P_M=21',\qquad H^1(21',V/T_L)=\mathbb{Z}_2$} \\[3pt]
        $P112/m$   & $R_{\pi},m_t$          & $1$
        & $A112/m$ & $R_{\pi},m_t$          & $1$ \\
        $P112_1/m$ & $R_{\pi}T_t^{1/2},m_t$ & $u_1$
        & $A112/a$ & $R_{\pi},g_t$          & $u_1$ \\
        $P112/a$   & $R_{\pi},g_t$          & $u_2$
        &          &                         &       \\
        $P112_1/a$ & $R_{\pi}T_t^{1/2},g_t$ & $u_1\cdot u_2$
        &          &                         &       \\
        \bottomrule
    \end{tabular}

    \label{tab:t_monoclinic}
    \label{tb:ptmnclnc}
    \label{tb:ctmncln}
\end{table}
\clearpage

\clearpage
\subsection{  R-monoclinic crystal system}

\subsubsection{Primitive lattice}
The primitive R-monoclinic lattice has two lattice basis vectors in the $x$--$t$ plane, and a third lattice basis vector along the spatial $y$ direction.
The basis vectors are shown in Fig. \ref{fig:R_monoclinic_crystal_system}(a).
We choose them as
\begin{equation}
    a_1^{\mathrm{RP}}=(x_1,0,t_1),\qquad
    a_2^{\mathrm{RP}}=(x_2,0,t_2),\qquad
    a_3^{\mathrm{RP}}=(0,y_0,0).
\end{equation}
There exist 3 ACCs $m^\prime P$, $mP$, and $m^\prime m2^\prime P$ (NO. $9\sim 11$), giving rise to 2, 2, and 4 non-equivalent space-time groups, respectively, summarized in the left column of Table~\ref{tb:prmnclnc}.

The maximal magnetic point group for the primitive R-monoclinic lattice is $m'm2'$.
The associated ACC is denoted as $m'm2'P$ (No.~11), which is generated by the reflection $m_y$ and the combined operation $m_xm_t$.
The non-symmorphic operations are listed as follows: $m_y$ may be combined with $a_1^{\text{RP}}/2$ or $a_2^{\text{RP}}/2$ to form a glide reflection; $m_xm_t$ is followed by $a_3^{\text{RP}}/2$ to form a twofold screw ``rotation" along the $y$-direction.
(Here, ``rotation" refers to the simultaneous reversal of time and one spatial axis.)
The 1st cohomology group is generated 
by
\be
\begin{aligned}
     & u_1^{(11)}[m_y]=0,
     &                                  & u_1^{(11)}[m_xm_t]=\frac12a_3^{\text{RP}}, \\
     & u_2^{(11)}[m_y]=\frac12a_1^{\text{RP}},
     &                                  & u_2^{(11)}[m_xm_t]=0,               \\
     & u_3^{(11)}[m_y]=\frac12a_2^{\text{RP}},
     &                                  & u_3^{(11)}[m_xm_t]=0.
\end{aligned}
\ee
The glide generators $u_{2,3}$ are equivalent under relabeling of the two lattice basis vectors in the $x$--$t$ plane, and we choose $u_2$ as the representative.

We consider the magnetic point subgroups of $m'm2'$ generated by $m_xm_t$ and $m_y$, respectively.
The corresponding ACCs 
are $m'P$ (No.~9) and $mP$ (No.~10),
respectively.
Their nonsymmorphic operations are inherited from the retained symmetry operations in $m'm2'$:
$m'P$ is followed by the twofold screw 
``rotation" around the $y$ direction;
$mP$ is associated with fractional translations in the $x$--$t$ plane.
Their cohomology-group generators are
\be
\begin{aligned}
     & u_1^{(9)}[m_xm_t]=\frac12a_3^{\text{RP}},                                    \\
     & u_1^{(10)}[m_y]=\frac12a_1^{\text{RP}},
     &                                    & u_2^{(10)}[m_y]=\frac12a_2^{\text{RP}}.
\end{aligned}
\ee
For $mP$, the three nonzero elements in the cohomology group are equivalent under changes of the lattice basis in the $x$--$t$ plane.



\subsubsection{Base-centered lattice}
\begin{figure}[h]
    \caption{The basis vectors (blue arrows) for the (a) primitive and (b) base-centered Bravais lattices in the space-time R-monoclinc crystal system.}
 
 \includegraphics[width=0.9\linewidth]{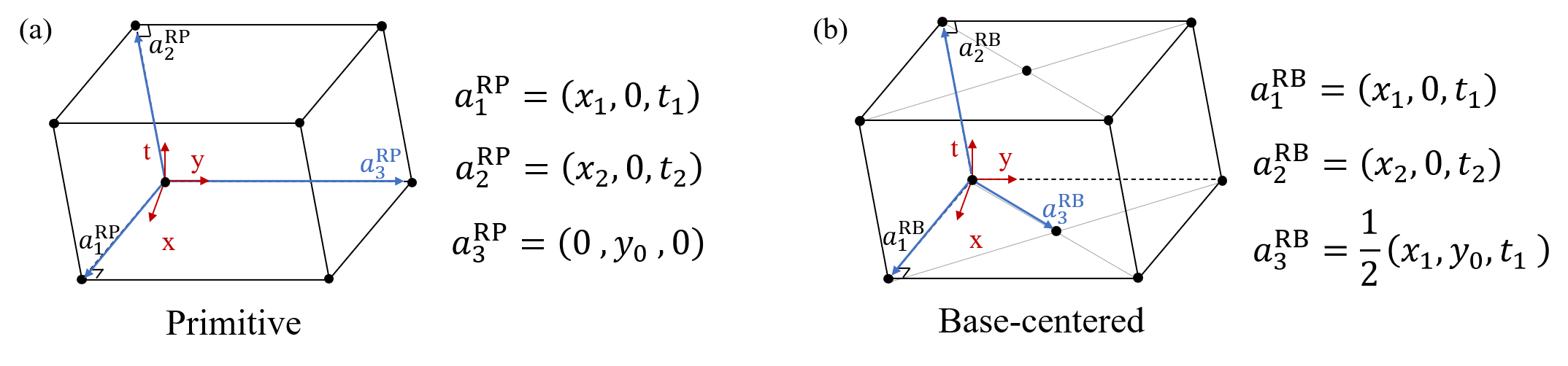}
    \label{fig:R_monoclinic_crystal_system}
\end{figure}

The base-centered R-monoclinic lattice is obtained by adding a center point on the face spanned by $a_1^{\text{RP}}$ and $a_3^{\text{RP}}$.
We choose the lattice basis vectors as shown in Fig. \ref{fig:R_monoclinic_crystal_system}(b).
The basis vectors, written in terms of the primitive basis, are
\begin{equation}
    a_1^{\mathrm{RB}}=  a_1^{\mathrm{RP}},\qquad
    a_2^{\mathrm{RB}}=a_2^{\mathrm{RP}},\qquad
    a_3^{\mathrm{RB}}=\tfrac12\bigl(a_1^{\mathrm{RP}}+a_3^{\mathrm{RP}}\bigr).
\end{equation}
There exist 3 ACCs $m'C$, $mC$, and $m'm2'C$ (NO. $12\sim 14$), giving rise to 1, 2, and 2 non-equivalent space-time groups, respectively, and are summarized in the right column of Table~\ref{tb:crmnclnc}.

The maximal magnetic point group for the base-centered R-monoclinic lattice remains $m'm2'$, and the associated ACC
corresponds to $m'm2'C$ (No.~14).
The additional lattice point implies
\begin{equation}
    \frac12a_1^{\text{RP}}\equiv-\frac12a_3^{\text{RP}}\pmod{T_L}.
\end{equation}
Consequently, when applying the screw translation $a_3^{\text{RP}}/2$ and the glide translation $a_1^{\text{RP}}/2$ for the primitive lattice to the base-centered case, they become symmorphic. 
The only nonsymmorphic operation associated with ACC $m'm2'C$ is the glide reflection carrying $a_2^{\text{RP}}/2$.

We consider the magnetic point subgroups of $m'm2'$ generated by $m_xm_t$ and $m_y$, respectively.
The corresponding ACCs are $m'C$ (No.~12) 
and $mC$ (No.~13), respectively.
The former case only leads to symmorphic space-time group, while the latter inherits the glide reflection associated with the ACC $m'm2'C$.
The corresponding cohomology-group generators for ACCs $mC$ (No.~13) and $m'm2'C$ (No.~14) are
\be
\begin{aligned}
     & u_1^{(13)}[m_y]=\frac12a_2^{\text{RP}},                         \\
     & u_1^{(14)}[m_y]=\frac12a_2^{\text{RP}},
     &                                  & u_1^{(14)}[m_xm_t]=0.
\end{aligned}
\ee

 
\begin{table}[h]
        \caption{2+1 dimensional space-time groups for the R-monoclinic crystal
    system. This crystal system contains 6 ACCs and 13 space-time groups. The glide reflection $g_y$ and fractional translation $T_y^{1/2}$ are defined as
    $g_y(x,y,t)=(x,-y,t)+a_1^{\mathrm{RP}}/2$ and
    $T_y^{1/2}(x,y,t)=(x,y,t)+a_3^{\mathrm{RP}}/2$, respectively.}
    \centering
    \setlength{\extrarowheight}{2pt}
    \setlength{\tabcolsep}{5pt}

    \begin{tabular}{@{}lll@{\hspace{3.5em}}lll@{}}
        \toprule
        \multicolumn{3}{c}{\textbf{Primitive R-monoclinic}}
        & \multicolumn{3}{c}{\textbf{Base-centered R-monoclinic}} \\[2pt]
        \cmidrule(r){1-3}\cmidrule{4-6}

        \multicolumn{3}{l}{\textbf{ACC No. 9:} $m'P$}
        & \multicolumn{3}{l}{\textbf{ACC No. 12:} $m'C$} \\
        \multicolumn{3}{l}{$P_M=m',\qquad H^1(m',V/T_L)=\mathbb{Z}_2$}
        & \multicolumn{3}{l}{$P_M=m',\qquad H^1(m',V/T_L)=\mathbb{I}$} \\[3pt]
        $P121$   & $m_xm_t$          & $1$
        & $C121$ & $m_xm_t$          & $1$ \\
        $P12_11$ & $m_xm_tT_y^{1/2}$ & $u_1$
        &        &                    &     \\

        \addlinespace[5pt]
        \multicolumn{3}{l}{\textbf{ACC No. 10:} $mP$}
        & \multicolumn{3}{l}{\textbf{ACC No. 13:} $mC$} \\
        \multicolumn{3}{l}{$P_M=m,\qquad H^1(m,V/T_L)=\mathbb{Z}_2^2$}
        & \multicolumn{3}{l}{$P_M=m,\qquad H^1(m,V/T_L)=\mathbb{Z}_2$} \\[3pt]
        $P1m1$ & $m_y$ & $1$
        & $C1m1$ & $m_y$ & $1$ \\
        $P1a1$ & $g_y$ & $u_1$
        & $C1c1$ & $g_y$ & $u_1$ \\

        \addlinespace[5pt]
        \multicolumn{3}{l}{\textbf{ACC No. 11:} $m'm2'P$}
        & \multicolumn{3}{l}{\textbf{ACC No. 14:} $m'm2'C$} \\
        \multicolumn{3}{l}{$P_M=m'm2',\qquad H^1(m'm2',V/T_L)=\mathbb{Z}_2^3$}
        & \multicolumn{3}{l}{$P_M=m'm2',\qquad H^1(m'm2',V/T_L)=\mathbb{Z}_2$} \\[3pt]
        $P12/m1$   & $m_y,m_xm_t$          & $1$
        & $C12/m1$ & $m_y,m_xm_t$          & $1$ \\
        $P12_1/m1$ & $m_y,m_xm_tT_y^{1/2}$ & $u_1$
        & $C12/c1$ & $g_y,m_xm_t$          & $u_1$ \\
        $P12/a1$   & $g_y,m_xm_t$          & $u_2$
        &          &                        &       \\
        $P12_1/a1$ & $g_y,m_xm_tT_y^{1/2}$ & $u_1\cdot u_2$
        &          &                        &       \\
        \bottomrule
    \end{tabular}
    \label{tab:r_monoclinic}
    \label{tb:prmnclnc}
    \label{tb:crmnclnc}
\end{table}
\clearpage

\clearpage
\subsection{ Orthorhombic crystal system}

\subsubsection{Primitive orthorhombic lattice}
The primitive orthorhombic lattice has three mutually orthogonal primitive directions aligned with the two spatial axes and the temporal one. 
We define the corresponding lattice vectors in Fig. \ref{fig:orthorhombic_crystal_system}(a).
Explicitly,
\begin{equation}
    a_1^{\mathrm{P}}=(x_0,0,0),\qquad
    a_2^{\mathrm{P}}=(0,y_0,0),\qquad
    a_3^{\mathrm{P}}=(0,0,t_0).
\end{equation}
There exist 4 ACCs $m'm'2P$, $mm2P$, $m1'P$, and $mm21'P$
(NO. $15\sim 18$), giving rise to 6, 10, 16, and 36 non-equivalent space-time groups, respectively, as summarized in Table~\ref{tb:po2}.


The maximal magnetic point group compatible with the primitive
orthorhombic Bravais lattice is
$mm21'$, generated by the two reflections $m_x$ and $m_y$ and time-reversal $m_t$.
The corresponding ACC is $mm21'P$ (No.~18).
The nonsymmorphic operations are listed below: $m_x$ can be combined with $a_2^{\text{P}}/2$ to form a glide reflection or with $a_3^{\text{P}}/2$ to form a time-glide reflection; 
similar arrangements also apply to 
$m_y$; time-reversal $m_t$ can carry $a_1^{\text{P}}/2$ or $a_2^{\text{P}}/2$, producing glide-time-reversal operations along the two spatial directions, respectively. 
Consequently, the cohomology group is generated by
\beaa
&u_{1}^{(18)}[m_x]=\frac{1}{2}{a^{\text{P}}_2},  &&u_{1}^{(18)}[m_y]=0, &&u_{1}^{(18)}[m_t]=0\\
&u_{2}^{(18)}[m_x]=\frac{1}{2}{a^{\text{P}}_3},  &&u_{2}^{(18)}[m_y]=0, &&u_{2}^{(18)}[m_t]=0\\
&u_{3}^{(18)}[m_x]=0,  &&u_{3}^{(18)}[m_y]=\frac{1}{2}{a^{\text{P}}_3}, &&u_{3}^{(18)}[m_t]=0\\
&u_{4}^{(18)}[m_x]=0,  &&u_{4}^{(18)}[m_y]=\frac{1}{2}{a^{\text{P}}_1}, &&u_{4}^{(18)}[m_t]=0\\
&u_{5}^{(18)}[m_x]=0,  &&u_{5}^{(18)}[m_y]=0, &&u_{5}^{(18)}[m_t]=\frac{1}{2}{a^{\text{P}}_1}\\
&u_{6}^{(18)}[m_x]=0,  &&u_{6}^{(18)}[m_y]=0, &&u_{6}^{(18)}[m_t]=\frac{1}{2}{a^{\text{P}}_2}.
\eeaa
The generators $u_{1,4}$ represent glide-reflection operations, $u_{2,3}$ represent time-glide-reflection operations, and $u_{5,6}$ represent glide-time-reversal operations. 
Switching two spatial directions exchanges $u_1$ and $u_4$, $u_2$ and $u_3$, and $u_5$ and $u_6$.
Hence, this ACC contains $36$ non-equivalent space-time groups.

The magnetic point group of ACC $m'm'2P$ (No.~15) is generated by the combined reflection and time-reversal operation $m_ym_t$ and the twofold spatial rotation $m_xm_y=R_\pi$. 
The former and the latter can be followed by $a_1^{\text{P}}/2$ and $a_3^{\text{P}}/2$,
respectively, which become screw ``rotation" and time-screw rotation,
respectively. 
Its 1st cohomology group is $\mathbb{Z}_2^3$, whose generators are
\beaa
&u_{1}^{(15)}[m_{y} m_{t}]=0, &&u_{1}^{(15)}[m_{x} m_{y}]=\frac{1}{2} {a_{3}^{\text{P}}} \\
&u_{2}^{(15)}[m_{y} m_{t}]=\frac{1}{2} {a_{1}^{\text{P}}}, &&u_{2}^{(15)}[m_{x} m_{y}]=0 \\
&u_{3}^{(15)}[m_{y} m_{t}]=0,  &&u_{3}^{(15)}[m_{x} m_{y}]=\frac{1}{2} {a_{2}^{\text{P}}}
\eeaa
The generators $u_{1,2,3}$ represent
time-screw rotation, screw ``rotations",
respectively. 
Since these directions can be interchanged by an allowed global transformation, $u_{2,3}$ are equivalent under the normalizer.


The magnetic subgroup $mm2$ is obtained from the magnetic group $mm21'$ by removing one generator, and so does $m1'$. 
$mm2$ retains the two reflections $m_x$ and $m_y$ but removes $m_t$, whereas $m1'$ retains $m_y$ and $m_t$ but removes $m_x$.
The corresponding ACCs are denoted 
ACC $mm2P$ (No.~16) and $m1'P$ (No.~17), respectively. 
Their possible nonsymmorphic 
operations are therefore inherited directly from those of $mm21'P$.

The corresponding cohomology-group generators for $mm2P$ are
\be
\begin{aligned}
     & u_{1}^{(16)}[m_x]=\frac{1}{2}{a^{\text{P}}_2}, &  & u_{1}^{(16)}[m_y]=0                  \\
     & u_{2}^{(16)}[m_x]=\frac{1}{2}{a^{\text{P}}_3}, &  & u_{2}^{(16)}[m_y]=0                  \\
     & u_{3}^{(16)}[m_x]=0,                  &  & u_{3}^{(16)}[m_y]=\frac{1}{2}{a^{\text{P}}_3} \\
     & u_{4}^{(16)}[m_x]=0,                  &  & u_{4}^{(16)}[m_y]=\frac{1}{2}{a^{\text{P}}_1} \\
\end{aligned}\ee
Interchanging the two spatial directions maps $u_1$ to $u_4$ and $u_2$ to $u_3$; hence the generators in each pair are equivalent under the normalizer.

The corresponding cohomology-group generators for $m1'P$ are
\beaa
&u_{1}^{(17)}[m_y]=\frac{1}{2}{a^{\text{P}}_3}, \ &&u_{1}^{(17)}[m_t]=0\\
&u_{2}^{(17)}[m_y]=\frac{1}{2}{a^{\text{P}}_1}, \ &&u_{2}^{(17)}[m_t]=0\\
&u_{3}^{(17)}[m_y]=0, \ &&u_{3}^{(17)}[m_t]=\frac{1}{2}{a^{\text{P}}_1}\\
&u_{4}^{(17)}[m_y]=0, \ &&u_{4}^{(17)}[m_t]=\frac{1}{2}{a^{\text{P}}_2}\\
\eeaa
No nontrivial normalizer identification occurs among the sixteen elements in the cohomology group, so they represent sixteen distinct space-time groups.


\subsubsection{R-base-centered orthorhombic lattice}
There are two non-equivalent base-centered orthorhombic Bravais lattices.
We first consider the R-base-centered one, whose centering point lies in the spatial $x$-$y$ face; consequently, the temporal axis is perpendicular to the centered face.
The basis vectors $a_{1,2,3 }^{\text{RB}}$ are defined in Fig. \ref{fig:orthorhombic_crystal_system}(b).
The basis vectors, written in terms of the primitive basis, are
\begin{equation}
    a_1^{\mathrm{RB}}=\tfrac12\bigl(a_1^{\mathrm{P}}-a_2^{\mathrm{P}}\bigr),\qquad
    a_2^{\mathrm{RB}}=\tfrac12\bigl(a_1^{\mathrm{P}}+a_2^{\mathrm{P}}\bigr),\qquad
    a_3^{\mathrm{RB}}=a_3^{\mathrm{P}}.
\end{equation}
There exist 4 ACCs $m'm'2C$, $mm2C$,
$m1'C$, $mm21'C$
(NO. $19\sim 22$), 
which give rise to 
2, 3, 4, and 6
non-equivalent space-time groups, respectively, and are summarized in the upper-left panel of Table~\ref{tb:rbase_orthg}.


Because of the centering point, nonsymmorphic operations carrying half-translations along two spatial directions differ only by a unit vector of the R-base-centered lattice:
\begin{equation}
\frac{1}{2}a_1^{\text{P}}-\frac{1}{2}a_2^{\text{P}}=a_1^{\text{RB}}.
\label{eq:R-base-center—orth}
\end{equation}
Therefore, the vector systems with these two half-translations are equivalent.

The maximal magnetic point group for the
R-base-centered orthorhombic 
lattice $mm21'$, generated by $m_x$, $m_y$, and $m_t$.
The associated ACC
corresponds to $mm21'C$ (No.~22).
By identifying the equivalent fractional translations in terms of 
Eq. (\ref{eq:R-base-center—orth}),
the remaining nonsymmorphic symmetries are two time-glide reflections, obtained by attaching $a_3^{\text{P}}/2$ to $m_x$ or $m_y$, and one glide-time-reversal operation, obtained by attaching $a_1^{\text{P}}/2$ to $m_t$.

The corresponding cohomology-group generators are
\be
\begin{aligned}
    u_{1}^{(22)}[m_x] & =\frac{1}{2}a_3^{\text{P}}, & u_{1}^{(22)}[m_y] & =u_{1}^{(22)}[m_t]=0, \\
    u_{2}^{(22)}[m_y] & =\frac{1}{2}a_3^{\text{P}}, & u_{2}^{(22)}[m_x] & =u_{2}^{(22)}[m_t]=0, \\
    u_{3}^{(22)}[m_t] & =\frac{1}{2}a_1^{\text{P}}, & u_{3}^{(22)}[m_x] & =u_{3}^{(22)}[m_y]=0.
\end{aligned}
\ee
The generators $u_{1,2}$ are equivalent under the normalizer that exchanges two spatial directions, which reduces the 8 elements in the cohomology group to the 6 non-equivalent space-time groups.

The ACC $m'm'2C$ (No.~19) keeps only the time-screw rotation obtained by attaching $a_3^{\text{P}}/2$ to $m_xm_y$. 
Its cohomology group generator behaves as
\be
\begin{aligned}
    u_{1}^{(19)}[m_y m_t]=0,\qquad u_{1}^{(19)}[m_x m_y]=\frac{1}{2}a_3^{\text{P}},
\end{aligned}
\ee
which gives rise to two space-time groups.

The ACCs $m1'C$ (No.~21) and $mm2C$ (No.~20) retain different subsets of the nonsymmorphic operations of $mm21'C$.
The former retains one time-glide reflection that combines $a_3^{\text{P}}/2$ and $m_y$, and one glide-time-reversal that attaches $a_1^{\text{P}}/2$ to $m_t$. 
The latter retains the two time-glide reflections obtained by adding $a_3^{\text{P}}/2$ to $m_x$ or $m_y$.
The cohomology-group generators for 
the ACC $m1'C$ are
\be
\begin{aligned}
& u_{1}^{(21)}[m_y]=\frac{1}{2}a_3^{\text{P}}, &  & u_{1}^{(21)}[m_t]=0,                \\
& u_{2}^{(21)}[m_y]=0,                &  & u_{2}^{(21)}[m_t]=\frac{1}{2}a_1^{\text{P}}. \\
\end{aligned}
\ee
These two generators are 
non-equivalent under the normalizer, so the 4 elements in the cohomology group generate 4 non-equivalent space-time groups. 
The cohomology-group generators for 
the ACC $mm2C$ are
\be
\begin{aligned}
     & u_{1}^{(20)}[m_x]=\frac{1}{2}a_3^{\text{P}}, &  & u_{1}^{(20)}[m_y]=0;                \\
     & u_{2}^{(20)}[m_x]=0,                &  & u_{2}^{(20)}[m_y]=\frac{1}{2}a_3^{\text{P}}. \\
\end{aligned}
\ee
These two generators are equivalent under the normalizer by exchanging the two spatial directions, hence its 4 elements in the cohomology group give 3 non-equivalent space-time groups.

\subsubsection{T-base-centered orthorhombic lattice}
We next consider the T-base-centered orthorhombic lattice, whose centering point lies in the $y$--$t$ face; consequently, the temporal axis is parallel to the centered face.
The basis vectors $a_{1,2,3 }^{\text{TB}}$ are defined in Fig. \ref{fig:orthorhombic_crystal_system}(c).
The basis vectors, written in terms of the primitive basis, are
\begin{equation}
    a_1^{\mathrm{TB}}=a_1^{\mathrm{P}},\qquad
    a_2^{\mathrm{TB}}=\tfrac12\bigl(a_2^{\mathrm{P}}+a_3^{\mathrm{P}}\bigr),\qquad
    a_3^{\mathrm{TB}}=\tfrac12\bigl(-a_2^{\mathrm{P}}+a_3^{\mathrm{P}}\bigr).
\end{equation}
There exist 5 ACCs $m'm'2A$, $m_y1'A$,
$m_x1'A$, $mm2A$,
and $mm21'A$
(NO. $23\sim 27$), 
which give rise to 
2, 4, 4, 4 and 8
non-equivalent space-time groups, respectively, and are summarized in the lower-left panel of Table~\ref{tb:tbase_orthg}.


Because of the location of the centering point, the nonsymmorphic operations carrying half-translations along $y$ and $t$ directions differ only by a unit vector of the T-base-centered lattice:
\begin{equation}
\frac{1}{2}a_3^{\text{P}}-\frac{1}{2}a_2^{\text{P}}=a_3^{\text{TB}}.
\end{equation}
Hence, vector systems with these two half-translations are equivalent.

The maximal magnetic point group for the T-base-centered orthorhombic 
lattice $mm21'$, which is the same
as the R-based case. 
Nevertheless, the associated ACC
becomes $mm21'A$ (No.~27).
By identifying the equivalent fractional translations, the remaining non-symmorphic symmetries are two glide reflections, obtained by attaching $a_2^{\text{P}}/2$ to $m_x$, or, $a_1^{\text{P}}/2$ to $m_y$, and one glide-time-reversal operation, obtained by attaching $a_1^{\text{P}}/2$ to $m_t$.
The corresponding cohomology-group generators for $mm21'A$ are
\begin{equation}
    \begin{aligned}
                u_{1}^{(27)}[m_x] & =\frac{1}{2}a_2^{\text{P}}, & u_{1}^{(27)}[m_y]=u_{1}^{(27)}[m_t] & =0,     \\
                u_{2}^{(27)}[m_y] & =\frac{1}{2}a_1^{\text{P}}, & u_{2}^{(27)}[m_x]=u_{2}^{(27)}[m_t] & =0,     \\
                u_{3}^{(27)}[m_t] & =\frac{1}{2}a_1^{\text{P}}, & u_{3}^{(27)}[m_x]=u_{3}^{(27)}[m_y] & =0, & .
    \end{aligned}
\end{equation}
These three generators are non-equivalent under the normalizer, 
hence, the 8 elements in the cohomology group corresopond to 
8 non-equivalent space-time groups.

The ACC $m'm'2A$ (No.~23) retains only a glide-time-reversal operation, obtained by attaching $a_1^{\text{P}}/2$ to $m_ym_t$. 
Its cohomology group generator is
\beaa
&u_{1}^{(23)}[m_y m_t]=\frac{1}{2}a_1^{\text{P}}, &&u_{1}^{(23)}[m_x m_y]=0.
\eeaa
Thus, $m'm'2A$ contains two space-time groups.

The ACCs of $m_y1'A$ (No.~24) and $m_x1'A$ (No.~25) correspond to 
two choices of the reflection plane to the T-base-centered lattice. 
The ACC $m_y1'A$ retains a glide-reflection operation and a glide-time-reversal operation, both carrying $a_1^{\text{P}}/2$, whereas $m_x1'A$ retains a glide reflection carrying $a_2^{\text{P}}/2$ and a glide-time-reversal operation carrying $a_1^{\text{P}}/2$. Their cohomology-group generators are
\beaa
&u_{1}^{(24)}[m_y]=0,&&u_{1}^{(24)}[m_t]=\frac{1}{2}a_1^{\text{P}}\\
&u_{2}^{(24)}[m_y]=\frac{1}{2}a_1^{\text{P}},&&u_{2}^{(24)}[m_t]=0\\[2pt]
&u_{1}^{(25)}[m_x]=\frac{1}{2}a_2^{\text{P}},&&u_{1}^{(25)}[m_t]=0\\
&u_{2}^{(25)}[m_x]=0,&&u_{2}^{(25)}[m_t]=\frac{1}{2}a_1^{\text{P}}.
\eeaa
For each ACC, the two generators are non-equivalent under the normalizer, so the 4 elements in the cohomology group lead to 4 non-equivalent space-time groups.

The ACC $mm2A$ (No.~26) retains two glide reflections, whose cohomology-group generators are
\beaa
&u_{1}^{(26)}[m_x]=\frac{1}{2}a_2^{\text{P}},&&u_{1}^{(26)}[m_y]=0\\
&u_{2}^{(26)}[m_x]=0,&&u_{2}^{(26)}[m_y]=\frac{1}{2}a_1^{\text{P}}.
\eeaa
Since two spatial directions are non-equivalent in the T-base-centered lattice, the two generators are non-equivalent under the normalizer, and the 4 elements in the cohomology group give rise to  4 non-equivalent space-time groups.

\subsubsection{Body-centered orthorhombic lattice}
The body-centered orthorhombic lattice contains an additional lattice point at the center of the conventional space-time unit cell. Each orthorhombic MPG admits a single compatible ACC assignment on this Bravais lattice. 
We choose the basis vectors as shown in Fig. \ref{fig:orthorhombic_crystal_system}(d).
The basis vectors, written in terms of the primitive basis, are
\begin{equation}
    a_1^{\mathrm{I}}=\tfrac12\bigl(-a_1^{\mathrm{P}}+a_2^{\mathrm{P}}+a_3^{\mathrm{P}}\bigr),\qquad
    a_2^{\mathrm{I}}=\tfrac12\bigl(a_1^{\mathrm{P}}-a_2^{\mathrm{P}}+a_3^{\mathrm{P}}\bigr),\qquad
    a_3^{\mathrm{I}}=\tfrac12\bigl(a_1^{\mathrm{P}}+a_2^{\mathrm{P}}-a_3^{\mathrm{P}}\bigr).
\end{equation}
There exist 4 ACCs $m'm'2I$, $mm2I$,
$m1'I$, $mm21'I$
(NO. $28\sim 31$), 
which give rise to 
2, 3, 4, and 6
non-equivalent space-time groups, respectively, and are summarized in the upper-right panel of Table~\ref{tb:body_orthg}.


The maximal magnetic point group for the body-centered orthorhombic lattice $mm21'$, which is the same as the T- and R-based cases. 
The corresponding ACC is 
$mm21'I$ (No.~31).
In the body-centered lattice, $m_x$ and $m_y$ become glide-reflections by carrying $a_2^{\text{P}}/2$ and $a_1^{\text{P}}/2$, respectively, while $m_t$ can become a glide-time-reversal operation by carrying $a_1^{\text{P}}/2$. 
The corresponding cohomology group generators are
\begin{equation}
    \begin{aligned}
                u_{1}^{(31)}[m_x] & =\frac{1}{2}a_2^{\text{P}}, & u_{1}^{(31)}[m_y] & =0,                & u_{1}^{(31)}[m_t] & =0,                \\
                u_{2}^{(31)}[m_x] & =0,                & u_{2}^{(31)}[m_y] & =\frac{1}{2}a_1^{\text{P}}, & u_{2}^{(31)}[m_t] & =0,                \\
                u_{3}^{(31)}[m_x] & =0,                & u_{3}^{(31)}[m_y] & =0,                & u_{3}^{(31)}[m_t] & =\frac{1}{2}a_1^{\text{P}}.
    \end{aligned}
\end{equation}
The first two generators are equivalent under the normalizer that exchanges the two spatial directions, whereas the third is distinct. 
Hence, the ACC $mm21'I$ contains 
6 non-equivalent space-time groups.

For the ACC $m'm'2I$ (No.~28), there is only one nonsymmorphic space-time group. 
In this group, $m_xm_t$ and $m_ym_t$ simultaneously carry $(a_2^{\text{P}}+a_3^{\text{P}})/2$ and $(a_1^{\text{P}}+a_2^{\text{P}})/2$, respectively. 
Their product gives a third $2_1$ screw rotation along the temporal direction. 
The corresponding cohomology group generator is
\beaa
&u_{1}^{(28)}[m_xm_t]=\frac{1}{2}(a_2^{\text{P}}+a_3^{\text{P}}),&&u_{1}^{(28)}[m_ym_t]=\frac{1}{2}(a_1^{\text{P}}+a_2^{\text{P}}).
\eeaa
Hence, the ACC $m'm'2I$ contains two space-time groups.

The ACCs $mm2I$ (No.~29) and $m1'I$ (No.~30) are obtained from the maximal magnetic point group $mm21'$ by removing $m_t$ and $m_x$, respectively. 
Their cohomology group generators can therefore be constructed based on those of the ACC $mm21'I$ by removing the corresponding ones.  
For the ACC of $mm2I$, its cohomology group generators are 
\beaa
&u_{1}^{(29)}[m_x]=0,&&u_{1}^{(29)}[m_y]=\frac{1}{2}a_1^{\text{P}},\\
&u_{2}^{(29)}[m_x]=\frac{1}{2}a_2^{\text{P}},&&u_{2}^{(29)}[m_y]=0.
\eeaa
These two generators represent glide reflections and are equivalent under the normalizer that exchanges the two spatial directions. 
Hence, the ACC $mm2I$ contains 3 non-equivalent space-time groups.

On the other hand, for the ACC of $m1'I$, its cohomology group generators are
\beaa
&u_{1}^{(30)}[m_y]=\frac{1}{2}a_1^{\text{P}},&&u_{1}^{(30)}[m_t]=0,\\
&u_{2}^{(30)}[m_y]=0,&&u_{2}^{(30)}[m_t]=\frac{1}{2}a_1^{\text{P}}.
\eeaa
They represent a glide reflection and a glide-time-reversal operation, respectively, and are non-equivalent.
Thus, the ACC of $m1'I$ contains 4 non-equivalent space-time groups.

\subsubsection{Face-centered orthorhombic lattice}
\begin{figure}[p]
        \caption{The basis vectors (blue arrows) for the (a) primitive, (b) R-base-centered, (c) T-base-centered, (d) body-centered and (e) face-centered  Bravais lattices in the space-time orthorhombic crystal system.}
    \centering
 \includegraphics[width=\linewidth]{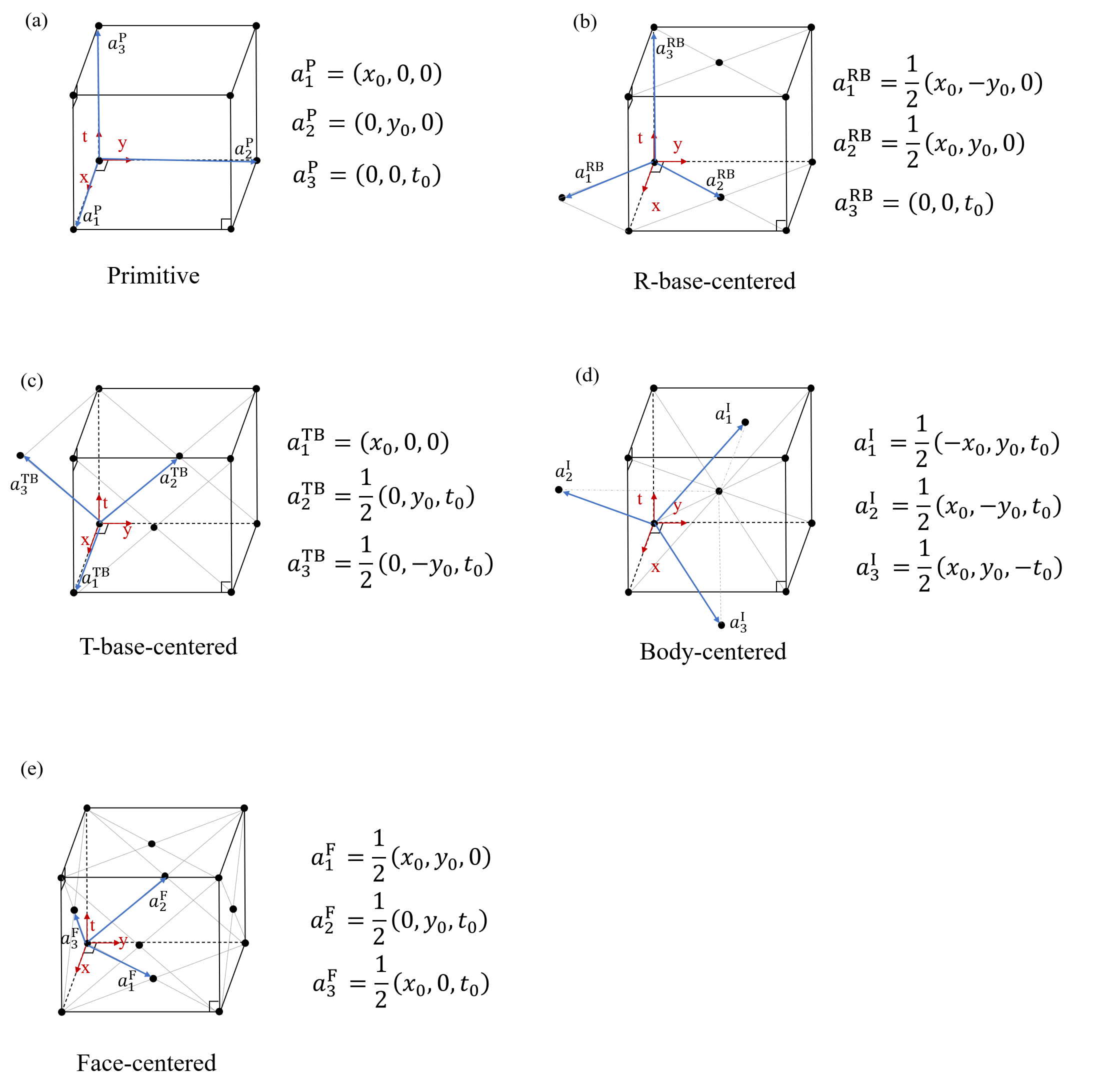}

    \label{fig:orthorhombic_crystal_system}
\end{figure}

The face-centered orthorhombic lattice contains additional lattice points at the centers of all three coordinate faces: the spatial $x$--$y$ face and the two space-time faces $x$--$t$ and $y$--$t$.
We choose the bases as shown in Fig. \ref{fig:orthorhombic_crystal_system}(e).
The basis vectors, written in terms of the primitive basis, are
\begin{equation}
    a_1^{\mathrm{F}}=\tfrac12\bigl(a_1^{\mathrm{P}}+a_2^{\mathrm{P}}\bigr),\qquad
    a_2^{\mathrm{F}}=\tfrac12\bigl(a_2^{\mathrm{P}}+a_3^{\mathrm{P}}\bigr),\qquad
    a_3^{\mathrm{F}}=\tfrac12\bigl(a_1^{\mathrm{P}}+a_3^{\mathrm{P}}\bigr).
\end{equation}
There exist 4 ACCs $m'm'2F$, $mm2F$,
$m1'F$, $mm21'F$
(NO. $32\sim 35$), 
which give rise to 
1, 2, 2, and 2
non-equivalent space-time groups, respectively, and are summarized in the lower-right panel of Table~\ref{tb:face_orthg}.


The nonsymmorphic symmetries of the face-centered lattice are different from those of the primitive and base-centered ones. 
Instead of carrying half of a primitive-lattice unit vector, they carry half of a face-centered unit vector, which corresponds to a 1/4-translation along a face diagonal.
Therefore, it is convenient to use the nonsymmorphic operations in terms of $a^{\text{F}}_{1,2,3}$.

The maximal magnetic point group for the face-centered orthorhombic lattice $mm21'$, which is the same as before. 
The corresponding ACC is $mm21'F$ (No.~35).
There is only one nonsymmorphic space-time group, in which $m_x$, $m_y$, and $m_t$ carry $a_2^{\text{F}}/2$, $a_3^{\text{F}}/2$, and $a_1^{\text{F}}/2$, respectively. These are 1/4-translations along the face-diagonals of $y$--$t$, $x$--$t$, and $x$--$y$, respectively, and
all of them occur simultaneously. 
The corresponding cohomology-group generator is
\begin{equation}
    u_{1}^{(35)}[m_x]=\frac{1}{2} {a_2^{\text{F}}},\quad
    u_{1}^{(35)}[m_y]=\frac{1}{2} {a_3^{\text{F}}},\quad
    u_{1}^{(35)}[m_t]=\frac{1}{2} {a_1^{\text{F}}}
\end{equation}
Hence, the ACC $mm21'F$ contains two space-time groups.

For the ACC $m'm'2F$ (No.~32), there is no nonsymmorphic operation, hence, only the symmorphic space-time group
exists. 

The ACCs $mm2F$ (No.~33) and $m1'F$ (No.~34) are obtained from the maximal magnetic point group $mm21'$ by removing $m_t$ and $m_x$, respectively. 
Their nonsymmorphic symmetries carry the same 1/4-translations as the corresponding operations in $mm21'F$. The corresponding cohomology-group generators are
\beaa
&u_{1}^{(33)}[m_x]=\frac{1}{2} {a_2^{\text{F}}},&&u_{1}^{(33)}[m_y] = \frac{1}{2} {a_3^{\text{F}}},\\
&u_{1}^{(34)}[m_y]=\frac{1}{2} {a_3^{\text{F}}},&&u_{1}^{(34)}[m_t] = \frac{1}{2} {a_1^{\text{F}}}.
\eeaa
Each ACC contains two space-time groups.

\clearpage

\clearpage
\begin{table}[p]
    \caption{2+1 dimensional space-time groups for the primitive orthorhombic
    lattice. This table contains 4 ACCs and 68 space-time groups. Here
    $T_x^{1/n}$, $T_y^{1/n}$, and $T_t^{1/n}$ denote
    translations by $a_1^{\mathrm{P}}/n$, $a_2^{\mathrm{P}}/n$, and
    $a_3^{\mathrm{P}}/n$, respectively.}
    \label{tab:primitive_orthorhombic}
    \centering
    \setlength{\extrarowheight}{0.8pt}
    \setlength{\tabcolsep}{3pt}
    \fontsize{6.2}{6.8}\selectfont
    \renewcommand{\arraystretch}{1.40}

    \begin{adjustbox}{max width=\dimexpr\linewidth+1em\relax,center}
    \begin{tabular}{@{}lll@{\hspace{2.5em}}lll@{}}
        \toprule
        \multicolumn{3}{c}{\textbf{Primitive orthorhombic: ACCs 15--17}}
        & \multicolumn{3}{c}{\textbf{Primitive orthorhombic: ACC 18}} \\[2pt]
        \cmidrule(r){1-3}\cmidrule{4-6}
        \multicolumn{3}{l}{\textbf{ACC No.15 }: $m'm'2P$} & \multicolumn{3}{l}{\textbf{ACC No.18}: $mm21'P$} \\
        \multicolumn{3}{l}{$P_M=m'm'2$, $H^1(m'm'2, V/T_L)=\mathbb{Z}_2^3$} & \multicolumn{3}{l}{$P_M=mm21'$, $H^1(mm21', V/T_L)=\mathbb{Z}_2^6$} \\[2pt]
        $P222$       & $m_ym_t,  m_xm_y$                              & 1 & $Pmmm$ & $m_x, m_y, m_t$                                                             & 1 \\
        $P222_1$     & $m_ym_t, m_xm_yT_t^{1/2}$                      & $u_1$ & $Pbmm$ & $m_xT_y^{1/2}, m_y, m_t$                                                    & $u_1$ \\
        $P2_122$     & $m_ym_tT_x^{1/2},m_xm_y$                       & $u_2$ & $Pcmm$ & $m_xT_t^{1/2},  m_y, m_t$                                                   & $u_2$ \\
        $P2_12_12$   & $m_ym_tT_x^{1/2} ,m_xm_yT_y^{1/2}$             & $u_2\cdot u_3$ & $Pmma$ & $m_x, m_y, m_tT_x^{1/2}$                                                    & $u_5$ \\
        $P2_122_1$   & $m_ym_tT_x^{1/2},m_xm_yT_t^{1/2}$              & $u_1 \cdot u_2$ & $Pcam$ & $m_xT_t^{1/2}, m_yT_x^{1/2}, m_t$                                           & $u_2\cdot u_4$ \\
        $P2_12_12_1$ & $m_ym_tT_x^{1/2},m_xm_yT_t^{1/2}T_y^{1/2}$     & $u_1\cdot u_2 \cdot u_3$ & $Pmcb$ & $m_x, m_yT_t^{1/2}, m_tT_y^{1/2}$                                           & $u_3\cdot u_6$ \\
        \multicolumn{3}{l}{\textbf{ACC No.16 }: $mm2P$} & $Pccm$ & $m_xT_t^{1/2}, m_yT_t^{1/2}, m_t$                                           & $u_2\cdot u_3$ \\
        \multicolumn{3}{l}{$P_M=mm2$, $H^1(mm2, V/T_L)=\mathbb{Z}_2^4$} & $Pmaa$ & $m_x, m_yT_x^{1/2}, m_tT_x^{1/2}$                                           & $u_4\cdot u_5$ \\[2pt]
        $Pmm2$       & $m_x, m_y$                                     & 1 & $Pbam$ & $m_xT_y^{1/2}, m_yT_x^{1/2}, m_t$                                           & $u_1\cdot u_4$ \\
        $Pmc2_1$     & $m_x, m_yT_t^{1/2}$                            & $u_3$ & $Pmca$ & $m_x, m_yT_t^{1/2}, m_tT_x^{1/2}$                                           & $u_3\cdot u_5$ \\
        $Pma2$       & $m_x, m_yT_x^{1/2}$                            & $u_4$ & $Pmab$ & $m_x, m_yT_x^{1/2},m_tT_y^{1/2}$                                            & $u_4\cdot u_6$ \\
        $Pcc2$       & $m_xT_t^{1/2}, m_yT_t^{1/2}$                   & $u_2\cdot u_3$ & $Pmmn$ & $m_x, m_y, m_tT_x^{1/2}T_y^{1/2}$                                           & $u_5\cdot u_6$ \\
        $Pca2_1$     & $m_xT_t^{1/2}, m_yT_x^{1/2}$                   & $u_2\cdot u_4$ & $Pnmm$ & $m_xT_y^{1/2}T_t^{1/2}, m_y, m_t$                                           & $u_1\cdot u_2$ \\
        $Pmn2_1$     & $m_x, m_yT_x^{1/2}T_t^{1/2}$                   & $u_3 \cdot u_4$ & $Pmna$ & $m_x, m_yT_x^{1/2}T_t^{1/2}, m_tT_x^{1/2}$                                  & $u_3\cdot u_4 \cdot u_5$ \\
        $Pba2$       & $m_xT_y^{1/2}, m_yT_x^{1/2}$                   & $u_1 \cdot u_4$ & $Pbmn$ & $m_xT_y^{1/2}, m_y, m_tT_x^{1/2}T_y^{1/2}$                                  & $u_1\cdot u_5 \cdot u_6$ \\
        $Pnc2$       & $m_xT_y^{1/2}T_t^{1/2}, m_yT_t^{1/2}$          & $u_1\cdot u_2 \cdot u_3$ & $Pncm$ & $m_xT_y^{1/2}T_t^{1/2}, m_yT_t^{1/2}, m_t$                                  & $u_1\cdot u_2 \cdot u_3$ \\
        $Pna2_1$     & $m_xT_y^{1/2}T_t^{1/2}$, $m_yT_x^{1/2}$        & $u_1 \cdot u_2 \cdot u_4$ & $Pcca$ & $m_xT_t^{1/2}, m_yT_t^{1/2}, m_tT_x^{1/2}$                                  & $u_2\cdot u_3 \cdot u_5$ \\
        $Pnn2$       & $m_xT_y^{1/2}T_t^{1/2}, m_yT_x^{1/2}T_t^{1/2}$ & $\prod_{i=1}^4 u_i$ & $Pcaa$ & $m_xT_t^{1/2}, m_yT_x^{1/2}, m_tT_x^{1/2}$                                  & $u_2\cdot u_4 \cdot u_5$ \\
        \multicolumn{3}{l}{\textbf{ACC No.17 }: $m1'P$} & $Pbaa$ & $m_xT_y^{1/2}, m_yT_x^{1/2}, m_tT_x^{1/2}$                                  & $u_1\cdot u_4 \cdot u_5$ \\
        \multicolumn{3}{l}{$P_M=m1'$, $H^1(m1', V/T_L)=\mathbb{Z}_2^4$} & $Pbca$ & $m_xT_y^{1/2}$, $m_yT_t^{1/2}$, $m_tT_x^{1/2}$                              & $u_1\cdot u_3 \cdot u_5$ \\[2pt]
        $P2mm$       & $m_y, m_t$                                     & 1 & $Pnma$ & $m_xT_y^{1/2}T_t^{1/2}, m_y, m_tT_x^{1/2}$                                  & $u_1\cdot u_2 \cdot u_5$ \\
        $P2cm$       & $m_yT_t^{1/2}, m_t$                            & $u_1$ & $Pbnm$ & $m_xT_y^{1/2},  m_yT_x^{1/2}T_t^{1/2}, m_t$                                 & $u_1\cdot u_3\cdot u_4$ \\
        $P2_1am$     & $m_yT_x^{1/2}, m_t$                            & $u_2$ & $Pmcn$ & $m_x, m_yT_t^{1/2}, m_tT_x^{1/2}T_y^{1/2}$                                  & $u_3  \cdot u_5 \cdot u_6$ \\
        $P2_1ma$     & $m_y, m_tT_x^{1/2}$                            & $u_3$ & $Pban$ & $m_xT_y^{1/2}, m_yT_x^{1/2}, m_tT_x^{1/2}T_y^{1/2}$                         & $u_1\cdot u_4 \cdot u_5\cdot u_6$ \\
        $P2mb$       & $m_y, m_tT_y^{1/2}$                            & $u_4$ & $Pncb$ & $m_xT_y^{1/2}T_t^{1/2}, m_yT_t^{1/2}, m_tT_y^{1/2}$                         & $u_1\cdot u_2 \cdot u_3 \cdot u_6$ \\
        $P2aa$       & $m_yT_x^{1/2}, m_tT_x^{1/2}$                   & $u_3\cdot u_2$ & $Pccn$ & $m_xT_t^{1/2}, m_yT_t^{1/2}, m_tT_x^{1/2}T_y^{1/2}$                         & $ u_2\cdot u_3 \cdot u_5 \cdot u_6$ \\
        $P2_1ab$     & $m_yT_x^{1/2}, m_tT_y^{1/2}$                   & $u_2\cdot u_4$ & $Pnaa$ & $m_xT_y^{1/2}T_t^{1/2}, m_yT_x^{1/2}, m_tT_x^{1/2}$                         & $ u_1\cdot u_2 \cdot u_4 \cdot u_5$ \\
        $P2_1ca$     & $m_yT_t^{1/2}, m_tT_x^{1/2}$                   & $u_3\cdot u_1$ & $Pnnm$ & $m_xT_y^{1/2}T_t^{1/2}, m_yT_x^{1/2}T_t^{1/2}, m_t$                         & $u_1\cdot u_2\cdot u_3 \cdot u_4$ \\
        $P2_1mn$     & $m_y, m_tT_x^{1/2}T_y^{1/2}$                   & $u_3\cdot u_4$ & $Pmnn$ & $m_x, m_yT_x^{1/2}T_t^{1/2}, m_tT_x^{1/2}T_y^{1/2}$                         & $u_3\cdot u_4\cdot u_5 \cdot u_6$ \\
        $P2_1nm$     & $m_yT_x^{1/2}T_t^{1/2}, m_t$                   & $u_2\cdot u_1$ & $Pbcn$ & $m_xT_y^{1/2}$, $m_yT_t^{1/2}$, $m_tT_x^{1/2}T_y^{1/2}$                     & $u_1\cdot u_3\cdot u_5\cdot u_6$ \\
        $P2cb$       & $m_yT_t^{1/2}, m_tT_y^{1/2}$                   & $u_4\cdot u_1$ & $Pnca$ & $m_xT_y^{1/2}T_t^{1/2}$, $m_yT_t^{1/2}$, $m_tT_x^{1/2}$                     & $u_1\cdot u_2\cdot u_3\cdot u_5$ \\
        $P2na$       & $m_yT_x^{1/2}T_t^{1/2}, m_tT_x^{1/2}$          & $u_3\cdot u_2\cdot u_1$ & $Pnab$ & $m_xT_y^{1/2}T_t^{1/2}$, $m_yT_x^{1/2}$, $m_tT_y^{1/2}$                     & $u_1\cdot u_2 \cdot u_4 \cdot u_6$ \\
        $P2an$       & $m_yT_x^{1/2}, m_tT_x^{1/2}T_y^{1/2}$          & $u_3\cdot u_2\cdot u_4$ & $Pnna$ & $m_xT_y^{1/2}T_t^{1/2}, m_yT_x^{1/2}T_t^{1/2}, m_tT_x^{1/2}$                & $u_1\cdot u_2\cdot u_3 \cdot u_4 \cdot u_5$ \\
        $P2_1nb$     & $m_yT_x^{1/2}T_t^{1/2}, m_tT_y^{1/2}$          & $u_2\cdot u_4\cdot u_1$ & $Pbnn$ & $m_xT_y^{1/2}, m_yT_x^{1/2}T_t^{1/2}, m_tT_x^{1/2}T_y^{1/2}$                & $u_1\cdot u_3\cdot u_4 \cdot u_5 \cdot u_6$ \\
        $P2_1cn$     & $m_yT_t^{1/2}, m_tT_x^{1/2}T_y^{1/2}$          & $u_3\cdot u_4\cdot u_1$ & $Pcnn$ & $m_xT_t^{1/2}, m_yT_x^{1/2}T_t^{1/2}, m_tT_x^{1/2}T_y^{1/2}$                & $u_2\cdot u_3\cdot u_4 \cdot u_5 \cdot u_6$ \\
        $P2nn$       & $m_yT_x^{1/2}T_t^{1/2}, m_tT_x^{1/2}T_y^{1/2}$ & $\prod_{i=1}^4 u_i$ & $Pnnn$ & {$m_xT_y^{1/2}T_t^{1/2}$, $m_yT_x^{1/2}T_t^{1/2}$, $m_tT_x^{1/2}T_y^{1/2}$} & $\prod_{i=1}^6u_i$ \\
        \bottomrule
    \end{tabular}
    \end{adjustbox}

    \label{tb:po2}
    \label{tb:po3}
\end{table}
\clearpage

\clearpage
\begin{table}[p]
       \caption{2+1 dimensional space-time groups for the remaining orthorhombic
    lattices. This table contains 17 ACCs and 59 space-time groups. Here
    $T_x^{1/n}$, $T_y^{1/n}$, and $T_t^{1/n}$ denote
    translations by $a_1^{\mathrm{P}}/n$, $a_2^{\mathrm{P}}/n$, and
    $a_3^{\mathrm{P}}/n$, respectively.}
    \label{tab:remaining_orthorhombic}
    \centering
    \setlength{\extrarowheight}{0.8pt}
    \setlength{\tabcolsep}{2.4pt}
    \fontsize{5.7}{6.2}\selectfont
    \renewcommand{\arraystretch}{1.40}

    \begin{adjustbox}{max totalsize={\dimexpr\linewidth+1em\relax}{0.93\textheight},center}
    \begin{tabular}{@{}lll@{\hspace{2.2em}}lll@{}}
        \toprule
        \multicolumn{3}{c}{\textbf{R-base-centered orthorhombic: ACCs 19--22}} &
        \multicolumn{3}{c}{\textbf{Body-centered orthorhombic: ACCs 28--31}} \\[1pt]
        \cmidrule(r){1-3}\cmidrule{4-6}
        \multicolumn{3}{l}{\textbf{ACC No.19}: $m'm'2C$} & \multicolumn{3}{l}{\textbf{ACC No.28}: $m'm'2I$} \\
        \multicolumn{3}{l}{$P_M=m'm'2$, $H^1(m'm'2,V/T_L)=\mathbb Z_2$} & \multicolumn{3}{l}{$P_M=m'm'2$, $H^1(m'm'2,V/T_L)=\mathbb Z_2$} \\
        $C222$ & $m_ym_t,m_xm_y$ & 1 & $I222$ & $m_xm_t,m_ym_t$ & 1 \\
        $C222_1$ & $m_ym_t,m_xm_yT_t^{1/2}$ & $u_1$ & $I2_12_12_1$ & $m_xm_tT_y^{1/2}T_t^{1/2},m_ym_tT_x^{1/2}T_y^{1/2}$ & $u_1$ \\
        \multicolumn{3}{l}{\textbf{ACC No.20}: $mm2C$} & \multicolumn{3}{l}{\textbf{ACC No.29}: $mm2I$} \\
        \multicolumn{3}{l}{$P_M=mm2$, $H^1(mm2,V/T_L)=\mathbb Z_2^2$} & \multicolumn{3}{l}{$P_M=mm2$, $H^1(mm2,V/T_L)=\mathbb Z_2^2$} \\
        $Cmm2$ & $m_x,m_y$ & 1 & $Imm2$ & $m_x,m_y$ & 1 \\
        $Cmc2_1$ & $m_x,m_yT_t^{1/2}$ & $u_1$ & $Ima2$ & $m_x,m_yT_x^{1/2}$ & $u_1$ \\
        $Ccc2$ & $m_xT_t^{1/2},m_yT_t^{1/2}$ & $u_1\cdot u_2$ & $Iba2$ & $m_xT_y^{1/2},m_yT_x^{1/2}$ & $u_1\cdot u_2$ \\
        \multicolumn{3}{l}{\textbf{ACC No.21}: $m1'C$} & \multicolumn{3}{l}{\textbf{ACC No.30}: $m1'I$} \\
        \multicolumn{3}{l}{$P_M=m1'$, $H^1(m1',V/T_L)=\mathbb Z_2^2$} & \multicolumn{3}{l}{$P_M=m1'$, $H^1(m1',V/T_L)=\mathbb Z_2^2$} \\
        $C2mm$ & $m_y,m_t$ & 1 & $I2mm$ & $m_y,m_t$ & 1 \\
        $C2cm$ & $m_yT_t^{1/2},m_t$ & $u_1$ & $I2am$ & $m_yT_x^{1/2},m_t$ & $u_1$ \\
        $C2_1ma$ & $m_y,m_tT_x^{1/2}$ & $u_2$ & $I2ma$ & $m_y,m_tT_x^{1/2}$ & $u_2$ \\
        $C2_1ca$ & $m_yT_t^{1/2},m_tT_x^{1/2}$ & $u_1\cdot u_2$ & $I2aa$ & $m_yT_x^{1/2},m_tT_x^{1/2}$ & $u_1\cdot u_2$ \\
        \multicolumn{3}{l}{\textbf{ACC No.22}: $mm21'C$} & \multicolumn{3}{l}{\textbf{ACC No.31}: $mm21'I$} \\
        \multicolumn{3}{l}{$P_M=mm21'$, $H^1(mm21',V/T_L)=\mathbb Z_2^3$} & \multicolumn{3}{l}{$P_M=mm21'$, $H^1(mm21',V/T_L)=\mathbb Z_2^3$} \\
        $Cmmm$ & $m_x,m_y,m_t$ & 1 & $Immm$ & $m_x,m_y,m_t$ & 1 \\
        $Cmcm$ & $m_x,m_yT_t^{1/2},m_t$ & $u_2$ & $Ibmm$ & $m_xT_y^{1/2},m_y,m_t$ & $u_1$ \\
        $Cmme$ & $m_x,m_y,T_x^{1/2}m_t$ & $u_3$ & $Imma$ & $m_x,m_y,m_tT_x^{1/2}$ & $u_3$ \\
        $Cmce$ & $m_x,m_yT_t^{1/2},m_tT_x^{1/2}$ & $u_2\cdot u_3$ & $Ibam$ & $m_xT_y^{1/2},m_yT_x^{1/2},m_t$ & $u_1\cdot u_2$ \\
        $Cccm$ & $m_xT_t^{1/2},m_yT_t^{1/2},m_t$ & $u_1\cdot u_2$ & $Imcb$ & $m_x,m_yT_t^{1/2},m_tT_y^{1/2}$ & $u_2\cdot u_3$ \\
        $Ccce$ & $m_xT_t^{1/2},m_yT_t^{1/2},m_tT_x^{1/2}$ & $u_1\cdot u_2\cdot u_3$ & $Ibca$ & $m_xT_y^{1/2},m_yT_t^{1/2},m_tT_x^{1/2}$ & $u_1\cdot u_2\cdot u_3$ \\
        \\[2pt]
        \midrule
        \multicolumn{3}{c}{\textbf{T-base-centered orthorhombic: ACCs 23--27}} &
        \multicolumn{3}{c}{\textbf{Face-centered orthorhombic: ACCs 32--35}} \\[1pt]
        \cmidrule(r){1-3}\cmidrule{4-6}
        \multicolumn{3}{l}{\textbf{ACC No.23}: $m'm'2A$} & \multicolumn{3}{l}{\textbf{ACC No.32}: $m'm'2F$} \\
        \multicolumn{3}{l}{$P_M=m'm'2$, $H^1(m'm'2,V/T_L)=\mathbb Z_2$} & \multicolumn{3}{l}{$P_M=m'm'2$, $H^1(m'm'2,V/T_L)=\mathbb I$} \\
        $A222$ & $m_ym_t,m_xm_y$ & 1 & $F222$ & $m_xm_t,m_ym_t$ & 1 \\
        $A2_122$ & $m_ym_tT_x^{1/2},m_xm_y$ & $u_1$ & \multicolumn{3}{l}{\textbf{ACC No.33}: $mm2F$} \\
        \multicolumn{3}{l}{\textbf{ACC No.24}: $m_y1'A$} & \multicolumn{3}{l}{$P_M=mm2$, $H^1(mm2,V/T_L)=\mathbb Z_2$} \\
        \multicolumn{3}{l}{$P_M=m1'$, $H^1(m1',V/T_L)=\mathbb Z_2^2$} & $Fmm2$ & $m_x,m_y$ & 1 \\
        $A2mm$ & $m_y,m_t$ & 1 & $Fdd2$ & $m_xT_y^{1/4}T_t^{1/4},m_yT_x^{1/4}T_t^{1/4}$ & $u_1$ \\
        $A2_1ma$ & $m_y,m_tT_x^{1/2}$ & $u_1$ & \multicolumn{3}{l}{\textbf{ACC No.34}: $m1'F$} \\
        $A2_1am$ & $m_yT_x^{1/2},m_t$ & $u_2$ & \multicolumn{3}{l}{$P_M=m1'$, $H^1(m1',V/T_L)=\mathbb Z_2$} \\
        $A2aa$ & $m_yT_x^{1/2},T_x^{1/2}m_t$ & $u_1\cdot u_2$ & $F2mm$ & $m_y,m_t$ & 1 \\
        \multicolumn{3}{l}{\textbf{ACC No.25}: $m_x1'A$} & $F2dd$ & $m_yT_x^{1/4}T_t^{1/4},m_tT_x^{1/4}T_y^{1/4}$ & $u_1$ \\
        \multicolumn{3}{l}{$P_M=m1'$, $H^1(m1',V/T_L)=\mathbb Z_2^2$} & \multicolumn{3}{l}{\textbf{ACC No.35}: $mm21'F$} \\
        $Am2m$ & $m_x,m_t$ & 1 & \multicolumn{3}{l}{$P_M=mm21'$, $H^1(mm21',V/T_L)=\mathbb Z_2$} \\
        $Ae2m$ & $m_xT_y^{1/2},m_t$ & $u_1$ & $Fmmm$ & $m_x,m_y,m_t$ & 1 \\
        $Am2a$ & $m_x,m_tT_x^{1/2}$ & $u_2$ & $Fddd$ & $m_xT_y^{1/4}T_t^{1/4},m_yT_x^{1/4}T_t^{1/4},m_tT_x^{1/4}T_y^{1/4}$ & $u_1$ \\
        $Ae2a$ & $m_xT_t^{1/2},m_tT_x^{1/2}$ & $u_1\cdot u_2$ &  & &  \\
        \multicolumn{3}{l}{\textbf{ACC No.26}: $mm2A$} &  & &  \\
        \multicolumn{3}{l}{$P_M=mm2$, $H^1(mm2,V/T_L)=\mathbb Z_2^2$} &  & &  \\
        $Amm2$ & $m_x,m_y$ & 1 &  & &  \\
        $Aem2$ & $m_xT_y^{1/2},m_y$ & $u_1$ &  & &  \\
        $Ama2$ & $m_x,m_yT_x^{1/2}$ & $u_2$ &  & &  \\
        $Aea2$ & $m_xT_y^{1/2},m_yT_x^{1/2}$ & $u_1\cdot u_2$ &  & &  \\
        \multicolumn{3}{l}{\textbf{ACC No.27}: $mm21'A$} &  & &  \\
        \multicolumn{3}{l}{$P_M=mm21'$, $H^1(mm21',V/T_L)=\mathbb Z_2^3$} &  & &  \\
        $Ammm$ & $m_x,m_y,m_t$ & 1 &  & &  \\
        $Aemm$ & $m_xT_y^{1/2},m_y,m_t$ & $u_1$ &  & &  \\
        $Amam$ & $m_x,m_yT_x^{1/2},m_t$ & $u_2$ &  & &  \\
        $Amma$ & $m_x,m_y,m_tT_x^{1/2}$ & $u_3$ &  & &  \\
        $Aema$ & $m_xT_y^{1/2},m_y,m_tT_x^{1/2}$ & $u_1\cdot u_3$ &  & &  \\
        $Aeam$ & $m_xT_y^{1/2},m_yT_x^{1/2},m_t$ & $u_1\cdot u_2$ &  & &  \\
        $Amaa$ & $m_x,m_yT_x^{1/2},m_tT_x^{1/2}$ & $u_2\cdot u_3$ &  & &  \\
        $Aeaa$ & $m_xT_y^{1/2},m_yT_x^{1/2},m_tT_x^{1/2}$ & $u_1\cdot u_2\cdot u_3$ &  & &  \\
        \bottomrule
    \end{tabular}
    \end{adjustbox}
    \label{tb:rbase_orthg}
    \label{tb:tbase_orthg}
    \label{tb:body_orthg}
    \label{tb:face_orthg}
\end{table}
\clearpage

\clearpage
\subsection{ Tetragonal crystal system}
Similar to the 3D tetragonal crystal system, 
the $(2+1)$D tetragonal space-time crystal system also has a 4-fold spatial rotation symmetry, in which the temporal direction corresponds to the principal rotation axis in the 3D case. 
Therefore, the tetragonal space-time groups are in one-to-one correspondence with the 3D tetragonal space groups under the replacement $z\rightarrow t$. 
Fractional translations along $z$ become temporal translations.

\subsubsection{Primitive tetragonal lattice}
A space-time primitive tetragonal lattice consists of a square base, spanned by two orthogonal basis vectors of equal length, together with a third basis vector along the temporal direction.
We choose the basis vectors as shown in Fig. \ref{fig:tetragonal_crystal_system}(a). 
Explicitly,
\begin{equation}
    a_1^{\mathrm{P}}=(x_0,0,0),\qquad
    a_2^{\mathrm{P}}=(0,x_0,0),\qquad
    a_3^{\mathrm{P}}=(0,0,t_0).
\end{equation}
There exist 8 ACCs: $4P$, $4'P$, $41'P$, $4m'm'P$, $4mmP$, $4'm'mP$, $4'mm'P$, and $4mm1'P$ (Nos.~36--43), giving rise to 4, 1, 4, 8, 8, 4, 4, and 16 non-equivalent space-time groups, respectively.
They are summarized in the left column of Table~\ref{tb:pttrg}.


The maximal magnetic point group for the primitive tetragonal lattice is $4mm1'$, generated by $R_{\pi/2}$, $m_x$, and $m_t$.
The associated ACC is $4mm1'P$ (No.~43).
After a suitable choice of the origin, these three
magnetic point group generators can be combined with the following fractional translations.
The rotation symmetry $R_{\pi/2}$ cannot carry a quarter of a temporal period because the resulting screw rotation is incompatible with the reflection symmetry of $m_x$.
It can only carry half a temporal period, giving a time-screw rotation.
$m_x$ can independently carry either a 1/2-translation along the spatial diagonal $a_1^{\text{P}}+a_2^{\text{P}}$ or half a temporal period, giving rise to a spatial glide reflection or a time-glide reflection, respectively.
Time-reversal $m_t$ can carry a 1/2-translation along the same spatial diagonal, leading to a glide-time-reversal operation.
There exist 4 cohomology-group generators, which behave as
\begin{equation}
    \begin{aligned}
                u_{1}^{(43)}[R_{\pi/2}] & =\frac{1}{2}a^{\text{P}}_3, & u_{1}^{(43)}[m_x] & =0,                        & u_{1}^{(43)}[m_t] & =0,                        \\
                u_{2}^{(43)}[R_{\pi/2}] & =0,                & u_{2}^{(43)}[m_x] & =\frac{1}{2}(a^{\text{P}}_1+a^{\text{P}}_2), & u_{2}^{(43)}[m_t] & =0,                        \\
                u_{3}^{(43)}[R_{\pi/2}] & =0,                & u_{3}^{(43)}[m_x] & =\frac{1}{2}a^{\text{P}}_3,         & u_{3}^{(43)}[m_t] & =0,                        \\
                u_{4}^{(43)}[R_{\pi/2}] & =0,                & u_{4}^{(43)}[m_x] & =0,                        & u_{4}^{(43)}[m_t] & =\frac{1}{2}(a^{\text{P}}_1+a^{\text{P}}_2).
    \end{aligned}
\label{eq:tetragonalACC43}
\end{equation}
They give rise to 16 non-equivalent space-time groups.

The ACCs $4P$ (No.~36), $41'P$ (No.~38), and $4mmP$ (No.~40) can be grouped 
together because their magnetic point groups retain different subsets of the generators of the same maximal magnetic point group $4mm1'$.
The magnetic point group $4$ retains only $R_{\pi/2}$, $41'$ retains $R_{\pi/2}$ and $m_t$, and $4mm$ retains $R_{\pi/2}$ and $m_x$.
For $41'P$ and $4mmP$, the allowed fractional translations are those already presented in Eq. (\ref{eq:tetragonalACC43}) for the case of $4mm1'P$.
For $4P$, removing both $m_x$ and $m_t$ additionally allows $R_{\pi/2}$ to carry a quarter of a temporal period, rather than only half a temporal period.
Their cohomology group generators, respectively, are
\begin{equation}
    \begin{aligned}
        u_{1}^{(36)}[R_{\pi/2}] & =\frac{1}{4}a^{\text{P}}_3, \\[2pt]
        u_{1}^{(38)}[R_{\pi/2}] & =\frac{1}{2}a^{\text{P}}_3, & u_{1}^{(38)}[m_t] & =0, \\
        u_{2}^{(38)}[R_{\pi/2}] & =0,                & u_{2}^{(38)}[m_t] & =\frac{1}{2}(a^{\text{P}}_1+a^{\text{P}}_2), \\[2pt]
        u_{1}^{(40)}[R_{\pi/2}] & =\frac{1}{2}a^{\text{P}}_3, & u_{1}^{(40)}[m_x] & =0, \\
        u_{2}^{(40)}[R_{\pi/2}] & =0,                & u_{2}^{(40)}[m_x] & =\frac{1}{2}(a^{\text{P}}_1+a^{\text{P}}_2), \\
        u_{3}^{(40)}[R_{\pi/2}] & =0,                & u_{3}^{(40)}[m_x] & =\frac{1}{2}a^{\text{P}}_3.
    \end{aligned}
\end{equation}
The normalizer does not produce further identifications. 
Hence, $4P$, $41'P$, and $4mmP$ give rise to 4, 4, and 8 space-time groups, respectively.

The ACC $4m'm'P$ (No.~39) can be regarded as adding $m_xm_t$ to the ACC $4P$. 
The rotation $R_{\pi/2}$ can carry a quarter of a temporal period, while $m_xm_t$ can carry half a diagonal translation. 
The cohomology group generators are
\beaa
&u_{1}^{(39)}[R_{\pi/2}]=0, &&u_{1}^{(39)}[m_xm_t]=\frac{1}{2} a^{\text{P}}_1+\frac{1}{2} a^{\text{P}}_2\\
&u_{2}^{(39)}[R_{\pi/2}]=\frac{1}{4} a^{\text{P}}_3, &&u_{2}^{(39)}[m_xm_t]=0\\
\eeaa
The normalizer does not produce any further identifications, and this ACC gives rise to 8 space-time groups.

The remaining three ACCs, $4'P$ (No.~37), $4'm'mP$ (No.~41), and $4'mm'P$ (No.~42), share the operation of $R_{\pi/2}m_t$, which cannot carry any fractional translation. 
Therefore, the ACC $4'P$, generated by $R_{\pi/2}m_t$, contains only its symmorphic space-time group.
The ACCs $4'm'mP$ and $4'mm'P$ can be obtained from $4mmP$ by attaching time-reversal to different sets of reflection symmetries: the reflections about the $x$ and $y$-axes for the ACC $4'm'mP$, and those about the two diagonals for the ACC $4'mm'P$. 
Their reflection generators are chosen as $m_xm_t$ and $m_x$, respectively. 
In both cases, their generators can carry half a temporal period or half a spatial-diagonal translation. 
Their cohomology-group generators are given as
\begin{equation}
    \begin{aligned}
        u_{1}^{(41)}[R_{\pi/2}m_t] & =0, & u_{1}^{(41)}[m_xm_t] & =\frac{1}{2}a^{\text{P}}_3, \\
        u_{2}^{(41)}[R_{\pi/2}m_t] & =0, & u_{2}^{(41)}[m_xm_t] & =\frac{1}{2}(a^{\text{P}}_1+a^{\text{P}}_2), \\[2pt]
        u_{1}^{(42)}[R_{\pi/2}m_t] & =0, & u_{1}^{(42)}[m_x]    & =\frac{1}{2}a^{\text{P}}_3, \\
        u_{2}^{(42)}[R_{\pi/2}m_t] & =0, & u_{2}^{(42)}[m_x]    & =\frac{1}{2}(a^{\text{P}}_1+a^{\text{P}}_2).
    \end{aligned}
\end{equation}
The normalizer does not produce any further identifications. Hence, $4'P$, $4'm'mP$, and $4'mm'P$ give rise to 1, 4, and 4 space-time groups, respectively.



\subsubsection{Body-centered tetragonal lattice}

\begin{figure}[h]
     \caption{The basis vectors (blue arrows) for the (a) primitive and (b) body-centered Bravais lattices in the space-time tetragonal crystal system.}
    \centering
 \includegraphics[width=\linewidth]{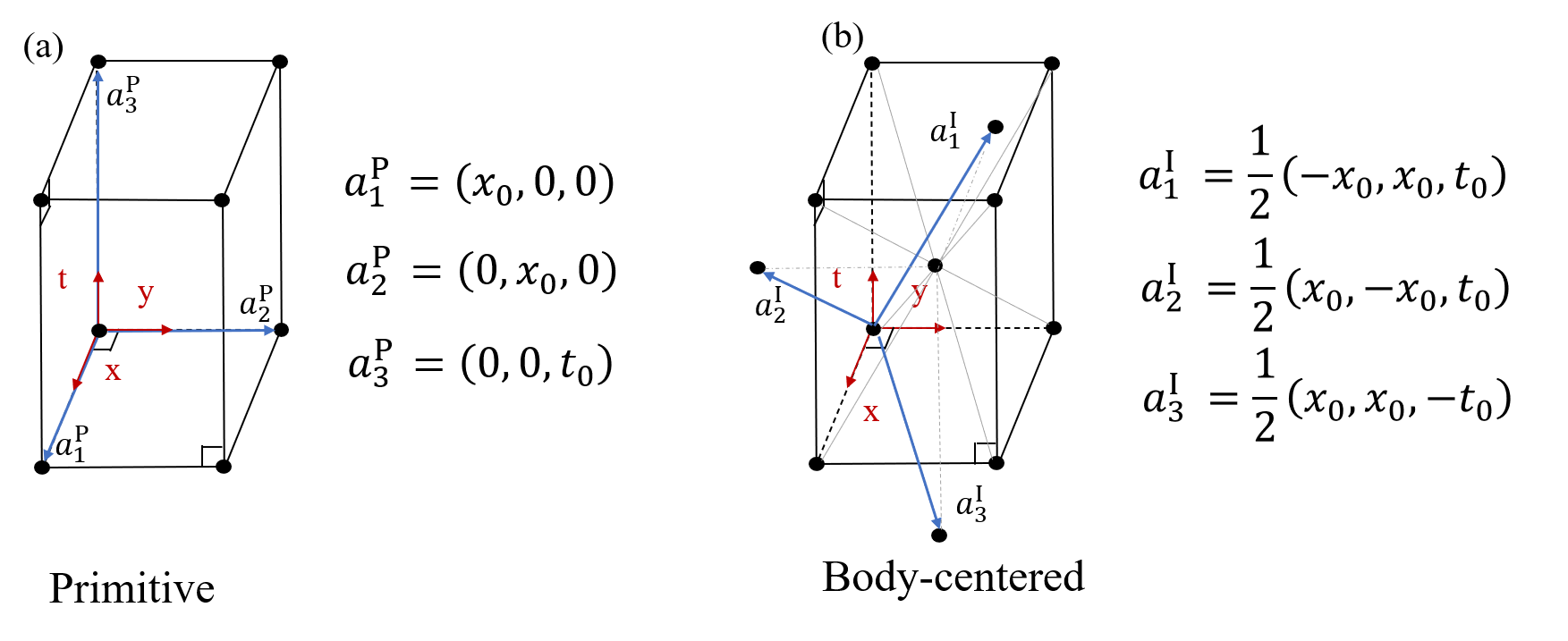}
    \label{fig:tetragonal_crystal_system}
\end{figure}

The body-centered tetragonal lattice is obtained by adding a lattice point at the center of the conventional tetragonal space-time cell while preserving the fourfold spatial axis. We choose the basis vectors as shown in Fig. \ref{fig:tetragonal_crystal_system}(b). The basis vectors, written in terms of the primitive basis, are
\begin{equation}
    a_1^{\mathrm{I}}=\tfrac12\bigl(-a_1^{\mathrm{P}}+a_2^{\mathrm{P}}+a_3^{\mathrm{P}}\bigr),\qquad
    a_2^{\mathrm{I}}=\tfrac12\bigl(a_1^{\mathrm{P}}-a_2^{\mathrm{P}}+a_3^{\mathrm{P}}\bigr),\qquad
    a_3^{\mathrm{I}}=\tfrac12\bigl(a_1^{\mathrm{P}}+a_2^{\mathrm{P}}-a_3^{\mathrm{P}}\bigr).
\end{equation}
There exist eight ACCs: $4I$, $4'I$, $41'I$, $4m'm'I$, $4mmI$, $4'mm'I$, $4'm'mI$, and $4mm1'I$ (Nos.~44--51), which give rise to 2, 1, 2, 2, 4, 2, 2, and 4 non-equivalent space-time groups, respectively, and are summarized in the right column of Table~\ref{tb:ittrg}.

The maximal magnetic point group for the body-centered tetragonal lattice is $4mm1'$, generated by $R_{\pi/2}$, $m_x$, and $m_t$.
The associated ACC is $4mm1'I$ (No.~51).
Its two cohomology-group generators are
\begin{equation}
    \begin{aligned}
                u_{1}^{(51)}[R_{\pi/2}] & =\frac{1}{4}a^{\text{P}}_3, & u_{1}^{(51)}[m_x] & =\frac{1}{2}a^{\text{P}}_1, & u_{1}^{(51)}[m_t] & =\frac{1}{2}a^{\text{P}}_1, \\
                u_{2}^{(51)}[R_{\pi/2}] & =0,                & u_{2}^{(51)}[m_x] & =\frac{1}{2}a^{\text{P}}_3, & u_{2}^{(51)}[m_t] & =0.
    \end{aligned}
\end{equation}
Unlike the ACC of $4mm1'P$ on the primitive tetragonal lattice, which has four independent cohomology-group generators and 16 space-time groups, the ACC of $4mm1'I$ has only two such generators and gives rise to 4 space-time groups. 
The reason is that the vector $(a_1^{\text{P}}+a_2^{\text{P}}+a_3^{\text{P}})/2$ is a lattice translation, making $(a_1^{\text{P}}+a_2^{\text{P}})/2$ and $a_3^{\text{P}}/2$ equivalent modulo basis vectors.
Therefore, $u_{2,3}^{(43)}$ in the primitive case are identified as $u_2^{(51)}$ in the body-centered case. 
Moreover, the cohomology-group generators $u_{1,4}^{(43)}$ in the primitive lattice become trivial since their translalation parts become integer. 
The remaining generator $u_1^{(51)}$ is a new assignment in which the 1/4-period translation attached to $R_{\pi/2}$ 
appears together with the 1/2-translation $a_1^{\text{P}}/2$ attached to both $m_x$ and $m_t$.

The magnetic point groups of the ACCs $4I$ (No.~44), $41'I$ (No.~46), and $4mmI$ (No.~48) are obtained by reducing symmetry operations from the ACC $4mm1'I$.
The fractional translations carried by the retained symmetry operations are obtained by applying the same restriction to those of the ACC $4mm1'I$.
The resulting cohomology-group generators are
\begin{equation}
    \begin{aligned}
        u_{1}^{(44)}[R_{\pi/2}] & =\frac{1}{4}a^{\text{P}}_3, \\[2pt]
        u_{1}^{(46)}[R_{\pi/2}] & =\frac{1}{4}a^{\text{P}}_3, & u_{1}^{(46)}[m_t] & =\frac{1}{2}a^{\text{P}}_1, \\[2pt]
        u_{1}^{(48)}[R_{\pi/2}] & =0,                & u_{1}^{(48)}[m_x] & =\frac{1}{2}a^{\text{P}}_3, \\
        u_{2}^{(48)}[R_{\pi/2}] & =\frac{1}{4}a^{\text{P}}_3, & u_{2}^{(48)}[m_x] & =\frac{1}{2}(a^{\text{P}}_2+a^{\text{P}}_3).
    \end{aligned}
\end{equation}
The normalizer does not produce any further identifications. Hence, the ACCs $4I$, $41'I$, and $4mmI$ give rise to 2, 2, and 4
space-time groups, respectively.

The ACC $4m'm'I$ (No.~47) can be obtained by adding $m_xm_t$ to the ACC $4I$. 
The operation $m_xm_t$ carries no fractional translation, and the cohomology group generator is
\beaa
&u_{1}^{(47)}[R_{\pi/2}]=\frac{1}{4}a^{\text{P}}_3, &&u_{1}^{(47)}[m_xm_t]=0.
\eeaa
The normalizer does not produce any further identification, and this ACC gives rise to 2 space-time groups.

The remaining three ACCs, $4'I$ (No.~45), $4'mm'I$ (No.~49), and $4'm'mI$ (No.~50), share the operation $R_{\pi/2}m_t$, which cannot carry a fractional translation. 
The magnetic point group of the ACC $4'I$ has no additional generator, so this ACC contains only its symmorphic space-time group. 
The reflection generators of $4'mm'I$ and $4'm'mI$ are $m_x$ and $m_xm_t$, respectively.
Their cohomology-group generators are
\beaa
&u_{1}^{(49)}[R_{\pi/2}m_t]=0, &&u_{1}^{(49)}[m_x]=\frac{1}{2}a^{\text{P}}_3\\
&u_{1}^{(50)}[R_{\pi/2}m_t]=0, &&u_{1}^{(50)}[m_xm_t]=\frac{1}{2}a^{\text{P}}_1+\frac{1}{4}a^{\text{P}}_3.
\eeaa
The normalizer does not produce any further identifications. 
Hence, the ACCs of $4'I$, $4'mm'I$, and $4'm'mI$ give rise to 1, 2, 2 space-time groups, respectively.


\clearpage
\begin{table}[p]
    \caption{2+1 dimensional space-time groups in the tetragonal crystal
    system. This crystal system contains 16 ACCs and 68 space-time groups. Here
    $T_x^{1/n}$, $T_y^{1/n}$, and $T_t^{1/n}$ denote translations by
    $a_1^{\mathrm{P}}/n$, $a_2^{\mathrm{P}}/n$, and
    $a_3^{\mathrm{P}}/n$, respectively.}
    \centering
    \setlength{\extrarowheight}{0pt}
    \setlength{\tabcolsep}{2.2pt}
    \fontsize{5.2}{5.3}\selectfont
    \renewcommand{\arraystretch}{1.15}
    \begin{adjustbox}{max totalsize={\dimexpr\linewidth+1em\relax}{0.95\textheight},center}
    \begin{tabular}{@{}lll@{\hspace{2.0em}}lll@{}}
        \toprule
        \multicolumn{3}{c}{\textbf{Primitive tetragonal: ACCs 36--43}} &
        \multicolumn{3}{c}{\textbf{Body-centered tetragonal: ACCs 44--51}} \\[1pt]
        \cmidrule(r){1-3}\cmidrule{4-6}
        \multicolumn{3}{l}{\textbf{ACC No.36 }: $4P$} & \multicolumn{3}{l}{\textbf{ACC No.44 }: $4I$} \\
        \multicolumn{3}{l}{$P_M= 4, \ H^1(4, V/T_L)=\mathbb{Z}_4$} & \multicolumn{3}{l}{$P_M=4, \  H^1(4, V/T_L)=\mathbb{Z}_2$} \\
        $P4$       & $R_{\pi/2}$                                    & $1$ & $I4$         & $R_{\pi/2}$                                               & $1$ \\
        $P4_1$     & $R_{\pi/2}T_t^{1/4}$                           & $u_1$ & $I4_1$       & $R_{\pi/2}T_t^{1/4}$                                      & $u_1$ \\
        $P4_2$     & $R_{\pi/2}T_t^{1/2}$                           & $u_1^2$ & \multicolumn{3}{l}{\textbf{ACC No.45 }: $4'I$} \\
        $P4_3$     & $R_{\pi/2}T_t^{3/4}$                           & $u_1^3$ & \multicolumn{3}{l}{$P_M=4', \  H^1(4', V/T_L)=\mathbb{I}$} \\
        \multicolumn{3}{l}{\textbf{ACC No.37 }: $4'P$} & $I\bar{4}$   & $R_{\pi/2} m_t$                                           & $1$ \\
        \multicolumn{3}{l}{$P_M=4', \  H^1(4', V/T_L)=\mathbb{I}$} & \multicolumn{3}{l}{\textbf{ACC No.46 }: $41'I$} \\
        $P\bar{4}$ & $R_{\pi/2}m_t$                                 & $1$ & \multicolumn{3}{l}{$P_M=41', \  H^1(41', V/T_L)=\mathbb{Z}_2$} \\
        \multicolumn{3}{l}{\textbf{ACC No.38 }: $41'P$} & $I4/m$       & $R_{\pi/2}, m_t$                                          & $1$ \\
        \multicolumn{3}{l}{$P_M=41', \  H^1(41', V/T_L)=\mathbb{Z}_2^2$} & $I4_1/a$     & $R_{\pi/2}T_t^{1/4}, m_tT_x^{1/2}$                        & $u_1$ \\
        $P4/m$     & $R_{\pi/2}, m_t$                               & $1$ & \multicolumn{3}{l}{\textbf{ACC No.47 }: $4m'm'I$} \\
        $P4_2/m$   & $R_{\pi/2}T_t^{1/2}, m_t$                      & $u_1$ & \multicolumn{3}{l}{$P_M=4m'm', \  H^1(4m'm', V/T_L)=\mathbb{Z}_2$} \\
        $P4/n$     & $R_{\pi/2}, m_tT_x^{1/2}T_y^{1/2}$             & $u_2$ & $I422$       & $R_{\pi/2}, m_xm_t$                                       & $1$ \\
        $P4_2/n$   & $R_{\pi/2}T_t^{1/2}, m_tT_x^{1/2}T_y^{1/2}$    & $u_1\cdot u_2$ & $I4_122$     & $R_{\pi/2}T_t^{1/4}, m_xm_t$                              & $u_1$ \\
        \multicolumn{3}{l}{\textbf{ACC No.39 }: $4m'm'P$} & \multicolumn{3}{l}{\textbf{ACC No.48 }: $4mmI$} \\
        \multicolumn{3}{l}{$P_M=4m'm', \  H^1(4m'm', V/T_L)=\mathbb{Z}_4\times\mathbb{Z}_2$} & \multicolumn{3}{l}{$P_M=4mm, \  H^1(4mm, V/T_L)=\mathbb{Z}_2^2$} \\
        $P422$     & $R_{\pi/2}, m_xm_t$                            & $1$ & $I4mm$       & $R_{\pi/2}, m_x$                                          & $1$ \\
        $P42_12$   & $R_{\pi/2}, m_xm_tT_x^{1/2}T_y^{1/2}$          & $u_1$ & $I4cm$       & $R_{\pi/2}, m_xT_t^{1/2}$                                 & $u_1$ \\
        $P4_122$   & $R_{\pi/2}T_t^{1/4}, m_xm_t$                   & $u_2$ & $I4_1md$     & $R_{\pi/2}T_t^{1/4}, m_xT_y^{1/2}T_t^{1/2}$               & $u_2$ \\
        $P4_222$   & $R_{\pi/2}T_t^{1/2}, m_xm_t$                   & $u_2^2$ & $I4_1cd$     & $R_{\pi/2}T_t^{1/4}, m_xT_y^{1/2}$                        & $u_1\cdot u_2$ \\
        $P4_322$   & $R_{\pi/2}T_t^{3/4}, m_xm_t$                   & $u_2^3$ & \multicolumn{3}{l}{\textbf{ACC No.49 }: $4'mm'I$} \\
        $P4_12_12$ & $R_{\pi/2}T_t^{1/4}, m_xm_tT_x^{1/2}T_y^{1/2}$ & $u_1\cdot u_2$ & \multicolumn{3}{l}{$P_M=4'mm', \  H^1(4'mm', V/T_L)=\mathbb{Z}_2$} \\
        $P4_22_12$ & $R_{\pi/2}T_t^{1/2}, m_xm_tT_x^{1/2}T_y^{1/2}$ & $u_1\cdot u_2^2$ & $I\bar{4}m2$ & $R_{\pi/2}m_t, m_x$                                       & 1 \\
        $P4_32_12$ & $R_{\pi/2}T_t^{3/4}, m_xm_tT_x^{1/2}T_y^{1/2}$ & $u_1\cdot u_2^3$ & $I\bar{4}c2$ & $R_{\pi/2}m_t, m_xT_t^{1/2}$                              & $u_1$ \\
        \multicolumn{3}{l}{\textbf{ACC No.40 }: $4mmP$} & \multicolumn{3}{l}{\textbf{ACC No.50 }: $4'm'mI$} \\
        \multicolumn{3}{l}{$P_M=4mm, \  H^1(4mm, V/T_L)=\mathbb{Z}_2^3$} & \multicolumn{3}{l}{$P_M=4'm'm, \  H^1(4'm'm, V/T_L)=\mathbb{Z}_2$} \\
        $P4mm$         & $R_{\pi/2}, m_x$                                                               & $1$ & $I\bar{4}2m$ & $R_{\pi/2}m_t, m_xm_t$                                    & 1 \\
        $P4_2mc$       & $R_{\pi/2}T_t^{1/2}, m_x$                                                      & $u_1$ & $I\bar{4}2d$ & $R_{\pi/2}m_t, m_xm_tT_x^{1/2}T_t^{1/4}$                  & $u_1$ \\
        $P4bm$         & $R_{\pi/2}, m_xT_x^{1/2}T_y^{1/2}$                                             & $u_2$ & \multicolumn{3}{l}{\textbf{ACC No.51 }: $4mm1'I$} \\
        $P4cc$         & $R_{\pi/2}, m_xT_t^{1/2}$                                                      & $u_3$ & \multicolumn{3}{l}{$P_M=4mm1', \  H^1(4mm1', V/T_L)=\mathbb{Z}_2^2$} \\
        $P4_2cm$       & $R_{\pi/2}T_t^{1/2}, m_xT_t^{1/2}$                                             & $u_1 \cdot u_3$ & $I4/mmm$     & $R_{\pi/2}, m_x, m_t$                                     & $1$ \\
        $P4nc$         & $R_{\pi/2}, m_xT_x^{1/2}T_y^{1/2}T_t^{1/2}$                                    & $u_2\cdot u_3$ & $I4_1/amd$   & $R_{\pi/2}T_t^{1/4}, m_xT_x^{1/2}, m_tT_x^{1/2}$          & $u_1$ \\
        $P4_2bc$       & $R_{\pi/2}T_t^{1/2}, m_xT_x^{1/2}T_y^{1/2}$                                    & $u_1\cdot u_2$ & $I4/mcm$     & $R_{\pi/2}, m_xT_t^{1/2}, m_t$                            & $u_2$ \\
        $P4_2nm$       & $R_{\pi/2}T_t^{1/2}, m_xT_x^{1/2}T_y^{1/2}T_t^{1/2}$                           & $u_1 \cdot u_2\cdot u_3$ & $I4_1/acd$   & $R_{\pi/2}T_t^{1/4}, m_xT_x^{1/2}T_t^{1/2}, m_tT_x^{1/2}$ & $u_1\cdot u_2$ \\
        \multicolumn{3}{l}{\textbf{ACC No.41 }: $4'm'mP$} &  & &  \\
        \multicolumn{3}{l}{$P_M=4'm'm, \  H^1(4'm'm, V/T_L)=\mathbb{Z}_2^2$} &  & &  \\
        $P\bar{4}2m$   & $R_{\pi/2}m_t, m_xm_t$                                                         & 1 &  & &  \\
        $P\bar{4}2c$   & $R_{\pi/2}m_t,  m_xm_tT_t^{1/2}$                                               & $u_1$ &  & &  \\
        $P\bar{4}2_1m$ & $R_{\pi/2}m_t, m_xm_tT_x^{1/2}T_y^{1/2}$                                       & $u_2$ &  & &  \\
        $P\bar{4}2_1c$ & $R_{\pi/2}m_t, m_xm_tT_x^{1/2}T_y^{1/2}T_t^{1/2}$                              & $u_1\cdot u_2$ &  & &  \\
        \multicolumn{3}{l}{\textbf{ACC No.42}: $4'mm'P$} &  & &  \\
        \multicolumn{3}{l}{$P_M=4'mm', \  H^1(4'mm', V/T_L)=\mathbb{Z}_2^2$} &  & &  \\
        $P\bar{4}m2$   & $R_{\pi/2}m_t, m_x$                                                            & 1 &  & &  \\
        $P\bar{4}c2$   & $R_{\pi/2}m_t, m_xT_t^{1/2}$                                                   & $u_1$ &  & &  \\
        $P\bar{4}b2$   & $R_{\pi/2}m_t, m_xT_x^{1/2}T_y^{1/2}$                                          & $u_2$ &  & &  \\
        $P\bar{4}n2$   & $R_{\pi/2}m_t, m_xT_x^{1/2}T_y^{1/2}T_t^{1/2}$                                 & $u_1\cdot u_2$ &  & &  \\
        \multicolumn{3}{l}{\textbf{ACC No.43 }: $4mm1'P$} &  & &  \\
        \multicolumn{3}{l}{$P_M=4mm1', \  H^1(4mm1', V/T_L)=\mathbb{Z}_2^4$} &  & &  \\
        $P4/mmm$       & $R_{\pi/2}, m_x, m_t$                                                          & 1 &  & &  \\
        $P4_2/mmc$     & $R_{\pi/2}T_t^{1/2}, m_x, m_t$                                                 & $u_1$ &  & &  \\
        $P4/mbm$       & $R_{\pi/2}, m_xT_x^{1/2}T_y^{1/2}, m_t$                                        & $u_2$ &  & &  \\
        $P4/mcc$       & $R_{\pi/2}, m_xT_t^{1/2}, m_t$                                                 & $u_3$ &  & &  \\
        $P4/nmm$       & $R_{\pi/2}, m_x, m_tT_x^{1/2}T_y^{1/2}$                                        & $u_4$ &  & &  \\
        $P4/nbm$       & $R_{\pi/2}, m_xT_x^{1/2}T_y^{1/2}, m_tT_x^{1/2}T_y^{1/2}$                      & $u_2\cdot u_4$ &  & &  \\
        $P4/mnc$       & $R_{\pi/2}, m_xT_x^{1/2}T_y^{1/2}T_t^{1/2}, m_t$                               & $u_2\cdot u_3$ &  & &  \\
        $P4/ncc$       & $R_{\pi/2}, m_xT_t^{1/2}, m_tT_x^{1/2}T_y^{1/2}$                               & $u_3\cdot u_4$ &  & &  \\
        $P4_2/nmc$     & $R_{\pi/2}T_t^{1/2}, m_x, m_tT_x^{1/2}T_y^{1/2}$                               & $u_1\cdot u_4$ &  & &  \\
        $P4_2/mcm$     & $R_{\pi/2}T_t^{1/2}, m_xT_t^{1/2}, m_t$                                        & $u_1\cdot u_3$ &  & &  \\
        $P4_2/mbc$     & $R_{\pi/2}T_t^{1/2}, m_xT_x^{1/2}T_y^{1/2}, m_t$                               & $u_1\cdot u_2$ &  & &  \\
        $P4_2/nbc$     & $R_{\pi/2}T_t^{1/2}, m_xT_x^{1/2}T_y^{1/2}, T_x^{1/2}T_y^{1/2}m_t$             & $u_1\cdot u_2 \cdot u_4$ &  & &  \\
        $P4/nnc$       & $R_{\pi/2}, m_xT_x^{1/2}T_y^{1/2}T_t^{1/2}, m_tT_x^{1/2}T_y^{1/2}$             & $u_2\cdot u_3 \cdot u_4$ &  & &  \\
        $P4_2/mnm$     & $R_{\pi/2}T_t^{1/2}, m_xT_x^{1/2}T_y^{1/2}T_t^{1/2}, m_t$                      & $u_1\cdot u_2\cdot u_3$ &  & &  \\
        $P4_2/ncm$     & $R_{\pi/2}T_t^{1/2}, m_xT_t^{1/2}, m_tT_x^{1/2}T_y^{1/2}$                      & $u_1\cdot u_3\cdot u_4$ &  & &  \\
        $P4_2/nnm$     & {$R_{\pi/2}T_t^{1/2}, m_xT_x^{1/2}T_y^{1/2}T_t^{1/2}$,$m_tT_x^{1/2}T_y^{1/2}$} & $\prod_{i=1}^4 u_i$ &  & &  \\
        \bottomrule
    \end{tabular}
    \end{adjustbox}
    \label{tab:tetragonal}
    \label{tb:pttrg}
    \label{tb:pttrg_c}
    \label{tb:ittrg}
\end{table}
\clearpage

\clearpage
\subsection{ Trigonal crystal system} 
In $(2+1)$ dimensions, only the temporal direction is left invariant by a threefold spatial rotation. It therefore corresponds to the principal rotation axis of a 3D trigonal crystal. Under the identification $z\leftrightarrow t$, the trigonal space-time groups are in one-to-one correspondence with the 3D trigonal space groups, with fractional translations along $z$ mapped to temporal translations.

\subsubsection{Primitive trigonal lattice}
The primitive trigonal lattice has two equal-length spatial basis vectors separated by $120^\circ$, forming a triangular lattice in the spatial plane. 
We choose the basis vectors as shown in Fig. \ref{fig:trigonal_crystal_system}(a).
Explicitly,
\begin{equation}
    a_1^{\mathrm{P}}=(l,0,0),\qquad
    a_2^{\mathrm{P}}=\left(-\tfrac12l,\tfrac{\sqrt3}{2}l,0\right),\qquad
    a_3^{\mathrm{P}}=(0,0,t_0).
\end{equation}
There exist 8 ACCs: $3P$, $6'P$, $312P$, $321P$, $3m1P$, $31mP$, $6'm'mP$, and $6'mm'P$ (Nos.~52--59), which give rise to 3, 1, 3, 3, 2, 2, 2, and 2 non-equivalent space-time groups, respectively, and are summarized in the left column of Table~\ref{tb:ptrig}.


The maximal magnetic point group for the primitive trigonal lattice is $6'm'm$, generated by $R_{\pi/3}m_t$ and $m_x$.
The associated ACC is $6'mm'P$ (No.~59).
The operation $R_{\pi/3}m_t$ carries no fractional translation. The only possible nonsymmorphic operation is the time-glide reflection, produced by combining $m_x$ with half a temporal period. The cohomology group generator is
\beaa
&u_{1}^{(59)}[R_{\pi/3}m_t]=0, &&u_{1}^{(59)}[m_x]=\frac{1}{2}a^{\text{P}}_3.
\eeaa
This ACC gives rise to 2 space-time groups.

The ACC $6'm'mP$ (No.~58) differs from the ACC $6'mm'P$ only in the direction of the reflection symmetry. 
Its symmetry operations are $R_{\pi/3}m_t$ and $m_y$. 
The operation $R_{\pi/3}m_t$ carries no fractional translation, whereas $m_y$ can carry half a temporal period. Its cohomology group generator is
\beaa
&u_{1}^{(58)}[R_{\pi/3}m_t]=0, &&u_{1}^{(58)}[m_y]=\frac{1}{2}a^{\text{P}}_3.
\eeaa
This ACC gives rise to 2 space-time groups.

Removing the reflection symmetry from either $6'mm'P$ or $6'm'mP$ gives rise to $6'P$ (No.~53). 
Its symmetry operation is $R_{\pi/3}m_t$, which cannot carry a fractional translation. 
Therefore, this ACC contains only the symmorphic space-time group.

For the ACC $3P$ (No.~52), the threefold rotation $R_{2\pi/3}$ can carry one third of a temporal period, giving rise to the following cohomology group generator:
\beaa
&u_{1}^{(52)}[R_{2\pi/3}]=\frac{1}{3}a^{\text{P}}_3.
\eeaa
This ACC gives rise to 3 space-time groups.

The magnetic point groups of the ACCs $312P$ (No.~54) and $321P$ (No.~55) are obtained by adding a reflection combined with time-reversal to that of $3P$. 
Their symmetry operations are $R_{2\pi/3}$ together with $m_xm_t$ and $m_ym_t$, respectively. 
The two magnetic point groups differ only in the direction of the reflection symmetry. 
The reflections carry no fractional translation, whereas $R_{2\pi/3}$ can carry 1/3 of a temporal period. Their cohomology-group generators are
\beaa
&u_{1}^{(54)}[R_{2\pi/3}]=\frac{1}{3}a^{\text{P}}_3, &&u_{1}^{(54)}[m_xm_t]=0,\\
&u_{1}^{(55)}[R_{2\pi/3}]=\frac{1}{3}a^{\text{P}}_3, &&u_{1}^{(55)}[m_ym_t]=0.
\eeaa
Each ACC gives rise to 3 space-time groups.

The symmetry operations of the ACCs $3m1P$ (No.~56) and $31mP$ (No.~57) are $R_{2\pi/3}$ together with $m_x$ and $m_y$, respectively. These magnetic point groups also differ only in the direction of the reflection symmetry. 
In both ACCs, the only possible nonsymmorphic operation is the time-glide reflection, produced by combining $m_x$ and $m_y$, respectively, with half a temporal period. 
Their cohomology-group generators are
\beaa
&u_{1}^{(56)}[R_{2\pi/3}]=0, &&u_{1}^{(56)}[m_x]=\frac{1}{2}a^{\text{P}}_3,\\
&u_{1}^{(57)}[R_{2\pi/3}]=0, &&u_{1}^{(57)}[m_y]=\frac{1}{2}a^{\text{P}}_3.
\eeaa
Each ACC gives rise to 2 space-time groups.

\subsubsection{Rhombohedral lattice}
\begin{figure}[h]
      \caption{The basis vectors (blue arrows) for the (a) primitive and (b) rhombohedral Bravais  lattices in the space-time trigonal crystal system.}
    \centering
 \includegraphics[width=\linewidth]{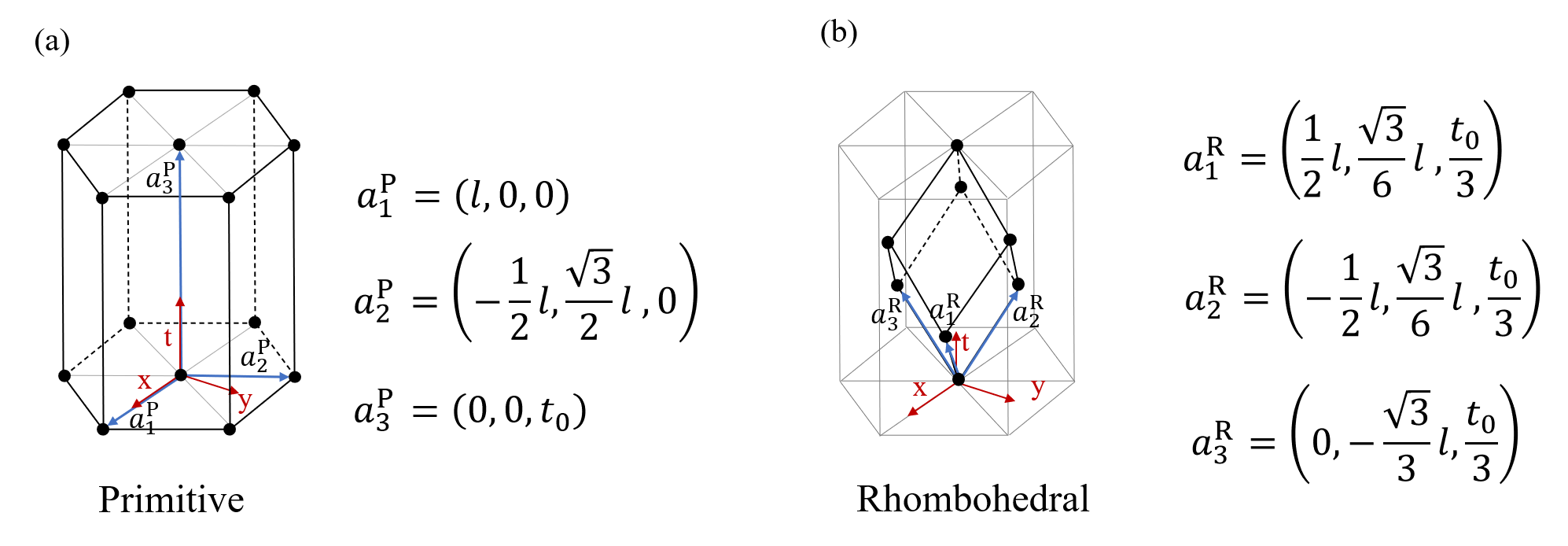}
    \label{fig:trigonal_crystal_system}
\end{figure}

A rhombohedral lattice is generated by 3 symmetry-equivalent basis vectors of equal length with equal mutual angles. Each basis vector mixes spatial and temporal components, while their sum defines a purely temporal lattice direction. The basis vectors shown in Fig. \ref{fig:trigonal_crystal_system}(b), written in terms of the primitive basis, are
\begin{equation}
    a_1^{\mathrm{R}}=\tfrac23a_1^{\mathrm{P}}+\tfrac13a_2^{\mathrm{P}}+\tfrac13a_3^{\mathrm{P}},\qquad
    a_2^{\mathrm{R}}=-\tfrac13a_1^{\mathrm{P}}+\tfrac13a_2^{\mathrm{P}}+\tfrac13a_3^{\mathrm{P}},\qquad
    a_3^{\mathrm{R}}=-\tfrac13a_1^{\mathrm{P}}-\tfrac23a_2^{\mathrm{P}}+\tfrac13a_3^{\mathrm{P}}.
\end{equation}
There exist 5 ACCs: $3R$, $6'R$, $3m'R$, $3mR$, and $6'm'mR$ (Nos.~60--64), which give rise to 1, 1, 1, 2, and 2 non-equivalent space-time groups, respectively, and are summarized in the right column of Table~\ref{tb:rtrig}.

The maximal magnetic point group for the rhombohedral lattice is $6'm'm$, generated by $R_{\pi/3}m_t$ and $m_x$.
The associated ACC is $6'm'mR$ (No.~64).
It is generated by $R_{\pi/3}m_t$ and $m_x$. 
The operation $R_{\pi/3}m_t$ carries no fractional translation. 
The only possible nonsymmorphic operation is the time-glide reflection, produced by combining $m_x$ with $a_3^{\text{P}}/2$. The cohomology group generator is
\beaa
&u_{1}^{(64)}[R_{\pi/3}m_t]=0, &&u_{1}^{(64)}[m_x]=\frac{1}{2}a_3^{\text{P}}.
\eeaa
This ACC gives rise to 2 space-time groups.

The magnetic point group of $3mR$ (No.~63) is generated by $R_{2\pi/3}$ and $m_x$. The only possible nonsymmorphic operation is the time-glide reflection, produced by combining $m_x$ with $a_3^{\text{P}}/2$. The cohomology group generator is
\beaa
&u_{1}^{(63)}[R_{2\pi/3}]=0, &&u_{1}^{(63)}[m_x]=\frac{1}{2}a_3^{\text{P}}.
\eeaa
This ACC gives rise to 2 space-time groups.

The remaining ACCs $3R$ (No.~60), $6'R$ (No.~61), and $3m'R$ (No.~62) contain no nonsymmorphic space-time groups. The reason is that each of the rhombohedral basis vectors contains a 1/3-period temporal component, so the fractional translation of the possible time-screw rotation can be removed by a suitable choice of origin. Therefore, each of these three ACCs contains only the symmorphic space-time group.

 
\begin{table}[h]
        \caption{2+1 dimensional space-time groups in the trigonal crystal
    system. This crystal system contains 13 ACCs and 25 space-time groups. $T_t^{1/n}$ denotes translation by
    $a_3^{\mathrm{P}}/n$, where $n=2,3$.}
    \centering
    \setlength{\extrarowheight}{1.1pt}
    \setlength{\tabcolsep}{4pt}
    \fontsize{7}{7.8}\selectfont
    \renewcommand{\arraystretch}{1.15}
    \begin{adjustbox}{max width=\dimexpr\linewidth+1em\relax,center}
    \begin{tabular}{@{}lll@{\hspace{3em}}lll@{}}
        \toprule
        \multicolumn{3}{c}{\textbf{Primitive trigonal: ACCs 52--59}} &
        \multicolumn{3}{c}{\textbf{Rhombohedral: ACCs 60--64}} \\[2pt]
        \cmidrule(r){1-3}\cmidrule{4-6}
        \multicolumn{3}{l}{\textbf{ACC No.52 }: $3P$} & \multicolumn{3}{l}{\textbf{ACC No.60 }: $3R$} \\
        \multicolumn{3}{l}{$P_M= 3, \  H^1(3, V/T_L)=\mathbb{Z}_3$} & \multicolumn{3}{l}{$P_M= 3, \  H^1(3, V/T_L)=\mathbb{I}$} \\
        $P3$         & $R_{2\pi/3}$                  & $1$ & $R3$        & $R_{2\pi/3}$                 & 1 \\
        $P3_1$       & $R_{2\pi/3}T_t^{1/3}$         & $u_1$ & \multicolumn{3}{l}{\textbf{ACC No.61 }: $6'R$} \\
        $P3_2$       & $R_{2\pi/3}T_t^{2/3}$         & $u_1^2$ & \multicolumn{3}{l}{$P_M= 6', \  H^1(6', V/T_L)=\mathbb{I}$} \\
        \multicolumn{3}{l}{\textbf{ACC No.53 }: $6'P$} & $R\bar{3}$  & $R_{\pi/3}m_t$               & 1 \\
        \multicolumn{3}{l}{$P_M= 6', \  H^1(6', V/T_L)=\mathbb{I}$} & \multicolumn{3}{l}{\textbf{ACC No.62 }: $3m'R$} \\
        $P\bar{3}$   & $R_{\pi/3}m_t$                & 1 & \multicolumn{3}{l}{$P_M= 3m', \  H^1(3m', V/T_L)=\mathbb{I}$} \\
        \multicolumn{3}{l}{\textbf{ACC No.54 }: $312P$} & $R32$       & $R_{2\pi/3}, m_ym_t$         & 1 \\
        \multicolumn{3}{l}{$P_M= 3m', \  H^1(3m', V/T_L)=\mathbb{Z}_3$} & \multicolumn{3}{l}{\textbf{ACC No.63 }: $3mR$} \\
        $P312$       & $R_{2\pi/3}, m_xm_t$          & 1 & \multicolumn{3}{l}{$P_M= 3m, \  H^1(3m, V/T_L)=\mathbb{Z}_2$} \\
        $P3_112$     & $R_{2\pi/3}T_t^{1/3}, m_xm_t$ & $u_1$ & $R3m$       & $R_{2\pi/3}, m_x$            & 1 \\
        $P3_212$     & $R_{2\pi/3}T_t^{2/3}, m_xm_t$ & $u_1^2$ & $R3c$       & $R_{2\pi/3}, m_xT_t^{1/2}$   & $u_1$ \\
        \multicolumn{3}{l}{\textbf{ACC No.55 }: $321P$} & \multicolumn{3}{l}{\textbf{ACC No.64 }: $6'm'mR$} \\
        \multicolumn{3}{l}{$P_M= 3m', \  H^1(3m', V/T_L)=\mathbb{Z}_3$} & \multicolumn{3}{l}{$P_M= 6'm'm, \  H^1(6'm'm, V/T_L)=\mathbb{Z}_2$} \\
        $P321$       & $R_{2\pi/3}, m_ym_t$          & 1 & $R\bar{3}m$ & $R_{\pi/3}m_t, m_x$          & 1 \\
        $P3_121$     & $R_{2\pi/3}T_t^{1/3}, m_ym_t$ & $u_1$ & $R\bar{3}c$ & $R_{\pi/3}m_t, m_xT_t^{1/2}$ & $u_1$ \\
        $P3_221$     & $R_{2\pi/3}T_t^{2/3}, m_ym_t$ & $u_1^2$ &  & &  \\
        \multicolumn{3}{l}{\textbf{ACC No.56 }: $3m1P$} &  & &  \\
        \multicolumn{3}{l}{$P_M= 3m, \  H^1(3m, V/T_L)=\mathbb{Z}_2$} &  & &  \\
        $P3m1$       & $R_{2\pi/3}, m_x$             & 1 &  & &  \\
        $P3c1$       & $R_{2\pi/3}, m_xT_t^{1/2}$    & $u_1$ &  & &  \\
        \multicolumn{3}{l}{\textbf{ACC No.57 }: $31mP$} &  & &  \\
        \multicolumn{3}{l}{$P_M= 3m, \  H^1(3m, V/T_L)=\mathbb{Z}_2$} &  & &  \\
        $P31m$       & $R_{2\pi/3}, m_y$             & 1 &  & &  \\
        $P31c$       & $R_{2\pi/3}, m_yT_t^{1/2}$    & $u_1$ &  & &  \\
        \multicolumn{3}{l}{\textbf{ACC No.58 }: $6'm'mP$} &  & &  \\
        \multicolumn{3}{l}{$P_M= 6'm'm, \  H^1(6'm'm, V/T_L)=\mathbb{Z}_2$} &  & &  \\
        $P\bar{3}1m$ & $R_{\pi/3}m_t, m_y$           & 1 &  & &  \\
        $P\bar{3}1c$ & $R_{\pi/3}m_t, m_yT_t^{1/2}$  & $u_1$ &  & &  \\
        \multicolumn{3}{l}{\textbf{ACC No.59 }: $6'mm'P$} &  & &  \\
        \multicolumn{3}{l}{$P_M= 6'mm', \  H^1(6'mm', V/T_L)=\mathbb{Z}_2$} &  & &  \\
        $P\bar{3}m1$ & $R_{\pi/3}m_t, m_x$           & 1 &  & &  \\
        $P\bar{3}c1$ & $R_{\pi/3}m_t, m_xT_t^{1/2}$  & $u_1$ &  & &  \\
        \bottomrule
    \end{tabular}
    \end{adjustbox}
    \label{tab:trigonal}
    \label{tb:ptrig}
    \label{tb:rtrig}
\end{table}
\clearpage

\clearpage
\subsection{ Hexagonal crystal system}
In $(2+1)$ dimensions, only the temporal direction is left invariant by a sixfold spatial rotation. It therefore corresponds to the principal rotation axis of a 3D hexagonal crystal. Under the identification $z\leftrightarrow t$, the hexagonal space-time groups are in one-to-one correspondence with the 3D hexagonal space groups, with fractional translations along $z$ mapped to temporal translations.

The Bravais lattice of the hexagonal crystal system is the same as the primitive lattice of the trigonal crystal system. 
The basis vectors can be chosen as shown in Fig. \ref{fig:trigonal_crystal_system}(a).
Thus, we use
\begin{equation}
    a_1^{\mathrm{P}}=(l,0,0),\qquad
    a_2^{\mathrm{P}}=\left(-\tfrac12l,\tfrac{\sqrt3}{2}l,0\right),\qquad
    a_3^{\mathrm{P}}=(0,0,t_0).
\end{equation}
These two crystal systems differ in the types of magnetic point groups they admit.
There exist 8 ACCs: $6P$, $31'P$, $61'P$, $6m'm'P$, $6mmP$, $3m_x1'P$, $3m_y1'P$, and $6mm1'P$ (Nos.~65--72), which give rise to 6, 1, 2, 6, 4, 2, 2, and 4 non-equivalent space-time groups, respectively, and are summarized in the Table~\ref{tb:p_hex}.

The maximal magnetic point group for the primitive hexagonal lattice is $6mm1'$, generated by the sixfold rotation $R_{\pi/3}$, the reflection $m_x$, and time-reversal $m_t$. 
The associated ACC is $6mm1'P$ (No.~72). 
The two possible nonsymmorphic operations are the time-glide reflection, produced by combining $m_x$ with half a temporal period, and the time-screw rotation, produced by combining $R_{\pi/3}$ with half a temporal period. 
Time-reversal cannot carry a nontrivial fractional translation. The two cohomology-group generators are
\beaa
&u_{1}^{(72)}[R_{\pi/3}]=0, &&u_{1}^{(72)}[m_x]=\frac{1}{2}a^{\text{P}}_3, &&u_{1}^{(72)}[m_t]=0,\\
&u_{2}^{(72)}[R_{\pi/3}]=\frac{1}{2}a^{\text{P}}_3, &&u_{2}^{(72)}[m_x]=0, &&u_{2}^{(72)}[m_t]=0.
\eeaa
This ACC gives rise to 4 space-time groups.

The magnetic point groups of $6P$ (No.~65), $61'P$ (No.~67), and $6mmP$ (No.~69) are obtained by reducing the symmetry operations from $6mm1'$. 
They are generated by $R_{\pi/3}$, by $R_{\pi/3}$ and $m_t$, and by $R_{\pi/3}$ and $m_x$, respectively. 
For the ACCs of $61'P$ and $6mmP$, the nontrivial fractional translation carried by $R_{\pi/3}$ can only be half a temporal period due to the presence of time-reversal and the reflection symmetry, respectively. The ACC $6mmP$ also allows the time-glide reflection, produced by combining $m_x$ with half a temporal period. For $6P$, removing both $m_x$ and $m_t$ allows $R_{\pi/3}$ to carry 1/6 of a temporal period. Their cohomology-group generators are
\beaa
&u_{1}^{(65)}[R_{\pi/3}]=\frac{1}{6}a^{\text{P}}_3,\\
&u_{1}^{(67)}[R_{\pi/3}]=\frac{1}{2}a^{\text{P}}_3, &&u_{1}^{(67)}[m_t]=0,\\
&u_{1}^{(69)}[R_{\pi/3}]=\frac{1}{2}a^{\text{P}}_3, &&u_{1}^{(69)}[m_x]=0,\\
&u_{2}^{(69)}[R_{\pi/3}]=0, &&u_{2}^{(69)}[m_x]=\frac{1}{2}a^{\text{P}}_3.
\eeaa
The ACCs $6P$, $61'P$, and $6mmP$ give 6, 2, and 4  space-time groups, respectively.

The magnetic point group of $6m'm'P$ (No.~68) is generated by $R_{\pi/3}$ and $m_xm_t$. The sixfold rotation can carry 1/6 of a temporal period, while $m_xm_t$ cannot carry a fractional translation. The cohomology group generator is
\beaa
&u_{1}^{(68)}[R_{\pi/3}]=\frac{1}{6}a^{\text{P}}_3, &&u_{1}^{(68)}[m_xm_t]=0.
\eeaa
This ACC gives rise to 6 space-time groups.

The magnetic point groups of $31'P$ (No.~66), $3m_x1'P$ (No.~70), and $3m_y1'P$ (No.~71) share the threefold rotation $R_{2\pi/3}$ and time-reversal $m_t$. 
They are generated by $R_{2\pi/3}$ and $m_t$, by $R_{2\pi/3}$, $m_x$, and $m_t$, and by $R_{2\pi/3}$, $m_y$, and $m_t$, respectively. 
For the ACC $31'P$, neither $R_{2\pi/3}$ nor $m_t$ can carry a nontrivial fractional translation, so this ACC contains only the symmorphic space-time group. 
For the ACCs $3m_x1'P$ and $3m_y1'P$, the only possible nonsymmorphic operation is the time-glide reflection, produced by combining $m_x$ and $m_y$, respectively, with half a temporal period. Their cohomology-group generators are
\beaa
&u_{1}^{(70)}[R_{2\pi/3}]=0, &&u_{1}^{(70)}[m_x]=\frac{1}{2}a^{\text{P}}_3, &&u_{1}^{(70)}[m_t]=0,\\
&u_{1}^{(71)}[R_{2\pi/3}]=0, &&u_{1}^{(71)}[m_y]=\frac{1}{2}a^{\text{P}}_3, &&u_{1}^{(71)}[m_t]=0.
\eeaa
The ACCs $3m_x1'P$ and $3m_y1'P$ each give 2 space-time groups.

\clearpage
\begin{table}[p]
        \caption{2+1 dimensional space-time groups in the hexagonal crystal
    system. This crystal system contains 8 ACCs and 27 space-time groups. Here
    $T_t^{1/n}$ denotes translation by $a_3^{\mathrm{P}}/n$, where
    $n=2,3,6$.}
    \centering
    \setlength{\extrarowheight}{1.5pt}
    \setlength{\tabcolsep}{4pt}
    \fontsize{7}{8}\selectfont
    \renewcommand{\arraystretch}{1.60}
    \begin{adjustbox}{max width=\dimexpr\linewidth+1em\relax,center}
    \begin{tabular}{@{}lll@{\hspace{3em}}lll@{}}
        \toprule
        \multicolumn{3}{c}{\textbf{Primitive hexagonal: ACCs 65--68}} &
        \multicolumn{3}{c}{\textbf{Primitive hexagonal: ACCs 69--72}} \\[2pt]
        \cmidrule(r){1-3}\cmidrule{4-6}
        \multicolumn{3}{l}{\textbf{ACC No.65 }: $6P$} & \multicolumn{3}{l}{\textbf{ACC No.69 }: $6mmP$} \\
        \multicolumn{3}{l}{$P_M= 6, \  H^1(6, V/T_L)=\mathbb{Z}_6$} & \multicolumn{3}{l}{$P_M= 6mm, \  H^1(6mm, V/T_L)=\mathbb{Z}_2^2$} \\
        $P6$       & $R_{\pi/3}$                  & $1$ & $P6mm$       & $R_{\pi/3}, m_x$                        & $1$ \\
        $P6_1$     & $R_{\pi/3}T_t^{1/6}$         & $u_1$ & $P6_3mc$     & $R_{\pi/3}T_t^{1/2}, m_x$               & $u_1$ \\
        $P6_2$     & $R_{\pi/3}T_t^{1/3}$         & $u_1^2$ & $P6cc$       & $R_{\pi/3}, m_xT_t^{1/2}$               & $u_2$ \\
        $P6_3$     & $R_{\pi/3}T_t^{1/2}$         & $u_1^3$ & $P6_3cm$     & $R_{\pi/3}T_t^{1/2}, m_xT_t^{1/2}$      & $u_1\cdot u_2$ \\
        $P6_4$     & $R_{\pi/3}T_t^{2/3}$         & $u_1^4$ & \multicolumn{3}{l}{\textbf{ACC No.70 }: $3m_x1'P$} \\
        $P6_5$     & $R_{\pi/3}T_t^{5/6}$         & $u_1^5$ & \multicolumn{3}{l}{$P_M= 3m1', \  H^1(3m1', V/T_L)=\mathbb{Z}_2$} \\
        \multicolumn{3}{l}{\textbf{ACC No.66 }: $31'P$} & $P\bar{6}m2$ & $R_{2\pi/3}, m_x, m_t$                  & $1$ \\
        \multicolumn{3}{l}{$P_M= 31', \  H^1(31', V/T_L)=\mathbb{I}$} & $P\bar{6}c2$ & $R_{2\pi/3}, m_xT_t^{1/2}, m_t$         & $u_1$ \\
        $P\bar{6}$ & $R_{2\pi/3}, m_t$            & 1 & \multicolumn{3}{l}{\textbf{ACC No.71 }: $3m_y1'P$} \\
        \multicolumn{3}{l}{\textbf{ACC No.67 }: $61'P$} & \multicolumn{3}{l}{$P_M= 3m1', \  H^1(3m1', V/T_L)=\mathbb{Z}_2$} \\
        \multicolumn{3}{l}{$P_M= 61', \  H^1(61', V/T_L)=\mathbb{Z}_2$} & $P\bar{6}2m$ & $R_{2\pi/3}, m_y, m_t$                  & $1$ \\
        $P6/m$     & $R_{\pi/3}, m_t$             & 1 & $P\bar{6}2c$ & $R_{2\pi/3}, m_yT_t^{1/2}, m_t$         & $u_1$ \\
        $P6_3/m$   & $R_{\pi/3}T_t^{1/2}, m_t$    & $u_1$ & \multicolumn{3}{l}{\textbf{ACC No.72 }: $6mm1'P$} \\
        \multicolumn{3}{l}{\textbf{ACC No.68 }: $6m'm'P$} & \multicolumn{3}{l}{$P_M= 6mm1', \  H^1(6mm1', V/T_L)=\mathbb{Z}_2^2$} \\
        \multicolumn{3}{l}{$P_M= 6m'm', \  H^1(6m'm', V/T_L)=\mathbb{Z}_6$} & $P6/mmm$     & $R_{\pi/3}, m_x, m_t$                   & $1$ \\
        $P622$     & $R_{\pi/3}, m_xm_t$          & 1 & $P6/mcc$     & $R_{\pi/3}, m_xT_t^{1/2}, m_t$          & $u_1$ \\
        $P6_122$   & $R_{\pi/3}T_t^{1/6}, m_xm_t$ & $u_1$ & $P6_3/mmc$   & $R_{\pi/3}T_t^{1/2}, m_x, m_t$          & $u_2$ \\
        $P6_222$   & $R_{\pi/3}T_t^{1/3}, m_xm_t$ & $u_1^2$ & $P6_3/mcm$   & $R_{\pi/3}T_t^{1/2}, m_xT_t^{1/2}, m_t$ & $u_1\cdot u_2$ \\
        $P6_322$   & $R_{\pi/3}T_t^{1/2}, m_xm_t$ & $u_1^3$ &  & &  \\
        $P6_422$   & $R_{\pi/3}T_t^{2/3}, m_xm_t$ & $u_1^4$ &  & &  \\
        $P6_522$   & $R_{\pi/3}T_t^{5/6}, m_xm_t$ & $u_1^5$ &  & &  \\
        \bottomrule
    \end{tabular}
    \end{adjustbox}
    \label{tab:hexagonal}
    \label{tb:p_hex}
    \label{tb:p_hex2}
\end{table}
\clearpage

\section{Real representations for the Floquet-Bloch states}
\noindent
The Floquet--Bloch theorem used in the main text originates from the representation of the space-time (ST) translation subgroup~\cite{xu2018space}.
Since it is Abelian, all of its complex irreducible representations are one-dimensional.
The Floquet-Bloch states at each frequency-momentum point $\kappa=(\mathbf k,\omega)$ transforms according to a one-dimensional representation $D_\kappa$.
For an ST lattice translation $g=(I|T^{n_1}_{ x}T^{n_2}_{ y}T_t^m)$, where $T^{n_1}_{ x}T^{n_2}_{ y}$ denotes a spatial translation by the lattice vector $\mathbf R_n$, for $n = (n_1,n_2)$ and $T_t^m$ denotes a temporal translation by $mT$, where $T$ is the driving period and $\Omega=2\pi/T$ is the driving angular frequency, the representation matrix is $D_\kappa(g)=e^{i\mathbf k\cdot\mathbf R_n-i\omega mT}$.

However, the physical solutions to the classical wave equation form a real vector space.
The action of translation subgroup on the physical solution space must therefore be described by a real representation rather than by a single complex representation $D_\kappa$.
Therefore, the Floquet-Bloch states at $\kappa$ and $-\kappa$, which carry the conjugate representations $D_\kappa$ and $D_{-\kappa}=D_\kappa^*$, must be combined to construct the corresponding real representation space:
\begin{equation}
    \begin{aligned}
        V_{\kappa}^{\mathrm{real}}
          =
        \left\{
        A_\kappa\psi_\kappa
        +
        A_\kappa^*\psi_{-\kappa}
        \;\middle|\;
        A_\kappa\in\mathbb C
        \right\}
          =
        \operatorname{span}_{\mathbb R}
        \left\{
        \operatorname{Re}\psi_\kappa,
        \operatorname{Im}\psi_\kappa
        \right\},
    \end{aligned}
    \label{eq:real_representation_space}
\end{equation}
where $\psi_{-\kappa}=\psi_\kappa^*$.
Only the elements of $V_{\kappa}^{\mathrm{real}}$ correspond to physical real-valued solutions.
The resulting space $V_{\kappa}^{\mathrm{real}}$ is a two-dimensional real representation space, obtained from the original two-dimensional complex coefficient space $\mathbb C^2$ by imposing the reality condition $A_{-\kappa}=A_\kappa^*$.
Writing $\psi_\kappa=\operatorname{Re}\psi_\kappa+i\operatorname{Im}\psi_\kappa$, the action of $g$ on its real and imaginary parts is
\begin{equation}
    \begin{aligned}
        \begin{pmatrix}
            \operatorname{Re}\psi_\kappa \\
            \operatorname{Im}\psi_\kappa
        \end{pmatrix}
        \longrightarrow
        \begin{pmatrix}
            \cos\theta_\kappa & -\sin\theta_\kappa \\
            \sin\theta_\kappa & \cos\theta_\kappa
        \end{pmatrix}
        \begin{pmatrix}
            \operatorname{Re}\psi_\kappa \\
            \operatorname{Im}\psi_\kappa
        \end{pmatrix},
    \end{aligned}
    \label{eq:real_translation_representation}
\end{equation}
with $\theta_\kappa
    =
    \mathbf{k}\cdot\mathbf {R_n}-\omega m T$.
The matrix above is the two-dimensional real representation of the ST translation group associated with the conjugate Floquet--Bloch pair at $\kappa$ and $-\kappa$.

For the degenerate point discussed in the main text, constructing real classical-wave solutions involves the four frequency-momentum points $(\pi,\pm k_y^0,\pm\Omega/2)$.
Since $\Omega/2$ and $-\Omega/2$ are identified at the boundary of the frequency-momentum Brillouin zone (FMBZ), these four points correspond to only two independent frequency-momentum points.
They can be represented by $\kappa_0=(\pi,k_y^0,\Omega/2)$ and its complex-conjugate partner $-\kappa_0$, represented by $(\pi,-k_y^0,-\Omega/2)$, where $k_x=\pi$ and $k_x=-\pi$ are also identified.
The degeneracy discussed in the main text is a twofold degeneracy of the complex Floquet-Bloch problem at $\kappa_0$.
Two linearly independent complex states, $\psi_{1,\kappa_0}$ and $\psi_{2,\kappa_0}$, occur at the same frequency-momentum point and carry the same one-dimensional complex representation $D_{\kappa_0}$ of the ST translation group.
To obtain real classical-wave solutions, each state at $\kappa_0$ must be combined with its complex-conjugate state at $-\kappa_0$.
The resulting real representation of the ST translation group is four-dimensional and consists of the physical fields
\begin{equation}
    \begin{aligned}
        \Phi
        =
        \sum_{a=1}^{2}
        \left(
        A_a\psi_{a,\kappa_0}
        +
        A_a^*\psi_{a,-\kappa_0}
        \right),
        \,
        \psi_{a,-\kappa_0}
        =
        \psi_{a,\kappa_0}^*.
    \end{aligned}
    \label{eq:real_physical_solution}
\end{equation}
The two independent complex amplitudes $A_1$ and $A_2$ provide the four real coordinates of this representation space.

\section{Space-time symmetry properties of heterodyne response tensors}

\noindent
In this section, the time-screw rotation $\mathcal{R}=(R_{\theta}|T_t^{\delta})$ is taken as an example to show how space-time symmetry constrains the response tensor, where $R_{\theta}$ is a rotation around the $z$-axis and 
$\delta$ is a time-shift of a fraction 
of the driving period.
Consider $\mathcal R^h$, for which the rotation angle is $2\pi$. 
Its translation component must then be an integer multiple of the driving period.
It is straightforward to show that $\theta=2\pi /h$ and $\delta= s/h $,
where $h$ and $s$ are integers and need not be coprime. 

The transformation of the response tensor
$\tilde{\sigma}^{\mathrm{ST}}_{ij}(n\Omega,\omega)$ is determined by those of the electric field $\mathbf{E}(t)$
and the induced current $\mathbf{J}(t)$.
Under the time-screw rotation $(R_{\theta}|T_t^{\delta})$, $\mathbf{E}(t)$ transforms according to $E_\alpha'(t)=D_{\alpha\beta}E_\beta(t-\delta T)$ where $D_{\alpha\beta}$ is the rotation matrix of $R_\theta$, and $\mathbf{J}(t)$ transforms in the same way. 
In the frequency domain, we arrive at
$\tilde{E}_\alpha'(\omega)=e^{i\omega\delta\, T}D_{\alpha\beta}\tilde{E}_\beta(\omega)$ and $\tilde{J}_\alpha'(\omega+n\Omega)=e^{i(\omega+n\Omega)\delta T}D_{\alpha\beta}\tilde{J}_\beta(\omega+n\Omega)$.
According to Eq. [\ref{eq:conductivity0}], it yields the transformation law
\begin{equation}
\begin{aligned}
\tilde{\sigma}'^{\mathrm{ST}}_{\alpha\beta}(n\Omega,\omega)
&= D_{\alpha i}(e^{-i n\delta \Omega  T} D_{\beta j})^* \,
\tilde{\sigma}^{\mathrm{ST}}_{ij}(n\Omega,\omega),
\end{aligned}
\label{eq:time_screw_conductivity_transformation}
\end{equation}
where the phase factor, $e^{-i n \delta\Omega T}=e^{-i n s \theta}$, arises from the fractional time translation.
As discussed below, it is interpreted as the quasi-angular-momentum contribution carried by the heterodyne response tensor.

The conductivity tensor $\tilde{\sigma}^{\mathrm{ST}}_{ij}(n\Omega,\omega)$ must remain invariant under the symmetry operation. 
According to Eq. [\ref{eq:circular}], the current density and electric field are decomposed into $\mb{J_{\pm,\omega^\prime}}(t)$ and $\mb{E_{\pm,\omega}}(t)$
, {\it i.e.}, right and left circular components, respectively, where 
$\omega^\prime=\omega+n\Omega$.
For $\mathcal{R}=(R_{\theta}|T_t^{\delta})$, both $R_{\theta}$ and $T_t^{\delta}$ introduce phase factors to $\mb{J_{\pm,\omega^\prime}}(t)$ and $\mb{E_{\pm,\omega}}(t)$, and their combination leaves the conductivity 
tensor invariant. 
Under $R_{\theta}$, $\mb{J_{\pm,\omega^\prime}}(t)$ acquire phases of
$e^{\pm i\theta}$, respectively, and 
so do $\mb{E_{\pm,\omega}}(t)$.
Under $T_t^{\delta}$, the heterodyne
frequency difference generates the phase 
difference of $e^{-in \delta\Omega T}$.
If the electric field and current have the same handedness, then $e^{-i n \delta\Omega T}=1$.
We represent $s/h=s^\prime/h^\prime$ with 
$s^\prime$ and $h^\prime$ coprime, and then the corresponding heterodyne response tensor 
$\tilde\sigma^{\mathrm{ST}}(n\Omega,\omega)$
takes $n=h^\prime p$, with $p$ running over all integer values.
Transforming back to the linear component bases of $\mb{J(E)}_{i,\omega}(t)$, 
we arrive at
\begin{equation}
\begin{aligned}
\tilde\sigma^{\mathrm{ST}}(h^\prime p,\omega) = 
a_{h^\prime p}(\omega)\tau_0+b_{h^\prime p}(\omega)i\tau_2,
\end{aligned}
\end{equation}
where $\tau_0$ and $i\tau_2$ are rotationally invariant, serving as the bases for the longitudinal and transverse responses, respectively. 

For the handedness-flipping response,
the phase difference generated by $R_\theta$ between 
$\mb{J_{\pm,\omega^\prime}}(t)$
and $\mb{E_{\pm,\omega^\prime}}(t)$
must be compensated by the phase generated by $T_t^\delta$.
Again we set $s'/h'=s/h$ with $s'$ and $h'$ coprime.
For the heterodyne response
tensor $\tilde{\sigma}^{\mathrm{ST}}_{ij}(n\Omega,\omega)$, $n$ is determined by the congruence equation $s n \equiv \pm 2 
(\mbox{mod}~h)$.
Solutions exist only when $\gcd(h,s)=1$ or $2$. 
The general solutions are expressed as $n= h^\prime p + r_\pm$ with $p$ taking integer values and 
$r_\pm$ being particular solutions.
Correspondingly, 
\begin{equation}
\begin{aligned}
\tilde{\sigma}^{\mathrm{ST}}
\bigl((h'p+r_\pm)\Omega,\omega\bigr)
&=
a_{h' p +r_\pm}(\omega)
\left(
\tau_3\pm i\tau_1
\right).
\end{aligned}
\end{equation}
The case studied in Eq. [\ref{eq:3p+1}] is an example of 
$s=1$ and $h=3$.


\section{ Group-theoretical proof of momentum degeneracy}

This section provides a proof of the symmetry-enforced degeneracy at the $M$ point for the system 
described by Eq. [\ref{eq:Fourier}].
For this purpose, the appropriate inner product is specified first. 

For electronic Bloch states with the same $\mathbf{k}$, the inner product is restricted to a single unit cell because their common Bloch-wave phase factors cancel:
$\langle \psi_{k} |\psi^\prime_{k} \rangle = \frac{1}{\mathbf{|\Omega|}} \int d\mathbf{r} u^*_{k}(\mathbf{r})u^\prime_{
k}(\mathbf{r})$
with $\bf{|\Omega|}$ the volume of a unit cell and
$u_{k}$ the periodic part of the Bloch wave. 
For classical systems governed by wave equations that are second order in time, we instead use a time-averaged inner product over one space–time unit cell:
\begin{equation}
\langle \psi_{\kappa} |\psi^\prime_{\kappa}\rangle =
\frac{1}{\mathbf{|\Omega|}T}
\int d\mathbf{r}dt u^*_{\kappa}(\mathbf{r},t)\,u^\prime_{\kappa}(\mathbf{r},t),
\label{eq:inner_prod}
\end{equation}
where $T$ is the driving period, and 
$u_{\kappa}$ is the periodic part of the wave function.
Under this definition, it is easy to show that both the time translation and $m_x$ preserve the inner product and are therefore unitary.
Their combination, {\it i.e.}, the time-glide reflection operation, is also unitary. 
The operation $m_t$ we use differs from the conventional anti-unitary time-reversal operation. 
Since both time-reversal and complex conjugation are anti-unitary, their product $m_t$ is unitary.


Consider the system described by Eq. [\ref{eq:EOM}].
The $M$-point with $\omega_0=\Omega/2$ possesses the group of frequency-momentum generated by $(m_x|T_t^{\frac{1}{2}})$ and $m_t$.
Let $\psi_{1,\kappa_0}$ be an eigenstate of $m_t$ 
satisfying $m_t \psi_{1,\kappa_0} =\eta\psi_{1,\kappa_0}$, and define $\psi_{2,\kappa_0}=(m_x|T_t^{\frac{1}{2}}) \psi_{1,\kappa_0}$.
According to Eq. [\ref{eq:comm}] together with $(I|T^{-1}_t)=-1$ at $\omega_0$, it yields 
$m_t \psi_{2,\kappa_0} = -\eta\psi_{2,\kappa_0}$.
Hence, $ \left\langle
\psi_{1,\kappa_0}
\middle|
\psi_{2,\kappa_0}
\right\rangle
=0$, and the two states are therefore linearly independent. 
Moreover, $\psi_{2,\kappa_0}$ is distinct from the complex conjugate partner of $\psi_{1,\kappa_0}$.
These two states with $\kappa_0$ establish the symmetry-enforced degeneracy at the $M$ point.

\section{Numerical solution to the horizontal cone} 

The equation of motion Eq. [\ref{eq:EOM}] can be transformed into a secular equation in the plane-wave basis, based on which the dispersion relation is solved. 
The periodic part $u_{\kappa}(\mathbf{r},t)$ of the 
Floquet-Bloch wave state $\psi_\kappa(\mb{r},t)$ is expanded as $u_{\kappa}(\mathbf{r},t)=\sum_n u_{\kappa}(\mathbf{r},n\Omega)e^{-in\Omega  t}$.
Eq. [\ref{eq:EOM}] becomes 
\begin{equation}
    \begin{aligned}    
    [ H\bigl(\mathbf{k},(m-n) \Omega\bigr) -(\omega+n\Omega)^2\delta_{m,n}]
    \ u_{\kappa}(\mathbf{r},n\Omega)=0,
    \end{aligned}
\label{eq:secular}
\end{equation}
with $H(\mathbf{k},n\Omega) =\frac{1}{T} \int dt H(\mathbf{k} ,t)e^{in\Omega t}$.
For Eq. [\ref{eq:Fourier}], its Fourier components, $H(\mathbf{k},n\Omega)$, are nonzero only for $n=0,\pm1$ and are given by
\begin{eqnarray}
H(\mathbf{k},0) 
&=&(\mu-2z_0\cos{k_y})+m\sigma_3
\nonumber\\
&+&[r_0(1+\cos{k_x})+w_0(\cos{k_y}+\cos{(k_y-k_x)})]\sigma_1 \nonumber \\
&+&[-r_0\sin{k_x}+w_0(\sin{k_y}+\sin{(k_y-k_x)})]\sigma_2, \nonumber \\
H(\mathbf{k},\pm\Omega) &=&  a(1-\cos{k_x})\sigma_1+a\sin{k_x}\sigma_2.
\label{eq:matrix equation}
\end{eqnarray}
The solutions based on Eq. [\ref{eq:secular}] have the following redundancy:  $u_\kappa(\mathbf{r},n\Omega)$ associated with the quasi-frequency $\omega$ is identical to $u_\kappa(\mathbf{r},(n+k)\Omega)$ with $\omega+k\Omega$ for an arbitrary integer $k$.
The Floquet indices are truncated within the range of $-5\le n,m \le 5$, which already contains all the physical solutions. 
The resulting dispersion is shown in Fig.~\ref{fig:momentum_cone}.




\end{appendix}


\end{document}